\documentclass{jfm}
\usepackage{graphicx}
\usepackage{epstopdf, epsfig}
\usepackage{comment}

\graphicspath{{figs/}}

\usepackage{XiGrpCommon}
\definechangesauthor[name={ZL}, color=green]{ZL}
\newcommand{\ZLcomment}[1]{{\footnotesize\color{violet}[\sl\uline{Comment:} #1]}}
\setremarkmarkup{\ZLcomment{#2}}

\renewcommand{\RevisedText}[1]{{#1}}

\shorttitle{VATIP: tracking three-dimensional vortices}
\shortauthor{L. Zhu and L. Xi}

\title{Vortex axis tracking by iterative propagation (VATIP): a method for analyzing three-dimensional turbulent structures}

\author{Lu Zhu\aff{1}
   and
  Li Xi\aff{1,2}\corresp{\email{xili@mcmaster.ca}}
}

\affiliation{
\aff{1}Department of Chemical Engineering, McMaster University, Hamilton, Ontario L8S 4L7, Canada
\aff{2}Kavli Institute for Theoretical Physics (KITP), University of California, Santa Barbara, California 93106-4030, U.S.A.
}

\begin{document}

\maketitle
\begin{abstract}
Vortex is a central concept in the understanding of turbulent dynamics.
\RevisedText{%
Objective algorithms for the detection and extraction of vortex structures can facilitate the physical understanding of turbulence regeneration dynamics by enabling automated and quantitative analysis of these structures.
}%
\RevisedText{%
Despite the wide availability of vortex identification criteria, they 
}%
only label spatial regions belonging to vortices, without any information on the identity, topology, and shape of individual vortices.
This latter information
\RevisedText{%
is stored in the axis-lines lining the contours of vortex tubes.
}%
In this study, a new tracking algorithm is proposed which propagates along the vortex axis-lines and iteratively search for new directions for growth.
The method is validated in flow fields from transient simulations where vortices of different shapes are controllably generated. It is then applied to statistical turbulence for the analysis of vortex configurations and distribution.
\RevisedText{%
It is shown to reliably extract axis-lines for complex three-dimensional vortices generated from the walls.
}%
A new procedure is also proposed that classifies vortices into commonly-observed shapes, including quasi-streamwise vortices, hairpins, hooks, and branches, based on their axis-line topology.
\RevisedText{%
Clustering analysis is performed on the extracted axis-lines to reveal vortex organization patterns and their potential connection with large-scale motions in turbulence.
}%

\end{abstract}


\section{Introduction}\label{Sec_intro}

The dynamics and physics of turbulent flows in wall-bounded geometries have been extensively studied for decades for their fundamental significance and practical implications.
The perpetual extraction of fluid kinetic energy from the mean flow to feed turbulent fluctuations (which is eventually lost to viscous dissipation) is a self-sustaining process, in the sense that continuing external disturbance is not required.
It is thus natural to ask how turbulence regenerates itself in parallel wall-bounded flows where the laminar state is linearly stable until Reynolds number ($\mathrm{Re}$) much (for the cases of Couette and pipe flows, infinitely) higher than the critical magnitude $\mathrm{Re}_\text{crit}$ for turbulence transition~\citep{mullin2011experimental}.
Much progress has been made over the decades but the detailed dynamics remain elusive because of the complexity of turbulent flow fields.
The concept of coherent structures, apparently repetitive flow patterns showing strong coherence in space and time frequently observed in the near-wall region of wall turbulence, is now the basis for understanding the physics of turbulence~\citep{cantwell1981organized,robinson1991coherent,panton2001overview}.
These structures are believed to play a central role in the self-sustaining dynamics and in the turbulent transport of mass and momentum~\citep{brooke1993origin,schoppa2002coherent,marusic2010wall}.

The concept of coherent structure was introduced some 80 years ago\RevisedText{~\citep{corrsin1943investigation,theodorsen1952mechanism,einstein1956viscous}}
and encompasses various types of flow structures.
\RevisedText{%
It is often reflected in well-recognizable patterns in turbulent velocity fields, such as the well-known low- and high-speed velocity streaks in near-wall turbulence~\citep{kline1967structure, offen1975proposed} intricately involved in the turbulence production and regeneration processes~\citep{kim1971production,Jimenez_JFM2018}.
Contributions of velocity variations to the shear component of the Reynolds stress (which describes turbulent momentum transport from the mean flow) are usually quantifiable through the quadrant analysis~\citep{wallace1972wall,willmarth1972structure}.
This approach was recently generalized by \citet{lozano2012three} to analyze three-dimensional flow structures most responsible for the Reynolds shear stress, defined as continuous regions with high $|v_x'v_y'|$ (the apostrophe indicates the fluctuation component of velocities). Therein, wall-attached structures were found to be self-similar in size and display increasing complexity with wall distance.
In addition to this Eulerian perspective, coherent structures are also studied using Lagrangian approaches, in which they are identified as either long-lived events with attracting or repelling materials lines or local maximums of finite-exponent Lyapunov exponents~\citep{haller2001distinguished}. This approach is particularly useful for applications such as mixing and scalar transport.
Given the large number of excellent review articles on the topic~\citep{blackwelder1976wall,robinson1991coherent,panton2001overview,adrian2007hairpin,haller2015lagrangian,Jimenez_JFM2018}, it is not our intention to provide a comprehensive overview of the entire field of coherent structure. Instead, we focus on the vortex structure,  
}%
which has been particularly instrumental in helping researchers conceptualizing turbulent structures and dynamics.
Despite its wide popularity, the concept of vortex is very difficult to precisely define.
Broadly, it describes the general class of revolving flow motions and the axis of fluid rotation is called the vortex axis or center-line: e.g.,
\citet{robinson1991coherent} defined vortex as motions with roughly circular or spiral instantaneous streamlines.
Although it is conceptually intuitive, the intrinsic flaw in this definition is that the topology of streamlines itself is not Galilean-invariant~\citep{Haller_JFM2005}. More precise criteria for vortex identification and how it impacts vortex analysis will be further discussed below.

Understanding how vortices are continuously produced and reproduced is thus the key to the fundamental inquiry into the turbulent self-sustaining dynamics.
Many mechanisms for vortex regeneration have been proposed.
These known mechanisms can be roughly summarized into two major categories according to \citet{schoppa2002coherent}.
In the first category, velocity streaks between streamwise vortices are susceptible to three-dimensional disturbances. This instability leads to the so-called ``break-down'' of the streaks, which, through nonlinear interactions, further feeds the generation of vortices~\citep{hamilton1995regeneration}. This was the basis for the first self-sustaining model for turbulent dynamics~\citep{Waleffe_POF1997} and has led to the discovery of various nonlinear traveling-wave solutions featuring the streak-vortex structure~\citep{Waleffe_PRL1998,Gibson_Cvitanovic_JFM2009}.
Streak breakdown is also found to play a pivotal role in the bypass transition to turbulence~\citep{Brandt_Henningson_JFM2002,Schlatter_Henningson_PoF2008}.
In the second category, as an existing vortex (the ``parent'') lifts up, its rotational motion leads to a strong spanwise shear layer underneath, from which new vortices (``off-springs'') can be generated~\citep{brooke1993origin,bernard1993vortex}.
Our recent study suggested that when sufficient drag-reducing polymer additives are introduced, the streak-instability mechanism can be greatly suppressed, exposing the parent-offspring mechanism as the primary pathway for vortex regeneration in viscoelastic fluids with high polymer elasticity~\citep{Zhu_Xi_JNNFM2018}.

Vortices in near-wall turbulence appear in distinct shapes.
The best-known type is quasi-linear: nearly straight
vortex tubes were frequently observed in experiments~\citep{smith1983observation, kim1971production} and simulations~\citep{bernard1993vortex}. These vortices align mostly along the streamwise direction with their downstream heads sometimes lifting up towards the upper layers. 
Quasi-streamwise vortices have been studied extensively: they are considered to be the dominant structure in the buffer layer ($5\lesssim y^+\lesssim 30$; where superscript $+$ indicates turbulent inner scaling -- see \cref{Sec:Method:DNS})~\citep{robinson1991coherent} and an essential element in both categories of self-sustaining mechanisms reviewed above~\citep{Waleffe_POF1997,hamilton1995regeneration,bernard1993vortex,schoppa2002coherent}.
On the other hand, vortices with more complex three-dimensional configuration\RevisedText{s} are often observed at larger $y^+$, from the log-law layer up to edge of the boundary layer~\citep{robinson1991coherent}. The axis of this type of vortices is often described as \textOmega- or \textLambda-shaped: the top of the arc of \textOmega\mbox{} is a spanwise segment that lifts up from the wall at the downstream end; the two legs extend towards the wall along the streamwise direction at the upstream end.
These so-called ``hairpin'' or ``horseshoe'' vortices were first conjectured in the conceptual model of \citet{theodorsen1952mechanism}. Their observations, in both experimental and numerical studies, remained anecdotal for decades~\RevisedText{\citep{willmarth1967structure,head1981new,perry1982mechanism,smith1984synthesized,adrian2000vortex}} until model hairpin structures were constructed via the conditional sampling of ejection events in direct numerical simulation (DNS)~\citep{adrian1989approximation,adrian1994stochastic}.
Direct evidence for the existence of clearly-shaped and well-organized hairpins in unfiltered statistical turbulence was not reported until fairly recently when \cite{wu2009direct} observed a ``forest'' of hairpins
\RevisedText{%
-- arrays of well-aligned near-perfectly shaped \textOmega\null-shaped vortex objects -- in the DNS of boundary layer flow. Notably, numerical traveling-wave solutions resembling a hairpin -- streamwise vortex pairs coalescing at the lifted-up downstream end -- were recently reported~\citep{Shekar_Graham_JFM2018}.
The \citet{wu2009direct} scenario was later challenged by \citet{schlatter2014near}, who, by analyzing a DNS dataset of boundary-layer flow extending to much higher $\mathrm{Re}$, showed that although the signature of a hairpin forest is clear near the transition to turbulence, hairpin vortices become increasingly insignificant as turbulence further develops.
Complete-shaped symmetric hairpin structures conforming to the canonical \textOmega\null-shape are never predominant in channel flow. Instead, hairpin-like structures are often highly asymmetric (e.g., one-legged) and fragmented, especially at high $\mathrm{Re}$~\citep{morris2007near,dennis2011experimental}.
}%

Compared with the relatively well-studied case of quasi-streamwise vortices, the role of hairpin vortices in turbulent dynamics is much less understood and often debated.
Because of their strong\RevisedText{er} presence in log-law and outer layers, much effort has been invested in unraveling their dynamics and relationship with turbulent self-sustaining cycles \RevisedText{there}~\citep{smith1984synthesized,zhou1999mechanisms,adrian2007hairpin}. 
\RevisedText{%
The most notable model was by \citet{adrian2007hairpin}, which proposed
}%
that hairpin regeneration is achieved through their quick reproduction and the formation of ``hairpin packets''.
\RevisedText{%
This conceptual model is related with \citet{townsend1980structure}'s attached eddy model and the alignment and grouping of hairpin vortex objects offers an appealing explanation to the experimentally observed large-scale motions (LSMs) and very large scale motions (VLSMs) in high-$\mathrm{Re}$ flows~\citep{jimenez1998largest,kim1999very}.
However, whether this picture is sufficient to describe the turbulent regeneration cycles in fully-developed turbulence is still up for debate~\citep{Jimenez_JFM2018}. In particular, it remains to be confirmed if hairpins are essential in the generation of turbulence or are they simply consequences of other primary coherent structures~\citep{del2006self,lozano2014time}?
After all, as noted by \citet{schlatter2014near}, at high $\mathrm{Re}$, it is unlikely that such well-defined structures can persist over the extended time period of their lift-up, without being disrupted by other turbulent motions.
The challenge of depicting a widely-accepted picture of hairpin dynamics is partially attributed to the lack of quantitative information on the evolution and conformation of these structures~\citep{marusic2010wall}.
}%
Compared with quasi-streamwise vortices in the buffer layer, hairpin vortices are not only more complex in shape, at higher $y^+$ they are also submerged in a more complex surroundings and their interaction with nearby structures becomes nontrivial.
A reliable method that objectively detects and extracts these structures from complex turbulent flow fields is required for their detailed statistical analysis.
In addition to the turbulent regeneration mechanism at higher $y^+$ and higher $\mathrm{Re}$, such a method will also be a valuable research tool in other areas.
One example is the bypass transition, where different modes of streak instability and streak interaction can lead to various breakdown pathways driven by different types of vortices~\citep{Schlatter_Henningson_PoF2008,Brandt_deLange_PoF2008,Wu_Moin_PNAS2015}.
Another is turbulent friction drag reduction, where reduced three-dimensional vortices and dominance of extended quasi-streamwise vortices are strongly associated with high levels of drag reduction~\citep{Xi_Graham_PRL2010,Xi_Graham_JFM2012}.

Objective vortex analysis must go beyond direct visual inspection and rely on quantifiable criteria and properly-designed algorithms for vortex auto-detection. Any such approach requires two steps: vortex identification and vortex tracking.
The first step goes back to the definition of a vortex and determines the quantitative criterion for identifying vortex regions in a flow field.
By instinct, one would most likely be drawn to the concept of vorticity $\mbf\omega\equiv\mbf\nabla\times\mbf v$.
However, its fundamental deficiency quickly becomes clear as it does not effectively differentiate 
between pure shear and real swirling flow motions.
Several more rigorous criteria for vortex identification have been proposed, all of which are Galilean-invariant and define vortex regions based on the quantitative magnitude of certain scalar quantities calculated from the flow field, or more specifically, the velocity gradient tensor $\mbf\nabla\mbf v$.
The earliest of them is the $Q$-criterion by \citet{hunt1988eddies}, which defines vortex zones as regions where the second invariant
of $\mbf{\nabla}\mbf{v}$ is positive. (The original \citet{hunt1988eddies} criterion also requires pressure to reach minimum within the vortex region, which is although not identical to the $Q$-criterion but practically equivalent in most cases~\citep{Jeong_Hussain_JFM1995}.)
The corresponding scalar criterion for vortices in incompressible fluid flow is
\begin{equation} 
	Q\equiv\frac{1}{2}(\|\mbf\Omega\|^2-\|\mbf S\|^2)>0
	\label{equ_Q}
\end{equation}
where $\mbf{S}\equiv\left(\mbf{\nabla}\mbf{v}+\mbf{\nabla}\mbf{v}^T\right)/2$ and $\mbf{\Omega}\equiv\left(\mbf{\nabla}\mbf{v}-\mbf{\nabla}\mbf{v}^T\right)/2$ are the rate of strain and vorticity tensors and $\Vert\cdot\Vert$ denotes the Frobenius tensor norm.
Other criteria have been proposed thenceforth.
For example, \citet{chong1990general} defined vortex zones as regions containing complex eigenvalues of $\mbf{\nabla}\mbf{v}$. For incompressible fluids, the corresponding scalar criterion is
\begin{gather}
	\Delta\equiv(R/2)^2+(Q/3)^2>0
\end{gather}
where $Q$ is given by \cref{equ_Q} and $R\equiv-\det(\mbf{\nabla}\mbf{v})$~\citep{chong1990general,chakraborty2005relationships}.
Another is the $\lambda_2$-criterion by \citet{Jeong_Hussain_JFM1995} which defines vortex zones as regions where
\begin{gather}
	\lambda_2\left({\mbf S}^2+{\mbf\Omega}^2\right)<0
\end{gather}
and $\lambda_2(\cdot)$ denotes the second largest eigenvalue of a tensor.
\RevisedText{%
These three criteria are most widely used in the literature and they all serve the same purpose:
}%
turning a velocity field into a scalar field that maps to the strength of vortex motion at different positions in the domain.
Taking the $Q$-criterion for example, $Q>0$ and $Q<0$ correspond to regions dominated by rotation and deformation (extension), respectively~\citep{hunt1988eddies} and a small absolute value of $Q$ ($|Q|\ll\Vert\mbf\nabla\mbf v\Vert^2/2$ according to \citet{Xi_Bai_PRE2016}) reflects simple shear.
\RevisedText{%
Despite their different mathematical origins, for application in real turbulent flows, they are shown to give comparable results with no practically significant differences~\citep{Dubief_Delcayre_JTURB2000,chakraborty2005relationships,chen2015comparison}.
A number of further attempts were made. For example, \citet{zhou1999mechanisms}'s swirling-strength criterion extends the $\Delta$-criterion to include information on the local strength and plane of swirling motions through the imaginary part of the complex eigenvalue of the velocity gradient tensor.
\citet{kida1998identification} developed a kinematic swirling condition to be used together with the pressure minimum criterion which avoids the arbitrariness in the choice of vortex identification threshold common to all major single scalar identifiers.
}%

\RevisedText{%
Choosing a minimum threshold of $Q$, $\Delta$, or $-\lambda_2$ for a given region to be identified as a vortex structure is non-trivial.
The original idea of using $0$ as the threshold would connect nearly all vortex regions into an indistinguishable percolating structure that is nearly impossible to decipher -- a value larger than $0$ is thus required~\citep{jeong1997coherent,blackburn1996topology,chong1998turbulence}.
}%
Obviously, both the size and configuration of the vortex regions identified depend on this threshold (see, e.g., fig.~19 of \citet{Zhu_Xi_JNNFM2018}).
\RevisedText{%
Although some arbitrariness is inevitable, \citet{lozano2012three} have demonstrated (for the quadrant quantity $|v_x'v_y'|$ in their case) that there is a well-identifiable threshold range in which individual structures are separated but not yet overly quenched.
Their so-called percolation analysis works equally well for vortex identifiers such as $Q$~\citep{Zhu_Xi_JNNFM2018}.
Details of this approach, which is also used in this study, will be discussed in \cref{Sec:ParamAnalys}.
Isosurfaces of the scalar identifier at the threshold value show the volumetric shapes of vortex structures. 
Jim{\'e}nez and coworkers have extensively studied the complex three-dimensional vortex structures in high $\mathrm{Re}$ turbulence~\citep{moisy2004geometry,del2006self}. In the case of channel flow, \citet{del2006self} found that using a threshold value (for the $\Delta$-criterion by \citet{chong1990general}) that varies with wall distance $y^+$ can fully reveal the complexity of outer-layer structures which deviate from the classical hairpin shape and are highly branched and often nearly isotropic.
These structures are clearly divided into the wall-attached and -detached classes and the former type shows self-similar dimensions with increasing $y^+$.
\citet{lozano2014time} then proposed an elegant method that, given sufficiently resolved DNS data, is able to track the temporal evolution of volumetric flow structures and document their life-time kinetics. 
}%

\RevisedText{%
Vortex identification criteria generate vortex-containing volumes without differentiating their individual identities (e.g., the analysis of \citet{del2006self} was based on vortex ``clusters'' -- interconnecting vortex regions -- instead of individual vortex objects).
By carefully adjusting the threshold, individual vortices can be visually spotted by direct inspection. However, this information is not easily passed on to a computer program for automated analysis.
The second step of the objective vortex analysis workflow -- i.e., vortex tracking -- is thus needed. 
This step turns volumetric vortex structures into line representations reflecting vortex conformation and topology, in which interconnected line segments represent a complete standalone vortex object.
(Note that in this study the word ``tracking'' refers to the extraction of such line representations from vortex volumes, which is to be differentiated from the temporal tracking of \citet{lozano2014time}.)
These vortex lines enable the direct quantitative measurements of the size, position, orientation, and conformation of vortices and are instrumental in understanding their roles in turbulent dynamics.
}%


Much less development has been made on this front. 
\RevisedText{%
The most intuitive approach is to represent vortices with their axis-lines -- the center line for the swirling motion of fluid elements in each vortex tube.
This is best exemplified by the vortex extraction scheme of \citet{jeong1997coherent} for conditional sampling.
The axis-line of a vortex tube is considered to cut through its each cross-sectional plane at its planar maximum. These two dimensional maximum points are labeled and then connected into the vortex axis-line through a so-called ``cone-detective'' method (see \cref{sec:VATIP}). The \citet{jeong1997coherent} approach was designed for streamwise vortices only, in which the axis-lines are constrained in the streamwise direction.
}%
The method was recently adapted for the conditional sampling of streamwise vortices in viscoelastic flows~\citep{Zhu_Xi_JPhysCS2018}. However, the observations were limited to the changes in the vortex dimension and lifting angle with the addition of drag-reducing polymers. The most important fundamental changes in vortex dynamics, i.e., the suppression of three-dimensional vortices and different vortex regeneration mechanisms, could not be tested because of the restriction of streamwise tracking.
\RevisedText{%
A similar approach was used in \citet{kida1998identification} which extracted the axis-line of each vortex in isotropic turbulence by connecting the two-dimensional pressure minimums (in regions satisfying their swirling condition) within planes that are normal to the direction of vorticity or the third eigenvector of the pressure Hessian matrix.
Tracking of three-dimensional vortex structures in inhomogeneous wall turbulence with line representations was only reported very recently by \citet{hack2018coherent}. They used a ``morphological thinning'' method which gradually trims the vortex volume while preserving its topology, until each tube is reduced to a line.
Different from the direct axis-line tracking approach of \citet{jeong1997coherent} and \citet{kida1998identification}, the \citet{hack2018coherent} approach does not always render vortex axis-lines. Indeed, it is designed to preserve the connectivity of vortex volumes at the line representation level: vortex tubes that have interconnection in their volumes but no intersection between axis-lines -- i.e. interacting vortices that are not topologically connected -- will result in interconnected representation lines.
In another closely related development, \citet{lee2014spatial} proposed and implemented a streak-tracking method -- which extracts line representations of velocity streaks by detecting the ridges in a smoothened surface capturing the velocity structure. Distribution of these ``spine'' lines reveals the spatio-temporal patterns of LSMs and VLSMs. 
}%

In this study, we propose a new algorithm -- vortex axis tracking by iterative propagation (VATIP) -- 
\RevisedText{%
for the axis-line tracking and analysis of three-dimensional vortices in wall turbulence.
}%
The method builds on the initial idea of \citet{jeong1997coherent} for tracking vortex axis-lines by sequentially connecting axis points (thus the word ``propagation'') but extends its target from simple quasi-linear vortex axes to complex three-dimensional configurations representative of generic hairpin-like vortices, including not only the strictly \textOmega- or \textLambda-shaped vortices, but also asymmetric, incomplete, distorted, and highly-branched ones.
For this purpose, the algorithm must also ``iteratively'' grow the propagating axis-line in all three dimensions.  
We will first test VATIP in transient flow fields in which well-organized hairpin vortices are generated in a controlled manner. It is then applied to flow fields of statistical turbulence at several different $\mathrm{Re}$ and the statistics of vortex configuration are analyzed.
In addition to vortex tracking, we also propose a procedure for vortex categorization based on the axis-line topology. Statistics of vortices of different topologies are thus also analyzed.
\RevisedText{%
Access to the detailed information about vortex conformation and position, enabled by the new method, allows us to analyze their clustering patterns, which offers direct insight into the organization of vortices and its potential connection with LSMs.
After presenting all major results, we will examine the robustness of the method with different parameters and settings.
Finally, a major assumption of the method is that vortices can be traced to well-aligned streamwise legs, which applies well to nearly all major vortices in the near-wall layer. However, it no longer holds for complex isotropic structures observed in the outer layer of turbulence at higher $\mathrm{Re}$. This limitation and future development will be discussed at the end.
}%

\section{Formulation and numerical details}\label{Sec_method}
\subsection{Direct Numerical Simulation (DNS)}\label{Sec:Method:DNS}
\begin{figure}
	\centering
	\includegraphics[width=.6\linewidth, trim=0mm 0mm 0mm 0mm, clip]{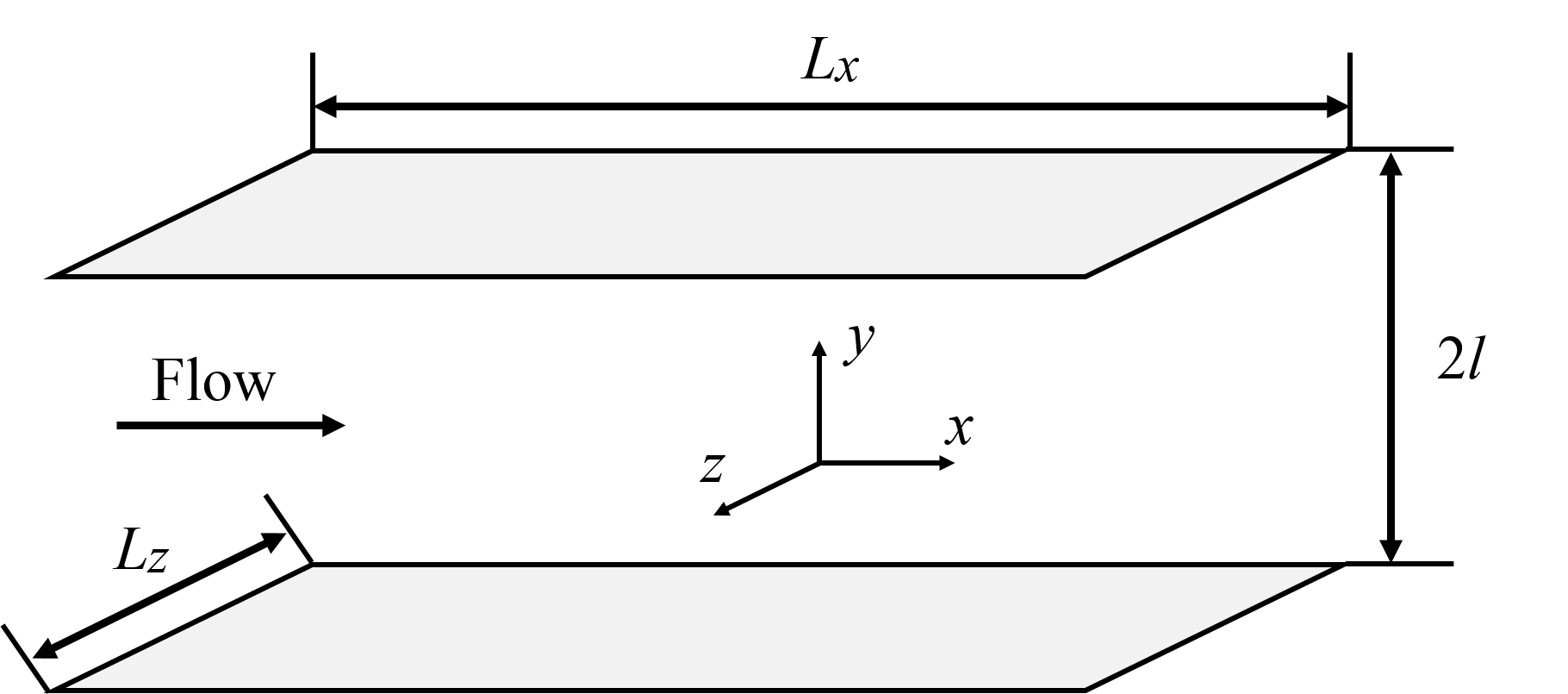}
	\caption{\label{fig:geometry} Conceptual plot of the plane Poiseuille flow geometry.}
\end{figure}
This study focuses on the plane Poiseuille flow. \Cref{fig:geometry} shows the geometry of the simulation domain.
A constant streamwise ($x$-direction) pressure gradient drives the flow between two infinite parallel plates. The periodic boundary condition is applied in the streamwise and spanwise ($z$-direction) directions with the period dimensions represented by $L_x$ and $L_z$. A no-slip boundary condition is applied to the walls in the $y$-direction (wall-normal).
By default, nondimensionalization using turbulent outer scales is applied to all variables: i.e., the half-channel height $l$ is used for the scaling of length, the laminar center-line velocity $U_c$ for velocity, $l/U_c$ for time, and $\rho U_c^2$ for pressure (where $\rho$ denotes the density of fluid).
The Reynolds number is thus defined as $\mathrm{Re}\equiv\rho U_cl/\eta$, where $\eta$ is the viscosity of the fluid.
Turbulent inner scales are used to report results of near-wall flow statistics and structure, for which the friction velocity \RevisedText{$u_\tau\equiv\sqrt{\tau_\text{w}/\rho}$} and viscous length (or wall unit) $\delta_\text{v}\equiv\eta/\rho u_\tau$ are used. Quantities so scaled are denoted with a superscript ``$+$''.
Under these definitions, the friction Reynolds number, $\mathrm{Re}_\tau\equiv\rho u_\tau l/\eta$, can be directly related to $\mathrm{Re}$ through $\mathrm{Re}_\tau=\sqrt{2\mathrm{Re}}$.
The governing equations of momentum and mass balances are
\begin{gather}
	\frac{\partial\boldsymbol{v}}{\partial t}+\boldsymbol{v}\cdot \mathbf{\nabla }\boldsymbol{v}=-\mathbf{\nabla }p+\frac{1}{\mathrm{Re}}\mathbf{\nabla }^2\boldsymbol{v},
	\label{equ_momentum}
	\\
	\mathbf{\nabla }\cdot\boldsymbol{v}=0.
	\label{equ_continuity}
\end{gather}

\begin{table}
	\begin{center}
		\def~{\hphantom{0}}	
		\begin{tabular}{llcccccccc}
			$\mathrm{Re}$	&	$\mathrm{Re}_\tau$	&	$\delta_t$	&	$L_x^+$	&	$L_z^+$	&	$\delta_x^+$	&	$\delta_z^+$	&	$N_y$	&	$\delta_{y,\min}^+$	&	$\delta_{y,\max}^+$ \\[5pt]
			3600	&	84.85	&	0.01	&	4000	&	800		&	9.09	&	5.44	&	97	&	0.046	&	2.81\\	
			14400	&	169.71	&	0.01	&	4000	&	800		&	9.09	&	5.44	&	195	&	0.022	&	2.79\\	
			80000	&	400		&	0.01	&	4000	&	800		&	9.09	&	5.44	&	437	&	0.011	&	3.03\\	
		\end{tabular}
		\caption{Summary of the numerical settings for the DNS of statistical turbulence.}
		\label{tab:numerical}
	\end{center}	
\end{table}

A Fourier ($x$)-Chebyshev ($y$)-Fourier ($z$) pseudo-spectral scheme is adopted for spatial discretization while a third-order semi-implicit backward-differentiation-Adams-Bashforth scheme~\citep{peyret2002spectral} is used for time integration.
DNS has been performed at three different
$\mathrm{Re}$, i.e., $3600$ ($\mathrm{Re}_\tau=84.85$), $14400$ ($\mathrm{Re}_\tau=169.71$), and $80000$ ($\mathrm{Re}_\tau=400$).
A summary of the numerical settings for the DNS of statistical turbulence is provided in \cref{tab:numerical}.
The simulation domain is kept the same in inner units ($L_x^+\times L_z^+$; and thus in outer units both $L_x$ and $L_z$ scale with \RevisedText{$1/\mathrm{Re}_\tau$}). Likewise, the grid sizes in transverse directions $\delta^+_x$ and $\delta^+_z$ are also kept constant in inner units.
The number of grid points in the $y$-direction increases with the $\mathrm{Re}$ to keep the wall-normal resolution approximately the same in inner units.
The numerical solver is implemented in a custom code parallelized with MPI based on the open source \texttt{ChannelFlow} package~\citep{gibson2012channelflow}; the code was first reported in \citet{Tuckerman_Schneider_PoF2014}.

\subsection{Streak Transient Growth (STG) Simulation}\label{Sec:method:STG}
\begin{table}
	\begin{center}
		\def~{\hphantom{0}}	
		\begin{tabular}{lccccccccccccccc}
			$\mathrm{Re}$	&	$\mathrm{Re}_\tau$	&	$\delta_t$	&	$L_x^+$	&	$L_z^+$	&	$\delta_x^+$	&	$\delta_z^+$	&	$N_y$	&	$\delta_{y,\min}^+$	&	$\delta_{y,\max}^+$	&	$\eta$	&	$A_s$	&	$\beta_s$	&	$v_{z,\textrm{rms}}^{\prime+}$	&	$A_p$	&	$\alpha_p$ \\[5pt]
			80000	&	400		&	0.005	&	400	&	200		&	8.33	&	8.33	&	291		&	0.023	&	4.333	&	200	&	4	&	100	&	0.4		&	0.016	&	400\\	
		\end{tabular}
		\caption{Numerical settings and initial condition parameters used for STG simulations.}
		\label{tab:numerical_STG}
	\end{center}	
\end{table}

In statistical turbulence, vortices are often irregular in shape, highly concentrated in space, and intricately positioned relative to (sometimes partially connected with) one another.
Meanwhile, for the initial test of our vortex tracking algorithm, a benchmark system that enables controllable generation of well-defined three-dimensional vortex structures is required.
We adopt the streak transient growth (STG) approach of \citet{schoppa2002coherent} for this purpose, which controls the vortex configuration by adjusting several parameters of the initial condition.
(As another option, one may as well follow the approach of \citet{Brandt_deLange_PoF2008} in which vortices of different configurations are generated from different modes of streak interactions.)

The initial condition for STG is constructed by superposing a base flow with a perturbation velocity. 
The base flow
\begin{gather}
	U_b(y,z)=U_m(y)+U_s(z)g(y), V_b=W_b=0.
\end{gather}
is quasi-two-dimensional ($U_b$, $V_b$, and $W_b$ are the $x$-, $y$-, and $z$-component, respectively) and itself a superposition of the mean velocity profile of statistical turbulence at the same $\mathrm{Re}$
\begin{gather}
	U_m(y)\equiv\int_0^\infty\int_0^{L_x}\int_0^{L_z}v_x(x,y,z,t)dzdxdt
\end{gather}
\RevisedText{%
(where $v_x$ is the instantaneous streamwise velocity component in the statistical turbulence)
}%
with a streamwise velocity streak adjustment $U_s(z)g(y)$.
 The latter is factorized into spanwise and wall-normal dependence terms 
\begin{gather}
	U_s(z)=A_s\cos\left(\beta_s\left(z-z_\beta\right)\right)\text{ and}\\
	g(y)=y\exp\left(-\eta y^2\right).
\end{gather}
Here, $A_s$ adjusts the amplitude of the spanwise undulation, $\beta_s$ adjusts the spanwise streak spacing, $z_\beta$ is the spanwise phase parameter which is set so that the low-speed streak is aligned to the middle of the domain, and $\eta$ is set to align the wall-normal maximum at $y^+=20$.
The perturbation velocity
\RevisedText{%
\begin{gather}
	v_x^\prime=v_y^\prime=0, v_z^\prime=A_p\sin(\alpha_p x)g(y)
\end{gather}
}%
(\RevisedText{$v_x^\prime$, $v_y^\prime$ and $v_z^\prime$} are the $x$-, $y$-, and $z$-component, respectively) adds streamwise dependence to the base flow, without which the instability would not grow~\citep{Waleffe_POF1997}.
Here, $A_p$ is the perturbation amplitude and $\alpha_p$ is the streamwise wave number.

STG parameters used in this study for transient vortex generation, along with the numerical settings of STG simulations, are listed in \cref{tab:numerical_STG}. Note that \RevisedText{$v_{z,\text{rms}}^{\prime +}$} is the root mean square (RMS) magnitude of the spanwise perturbation velocity.
A small simulation domain close to a minimal flow unit (MFU)~\citep{Jimenez_Moin_JFM1991} is used because we only need to focus on a small set of vortex structure for algorithm testing purpose.
Vortices are generated by STG only in half of the channel: i.e., both the streak velocity and perturbation velocity are only applied at the $y<0$ side of the domain while for $y>0$, $g(y)=0$ and the initial velocity is simply $U_\text{m}(y)$.

\section{The algorithm: vortex tracking by VATIP}\label{sec:VATIP}
\begin{figure}
		\centering
		\includegraphics[width=.48\linewidth, trim=1mm 1mm 1mm 1mm, clip]{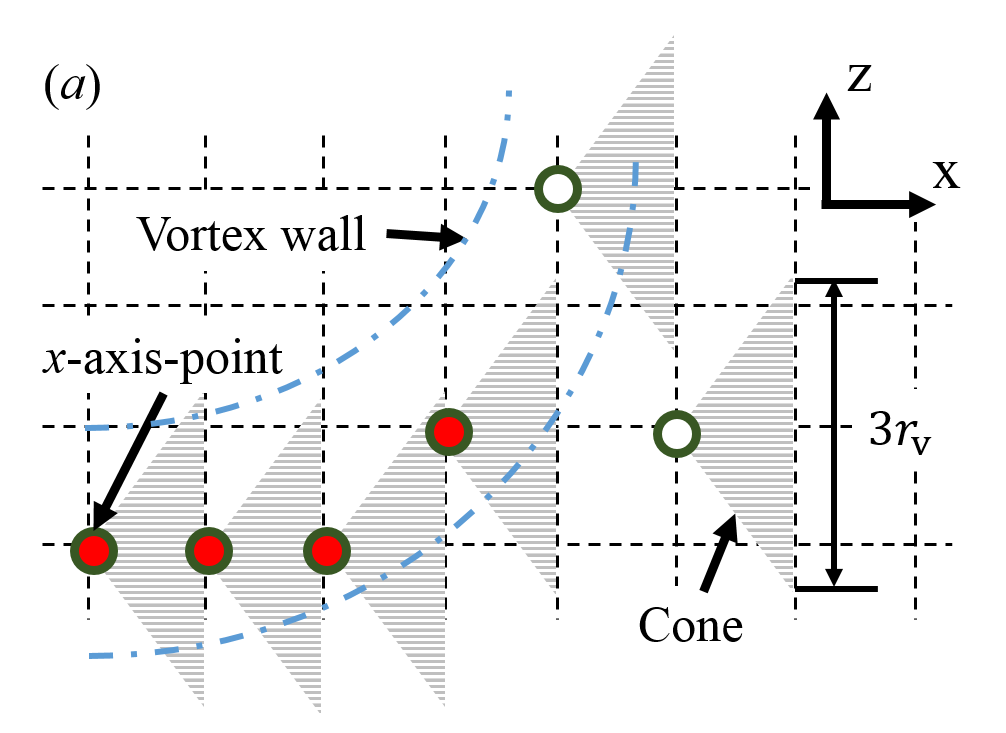}
		\includegraphics[width=.48\linewidth, trim=1mm 1mm 1mm 1mm, clip]{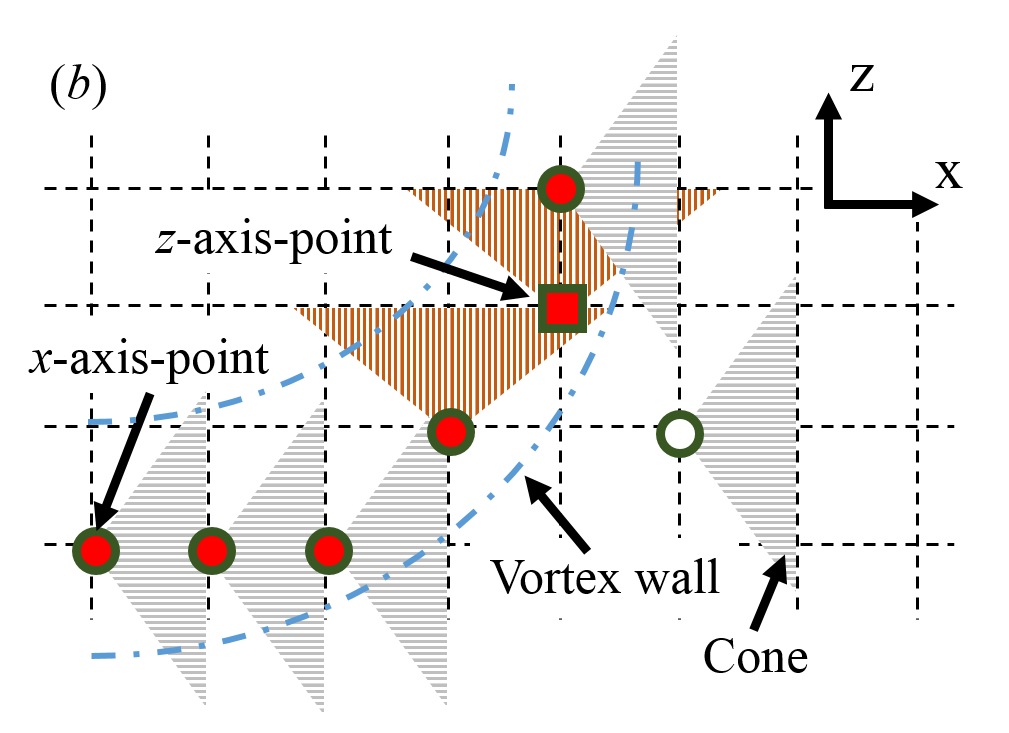}
	\caption{Conceptual plots of the vortex tracking algorithms for (\textit{a}) quasi-streamwise vortices (by \citet{jeong1997coherent}) and (\textit{b}) three-dimensional vortices (by VATIP).}
	\label{fig:track_method}
\end{figure}
 
We first review the original method by \citet{jeong1997coherent} for quasi-streamwise vortex tracking (illustrated in \cref{fig:track_method}(a)).
In their study, the $-\lambda_2$ isosurfaces are used to identify vortex shells in the three-dimensional flow domain and local maxima of $-\lambda_2$ in $yz$-planes are considered to be on vortex axes and labeled as vortex axis-points (circle markers).
The key element of the algorithm is a cone-detective procedure which groups individual vortex axis-points into the axis-lines for different vortices.
Starting from one axis-point, a cone is drawn toward the downstream direction. If another axis-point at the adjacent downstream $yz$-plane is found within the cone, i.e. the $yz$-projection of the vector connecting the two points is shorter than the cone diameter $d_\text{max}$, the two axis-points are grouped to the same vortex (red/solid makers). 
Because the search is limited to $yz$-planar maxima and the tracking cone extends in the downstream direction only, the method is only suitable for vortices staying closely aligned with the $x$-axis.
For significantly curved vortices, the tracking stops as soon as the axis-line steers towards other directions (hollow marker near the top of \cref{fig:track_method}(a)).
 
Building on the idea of extending an vortex axis-line by connecting new points in its direction of propagation, the new VATIP algorithm introduces two major changes to accommodate complex three-dimensional vortices typically observed at larger $y^+$ and higher $\mathrm{Re}$.
First, identification of axis-points goes beyond the $yz$-planar maxima (hereinafter referred to as ``$x$-axis-point'' in which ``$x$'' indicates the primary direction of the vortex axis) and also includes two-dimensional maxima on $xz$- and $xy$-planes ($y$- and $z$-axis-points).
Second (and more substantially), vortex axis propagation is no longer restricted to the $x$ direction and the search must explore all three dimensions iteratively until all possible directions of axis extension are exhausted.
For canonical hairpins, the vortex axis runs from the $x$ (legs), to the $y$ (lift up), and then to the $z$ (the arch) direction.
Other complicated (fragmented or highly branched) vortex configures are also observed, which requires the search algorithm to reexamine the $x$ direction after the $y$ and $z$ searches reach the end(\cref{fig:track_method}\textit{b}).

\begin{table}
	\begin{center}
		\begin{tabular}[t]{p{0.07\textwidth}p{0.39\textwidth}}
			Symbol			&	Description\\[5pt]
			$i$				&	index of planes normal to the search direction\\
			$j$				&	index of candidate points in plane $i$\\
			$j'$			&	index of candidate points in plane $i+1$\\
			$\theta$		&	index of individual vortices\\
			$\theta(j)$		&	vortex containing point $j$
		\end{tabular}
		\begin{tabular}[t]{p{0.07\textwidth}p{0.39\textwidth}}
			Symbol			&	Description\\[5pt]
			$\mathbb{D}$	&	distance set\\
			$D_{jj'}$		&	distance between points $j$ and $j'$\\	
			$D_\text{min}$	&	minimum distance in $\mathbb{D}$\\	
			$E_\theta$		&	propagation point of vortex $\theta$\\
			$\mathbb{P}_i$			&	set containing all candidate points on plane $i$
		\end{tabular}
		\caption{Nomenclature for VATIP flow charts in \cref{fig:track_flowchart}.}
		\label{tab:sym_desc}
	\end{center}	
\end{table}

\begin{figure}
	\centering
	\includegraphics[width=.45\linewidth, trim=0mm 0mm 0mm 0mm, clip]{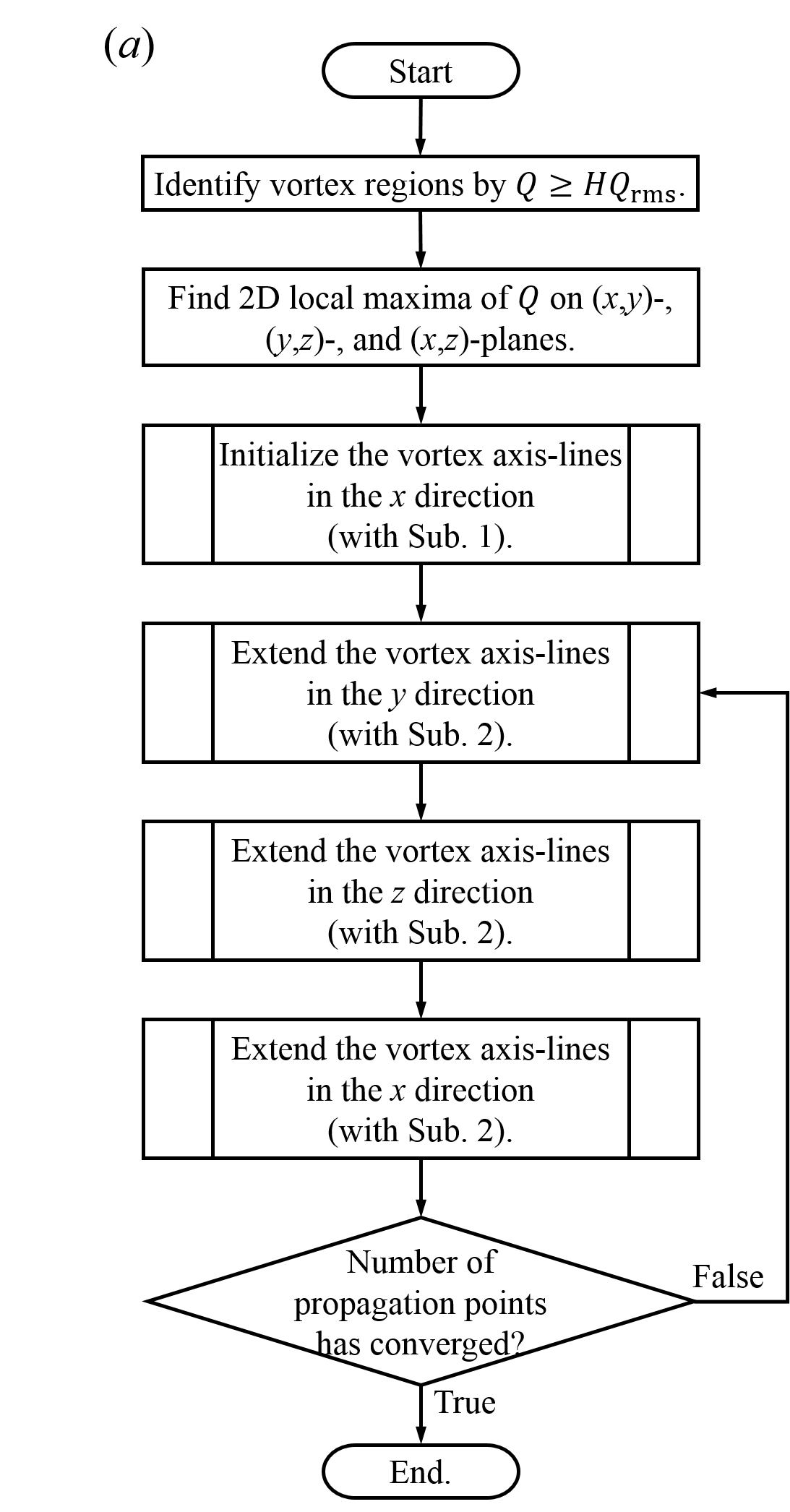}
	\caption{Flow charts for the VATIP algorithm: (\textit{a}) main routine, (\textit{b}) subroutine 1 for the initial tracking in the $x$ \RevisedText{direction}, and (\textit{c}) subroutine 2 for the continued tracking by iterative propagation in all directions. For the last, the loop over planes is unidirectional for the $x$- (downstream) and $y$- (wall$\to$center) directions and bidirectional for the $z$ direction -- see text. (To be continued).}
	\label{fig:track_flowchart}
\end{figure}

\begin{figure}
	\centering
	\ContinuedFloat
	\includegraphics[width=.85\linewidth, trim=0mm 0mm 0mm 0mm, clip]{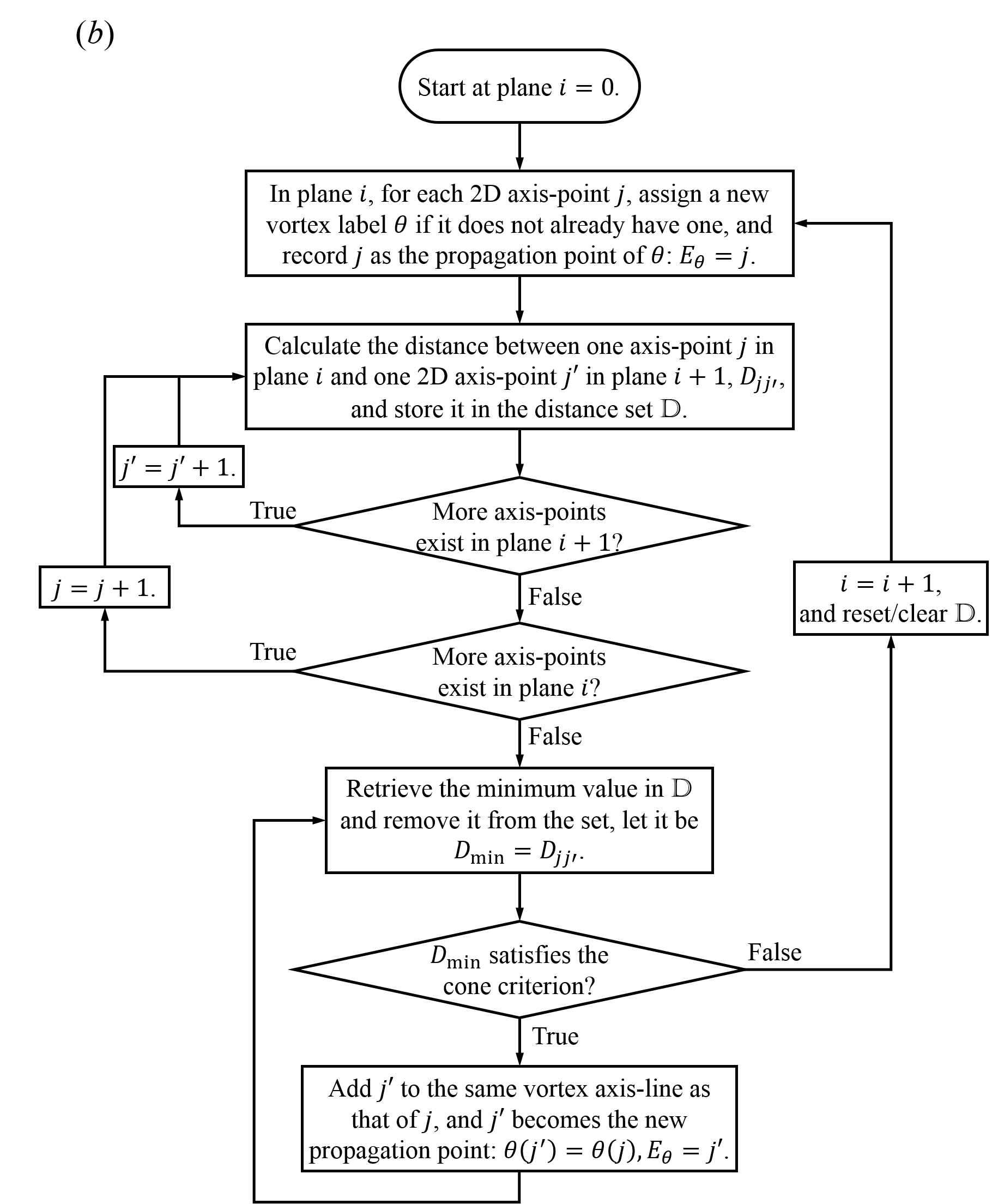}
	\caption{(Continued).}
\end{figure}

\begin{figure}
	\centering
	\ContinuedFloat
	\includegraphics[width=0.9\linewidth, trim=0mm 0mm 0mm 0mm, clip]{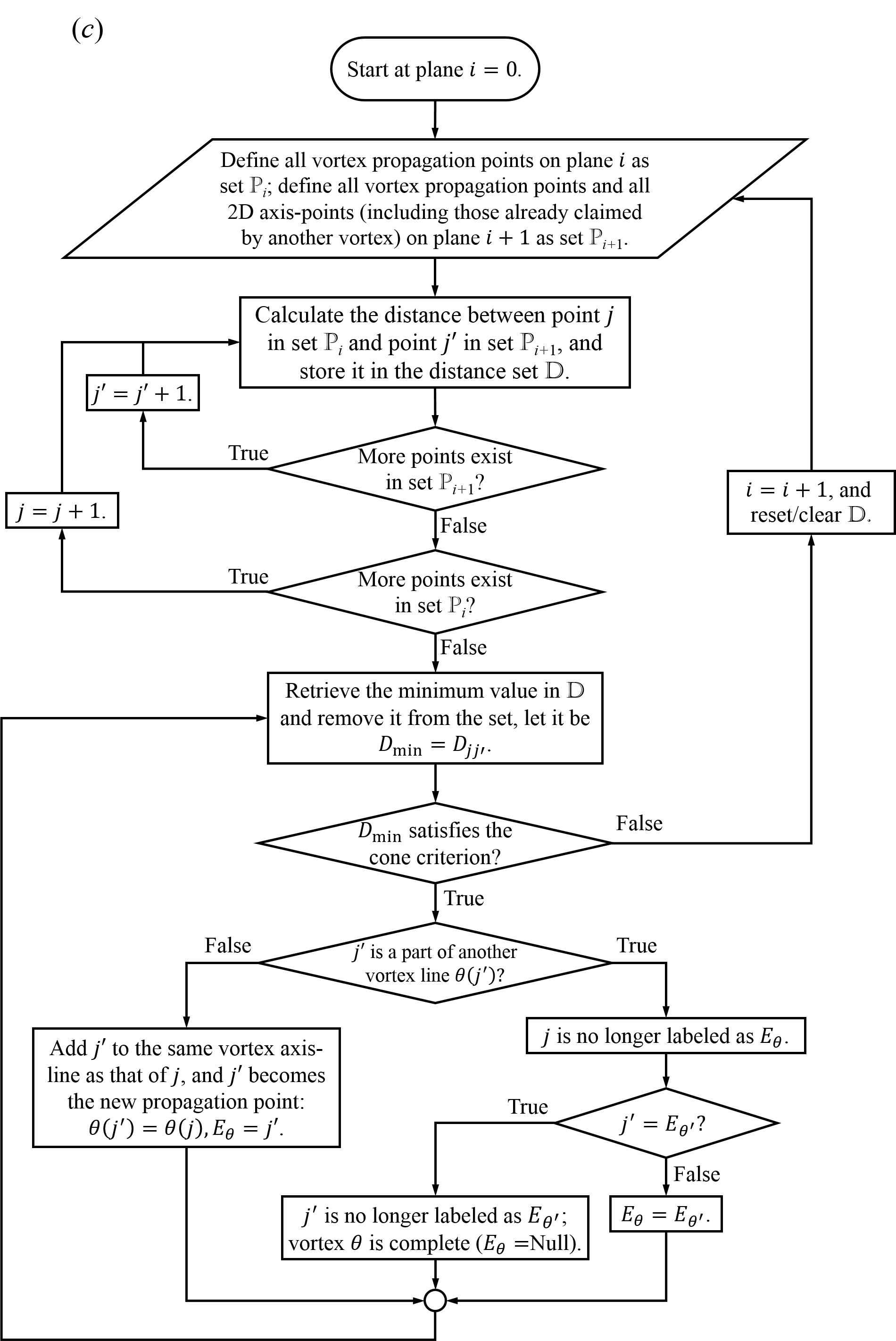}
	\caption{(Continued).}
\end{figure}

The approach of iterative propagation over all three dimensions is thus proposed to allow for more general topologies of vortex axis-lines. The resulting algorithm is much more complex than the original \citet{jeong1997coherent} method.
Flow charts illustrating all detailed steps in VATIP are presented in \cref{fig:track_flowchart} and the symbols used are explained in \cref{tab:sym_desc}.
The main algorithm is illustrated in \cref{fig:track_flowchart}(a) in which two specific subroutines are called: subroutine~1 (\cref{fig:track_flowchart}(b)) is used to initiate the vortex centrelines and subroutine~2 (\cref{fig:track_flowchart}(c)) is used to extend existing centrelines in a new direction.
The latter is repeatedly called in a loop to allow the vortex axis-lines to explore different directions of propagation.

A three-dimensional velocity field is first converted to a scalar field of the vortex identifier using one of the criteria reviewed above in \cref{Sec_intro}.
Without loss of generality, the $Q$-criterion is used here for illustration.
\RevisedText{%
(One may adapt the VATIP algorithm to any other vortex-identification criterion as long as the maximum -- or minimum -- of the scalar identifier marks the vortex axis.)
Regions with $Q$ larger than the threshold value of $0.4Q_\text{rms}$ ($Q_\text{rms}$ is the RMS value of Q; the threshold choice is discussed in \cref{Sec:ParamAnalys}) for statistical turbulence or $1.4Q_\text{rms}$ for STG (a higher threshold is needed because turbulent structures from STG are localized and $Q_\text{rms}$ is diluted by large non-turbulent regions) are selected, within which local maxima in two-dimensional grid planes of all three dimensions are recorded (\cref{fig:track_flowchart}(a)).
}%
Maximum points found on $yz$-, $xz$-, and $xy$-planes are labeled as $x$-, $y$-, and $z$-axis-points, respectively.
\RevisedText{%
Regions with lower $Q$ are not considered
}%
to avoid the interference from small-magnitude fluctuations in $Q$.

These scattered axis-points are connected to form vortex axis-lines through a multistep iterative vortex tracking process.
All axis-lines are initialized with subroutine~1 in the $x$-direction only (\cref{fig:track_flowchart}(b)).
\RevisedText{%
This choice is based on the conceptual model that vortices generated from the walls initially align along the streamwise direction in the buffer layer. Many of them can then lift up at the downstream end which rises into upper flow layers and form hairpins, branches, or other complex configurations.
This model well describes the vortex dynamics in the near-wall boundary layer~\citep{robinson1991coherent,zhou1999mechanisms,panton2001overview} (see \cref{fig:ins_vortex} for example). Consequently, as shown below, VATIP can reliably detect and extract the axis-lines for these vortices.
However, recent advances in the field revealed that vortex structures can also be generated independent of the walls as long as there is sufficient mean shear~\citep{jimenez2013near,del2006self}. These ``detached'' vortex structures are often more isotropic and complex in shape -- it is thus expected that at higher $\mathrm{Re}$ where these structures become more prominent, this bias towards $x$-lying legs will restrict the applicability of VATIP mainly to near-wall regions.
Further discussion about the necessity of this choice and limitations resulting therefrom is deferred to \cref{Sec:limitation}.
}%

Starting from the first $yz$-plane (at $x=0$ and labelled as plane $i=0$;
\RevisedText{%
as shown in \cref{Sec:ParamAnalys}, one can choose to start at any other $yz$-plane, which gives no real difference in the results%
}%
), all $x$-axis-points on the plane are initially assigned different vortex labels $\theta$.
Each growing vortex axis-line must have an open connection point -- referred to as the propagation point -- to which new axis-points can be added. The propagation point of the axis-line of vortex $\theta$ is denoted as $E_\theta$. At the very beginning ($i=0$), since each axis-line only has one point, it is automatically labeled as the propagation point.
For every propagation point on plane $i$, the closest axis-point on plane $i+1$ is found and if their distance is shorter than the slant edge of the cone (\cref{fig:track_method}), the new axis-point is connected to the existing vortex axis-line and designated as its new propagation point.
If an axis-point on plane $i+1$ is eligible for connection with multiple existing propagation points on plane $i$, the closest one is chosen. In practice, this is implemented by first calculating all distances between propagation points on plane $i$ and axis-points on plane $i+1$ and storing the results in a set $\mathbb{D}$.
Potential connections are processed from the shortest distance in $\mathbb{D}$ up to the cutoff distance (cone slant edge length; see \cref{fig:track_flowchart}(b)).
After all eligible connections are made, the process is repeated for the next $yz$-plane. On plane $i+1$, if an $x$-axis-point is not already designated as the propagation point of an existing vortex (in step $i$), it is labeled as the propagation point of a new vortex initiating from plane $i+1$.
All these propagation points on plane $i+1$ are then tested for connection with $x$-axis-points on plane $i+2$ following the same procedure as the previous step.
The iteration continues until all $yz$-planes are processed.
The resulting vortex axis-lines from this step (subroutine~1) is equivalent to the outcome of the \citet{jeong1997coherent} method.

Extension to three-dimensional vortex tracking requires the continuation of the search in other directions after the initial $x$-direction tracking stage.
As shown in \cref{fig:track_flowchart}(a), the search continues in the $y$- and then $z$-direction. This order is chosen considering the typical configuration of hairpin-like vortices (see, e.g., the $t=60$ image of \cref{fig:tran_grow}): the legs of the \textOmega-shaped axis-line align in the $x$-direction and as they extend downstream, the vortex contour lifts up ($y$-direction axis-line) before they merge to form a spanwise arch ($z$-direction axis-line).
\RevisedText{%
However, the vortex does not have to conform to this canonical shape: e.g., for a vortex without a clear lift-up (alighed in the $y$-direction) segment,
}%
the search will continue to the $z$-direction without interruption.
The tracking method for axis-line extension (subroutine~2 and \cref{fig:track_flowchart}(c)) is very similar to that of axis-line initialization (subroutine~1 and \cref{fig:track_flowchart}(b)) with two major modifications.
First, it only extends existing vortex axis-lines by adding to their propagation points and no new vortex will be initiated from any loose axis-point. 
Limiting vortex initiation to subroutine~1 (which is only called before the iteration of search directions) ensures that vortex segments in different directions are only grouped when they are topologically related: e.g., a $y$-segment happens to start where an $x$-segment ends. Planar maximum points of $Q$ that are spatially adjacent but showing no clear topological connection are not included. This is necessary to
\RevisedText{%
minimize false connections in complex flow fluids densely populated with vortex structures. Its impact on the generality of the method will be discussed in \cref{Sec:limitation}.
}%
Second, when the axis-point on the next plane (plane $i+1$) selected for connection is already part of another vortex, these two \RevisedText{vortices} must be properly merged.

For canonical hairpins, the steps of initial tracking (subroutine~1) in the $x$-direction followed by continued tracking (subroutine~2) in the $y$- and then $z$-direction would suffice. In order to capture more general three-dimensional vortex configurations, especially disfigured, highly-branched, and partially merged vortices, the loop containing subroutine~2 over all three dimensions must be continued until the number of vortices (measured by the number of propagation points; \cref{fig:track_flowchart}(a)) has converged.
The specific algorithm of subroutine~2 is nearly identical for different directions with proper adjustment for the directionality: for, e.g., the $y$- (or $z$- or $x$-) direction search, it moves over all $xz$- (or $xy$- or $yz$-) planes and connects $y$- (or $z$- or $x$-) axis-points to the propagating axis-lines.
The only difference is that the vortex axis-line propagation is unidirectional in the $x$- and $y$-tracking and bidirectional in the $z$-tracking.
The $x$-direction propagation proceeds in the flow direction (i.e., plane $i+1$ is immediately downstream of plane $i$) because of the convective asymmetry: vortex structures are always carried downstream by the flow.
\RevisedText{%
The $z$-direction should be statistically symmetric and thus the propagation must sweep both directions.
As shown in \cref{Sec:ParamAnalys}, the VATIP tracking result is practically unaffected by the choice of start planes in these two dimensions, indicating that these sweeping directions can well account for the translational symmetry in $x$ and $z$.
}%
The $y$-direction propagation always starts from the walls towards the channel center (i.e., plane $i+1$ is father away from the wall than plane $i$).
\RevisedText{%
This choice, again, restricts VATIP to wall-generated vortices which generally grow from the buffer layer to higher $y^+$.
}%

\RevisedText{T}he size of the detection cone is determined based on the average cross-sectional radius of vortex tubes.
The average streamwise vortex radius
\begin{gather}
	r_\text{v}=\sqrt{\frac{A_\text{v,total}}{\pi N_\text{v}}}
	\label{Eq:estiRadus}
\end{gather}
is used as the estimated vortex tube size.
Here regions with \RevisedText{$Q>Q_\text{threshold}$} on all $yz$-planes are grouped according to spatial adjacency: for a given $yz$-plane, grid points satisfying the $Q$-criterion that are immediate neighbors are grouped into the same vortex cross-section.
The total area of all vortex regions on these planes $A_\text{v,total}$ divided by the number of separate vortex cross-sections $N_\text{v}$ gives the average cross-sectional area of vortex tubes, from which an average radius is deduced.
In this study, the detection core is chosen so that it extends from plane $i$ to plane $i+1$ with a base (on plane $i+1$) radius of
\RevisedText{%
$1.5r_\text{v}$ (\cref{fig:track_method}). The choice of this parameter will again be examined and discussed in \cref{Sec:ParamAnalys}.
}%
In addition, $r_\text{v}$ is also used as the minimum separation between identified axis-points on each two-dimensional plane. If two or more local $Q$ maxima are separated by less than $r_\text{v}$ on the plane, they are considered to belong to the same vortex tube and the one with higher $Q$ value is kept as an axis-point.

\RevisedText{%
The computational cost of VATIP is negligible compared with DNS. To analyze a typical DNS flow field image in this study (domain size and resolutions are provided in \cref{tab:numerical}), the whole algorithm takes $\approx$\SIlist{100; 370; 1600}{\second} (running as a serial program on an Intel\textsuperscript\textregistered\null E5-2683 v4 \SI{2.10}{GHz} processor) for $\mathrm{Re}_\tau = 84.85$, $169.71$, and $400$, respectively.
To imitate the original algorithm of \citet{jeong1997coherent}, we turned off the whole iteration loop (see \cref{fig:track_flowchart}). The computational time of this streamwise-only search is comparable to that of the full VATIP algorithm.
This is because the calculation of $Q$ field and finding its planar maximum points are both computationally intensive within the program.
For the search and propagation steps, the first $x$-search step (subroutine~1) is also more expensive than the following iterative propagation steps (because of its larger number of distance calculations).
In all cases, the tracking result converges after 3 iterations or less.
The memory requirement of the program is \SIlist{1.02; 2.25; 5.10}{GB} (in the order of increasing $\mathrm{Re}$).
}%

\section{Results and Discussion}\label{Sec_results}

\subsection{Test of VATIP with STG-generated vortices}\label{Sec_STG}

\begin{figure}
	\centering
	\includegraphics[width=.6\linewidth, trim=0mm 0mm 0mm 0mm, clip]{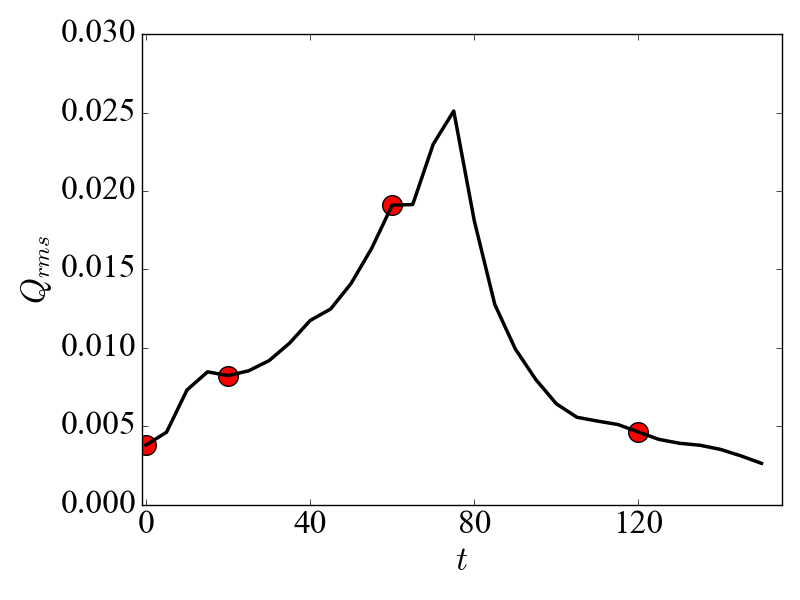}
	\caption{\label{fig:tim_STG} Time series of the root-mean-square of $Q$ in the STG simulation. Moments of the flow fields shown in \cref{fig:tran_grow} are marked with red circles.}
\end{figure}

\begin{figure}
	\centering
	\includegraphics[width=.95\linewidth, trim=0mm 0mm 0mm 0mm, clip]{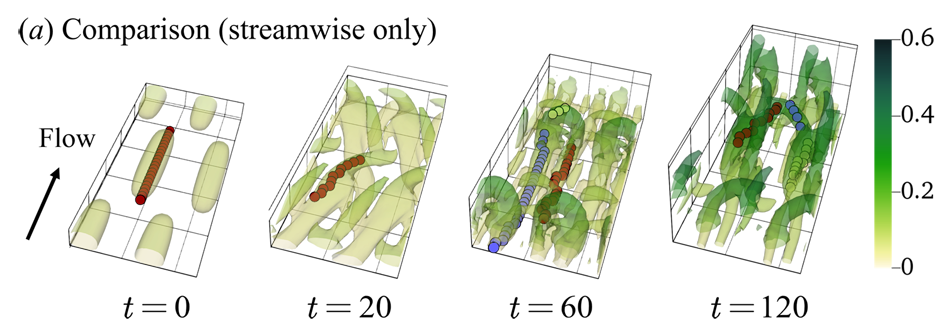}
	\includegraphics[width=.95\linewidth, trim=0mm 0mm 0mm 0mm, clip]{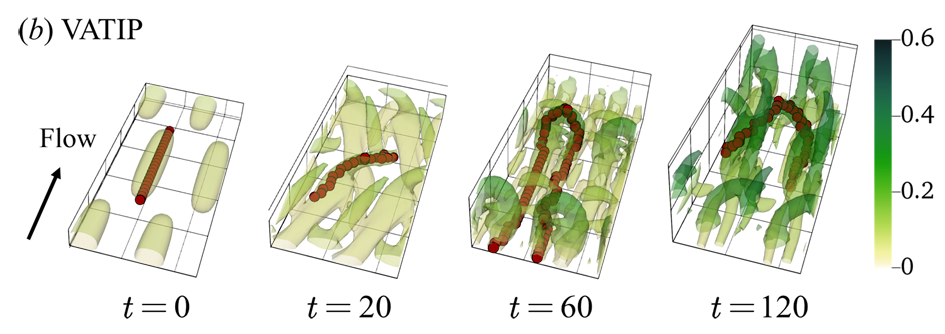}
	\caption{Vortex configurations of selected moments in the STG simulation. The isosurfaces are $Q=0.015$ for $t=0,20,$ and $60$ and $Q=0.01$ for $t=120$. The color scale maps to the distance from the wall in the outer unit. Vortex axis-lines from (a) a streamwise-only tracking approach (equivalent to the \citet{jeong1997coherent} method) and (b) VATIP are compared (circular marks; different colors are  used for different vortices as identified by the method).}
	\label{fig:tran_grow}
\end{figure}

We start by testing the effectiveness of VATIP in STG flow fields, where vortex generation is controllable by the parameters of initial disturbance~\citep{schoppa2002coherent}.
\Cref{fig:tim_STG} shows the time series of the root-mean-square of $Q$ in our STG simulation (numerical settings given in \cref{Sec:method:STG}) and vortex configurations of selected moments are shown in \cref{fig:tran_grow}.
The initial disturbance flow field ($t=0$)  contains strictly streamwise vortices with a spanwise phase shifts between upstream and downstream vortex sets.
At the beginning of STG, the $Q_\text{rms}$ profile starts to grow and reaches the first plateau at around $t=15$.
At this stage, the quasi-streamwise vortices tilt and bend sideways to the spanwise direction but wall-normal lifting up remains small ($t=20$ in \cref{fig:tran_grow}).
After the first plateau, the $Q_\text{rms}$ profile continuously increases and reaches its peak at $t=75$. During this period, neighboring tilted-streamwise vortices lift up and conjoin to form well-defined hairpin vortices ($t=60$ in \cref{fig:tran_grow}). 
The value of $Q_\text{rms}$ gradually decreases after $t=75$. Vortices in this period have a high lifting tendency despite their lower strength ($t=120$ in \cref{fig:tran_grow}). 

Typical vortex configurations in these moments, including the strictly streamwise ($t=0$), titled-streamwise ($t=20$), lifted-up hairpin ($t=60$) and decaying hairpin ($t=120$) vortices, are used as our benchmark systems for vortex tracking.
The original method of \citet{jeong1997coherent} is recovered when the algorithm of \cref{fig:track_flowchart}(a) is truncated right after subroutine~1 (i.e., no iterative propagation in other directions). This would be sufficient if the target was limited to streamwise ($t=0$) or quasi-streamwise vortices (as in the case of \citet{jeong1997coherent}). Its inadequacy starts to surface in titled-streamwise vortices ($t=20$) where the spanwise segment of the vortex is not fully captured in the axis-line obtained (circular markers).
For hairpin vortices ($t=60$ and $120$) not only are some of the axis-points missing (because they are not $yz$-plane maxima of $Q$), the method also breaks the axis-line of a well-defined hairpin into separate pieces.
The new VATIP algorithm successfully identified the complete axis-lines of vortices of all shapes and correctly grouped axis-points of the same vortex into one axis-line.

\subsection{DNS: flow statistics and visualization}

\begin{figure}
	\centering
	\includegraphics[width=.48\linewidth, trim=5mm 0mm 8mm 0mm, clip]{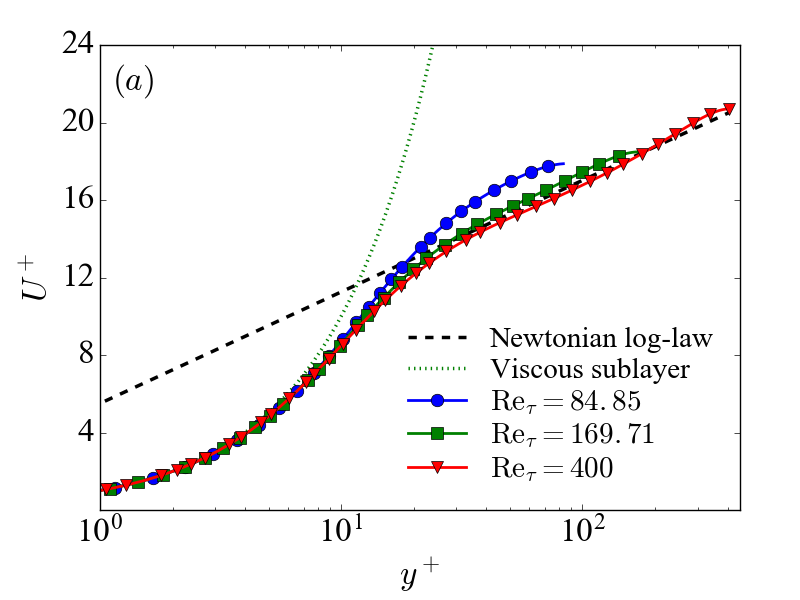}
		\includegraphics[width=.48\linewidth, trim=5mm 0mm 8mm 0mm, clip]{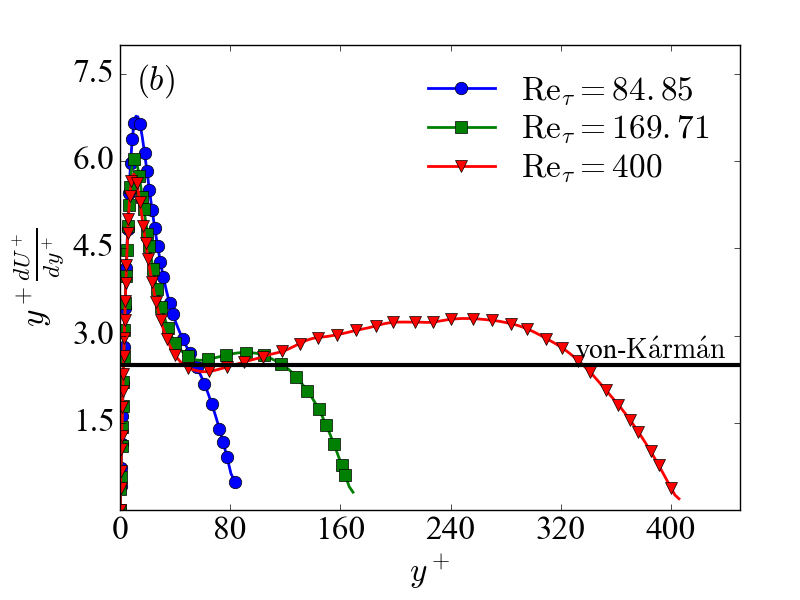}
	\caption{\label{fig:mean_vel} (\textit{a}) Mean velocity profiles ($U^+$ vesus $y^+$) and (\textit{b}) log-law indicator functions ($y^+dU^+/dy^+$ vesus $y^+$) of the statistical turbulence at $\mathrm{Re}_\tau=84.85$, $169.71$ and $400$}
\end{figure}

We now give an overview of the DNS results of statistical turbulence at three different $\mathrm{Re}$ ($\mathrm{Re}_\tau = 84.85$, $169.71$ and $400$) in this section. Application of VATIP to these flow fields will be discussed in \cref{Sec:res:DNSVATIP}.
The mean velocity $U^+$ as a function of $y^+$ is plotted in \cref{fig:mean_vel}(\textit{a}). As $\mathrm{Re}$ increases, the profile outside the buffer layer ($y^+ \ge 30$~\citep{Pope_2000}) gradually approaches the von-K{\'a}rm{\'a}n log-law~\citep{Kim_Moin_JFM1987,Pope_2000}
\RevisedText{%
\begin{gather}
	U^+=2.5\ln y^++5.5.
	\label{Eq:log-law}
\end{gather}
}%
At the lowest $\mathrm{Re}_\tau=84.85$, the profile is slightly higher than the von-K{\'a}rm{\'a}n asymptote, indicating that the log-law layer is not fully developed. The agreement is much better at the two higher $\mathrm{Re}$ and at the highest $\mathrm{Re}_\tau=400$, it nearly completely collapses on to \cref{Eq:log-law} for a wide range of $y^+$ (until the channel center). 

From a generic logarithmic profile
\begin{gather}
	U^+=A\ln y^++B,
	\label{eq:loglawgen}
\end{gather}
the log-law slope can be expressed as
\begin{gather}
	A=y^+\frac{dU^+}{dy^+}.
	\label{eq:indicator}
\end{gather}
When the profile does not strictly follow a logarithmic dependence (\cref{eq:loglawgen}), $A$ becomes a function of $y^+$ -- its variation indicates the departure of the log law.
This quantity (\cref{eq:indicator}), which is thus sometimes referred to as the log-law indicator or diagnostic function~\citep{hoyas2006scaling,marusic2010wall}, is plotted in \cref{fig:mean_vel}(b) for our DNS results.
For the lowest $\mathrm{Re}_\tau=84.85$, the function goes nearly straight down with no discernible flat region, indicating the lack of a well-defined log-law layer (despite that the profile is seemingly parallel to the von-K{\'a}rm{\'a}n asymptote in \cref{fig:mean_vel}(a)).
For the two higher $\mathrm{Re}$ ($\mathrm{Re}_\tau=169.71$ and $400$), an inflection point shows up at $y^+ \approx 50$ with nearly the same value of $2.5$, which
\RevisedText{%
agrees well with the von-K{\'a}rm{\'a}n log-law slope reported in \citet{Kim_Moin_JFM1987} and is also
}%
consistent with the observations of \citet{moser1999direct} and \citet{jimenez2007we}.
After the inflection point, the profile is not strictly flat but its variation is small for a distinct range of $y^+$ ($50\lesssim y^+\lesssim100$ for $\mathrm{Re}_\tau=169.71$ and $50\lesssim y^+\lesssim 320$ for $\mathrm{Re}_\tau=400$),
indicating that these $\mathrm{Re}$ are sufficiently close to and have already shared some common features with fully developed turbulence~\citep{moser1999direct,hoyas2006scaling}.

\begin{figure}
	\centering
	\includegraphics[width=.95\linewidth, trim=0mm 0mm 0mm 0mm, clip]{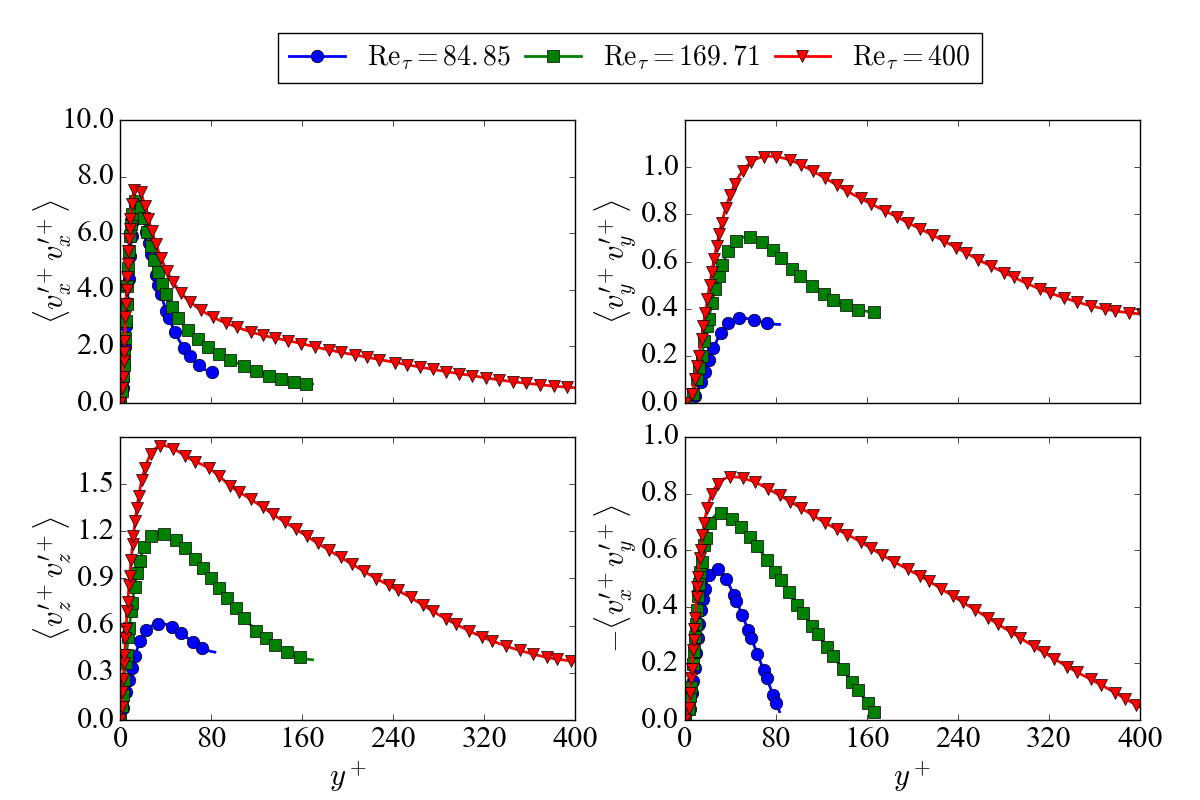}
	\caption{\label{fig:Newt_Re_stress} Reynolds stress profiles for $\mathrm{Re}_\tau=84.85$, $169.71$ and $400$}
\end{figure}

\Cref{fig:Newt_Re_stress} shows the four components of Reynolds stress, $\langle v_x^{\prime+}v_x^{\prime+} \rangle$, $\langle v_y^{\prime+}v_y^{\prime+} \rangle$, $\langle v_z^{\prime+}v_z^{\prime+} \rangle$ and $-\langle v_x^{\prime+}v_y^{\prime+} \rangle$, as functions of $y^+$. 
Consistent with the literature~\citep{abe2001direct,moser1999direct}, the profiles of all components rise with $\mathrm{Re}$ at $y^+$ above $\approx30$ while the peak shifts towards the center of channel. 
$\mathrm{Re}$-dependence is stronger in the transverse components $\langle v^{\prime+}v^{\prime+} \rangle$ and $\langle w^{\prime+}w^{\prime+} \rangle$, which reflects increasing energy redistribution~\citep{abe2001direct}, and the dependence in the streamwise component $\langle u^{\prime+}u^{\prime+} \rangle$ is much weaker.

\begin{figure}
	\centering
	\includegraphics[width=.75\linewidth, trim=0mm 0mm 0mm 0mm, clip]{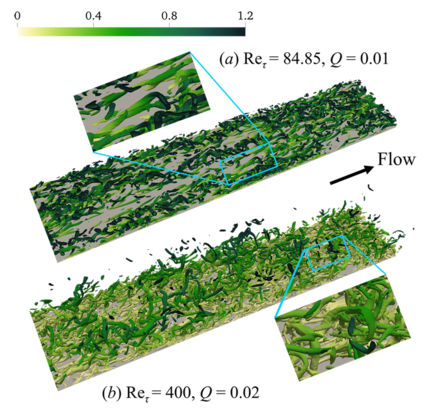}
	\caption{Instantaneous vortex structures at (\textit{a}) $\mathrm{Re}_\tau=84.85$ and (\textit{b}) $\mathrm{Re}_\tau=400$ cases.
	Isosurfaces are identified by the Q-criterion and in the wall normal direction only the bottom half and 20\% of the top half of the channel are shown (i.e., $0<y^+<1.2\mathrm{Re}_\tau$).
	The color shade (from light to dark) maps to the distance from the bottom wall in outer units.}
	\label{fig:ins_vortex} 
\end{figure}

\Cref{fig:ins_vortex} shows the vortex structures identified by the $Q$-criterion in typical snapshots at the lowest ($\mathrm{Re}_\tau=84.85$) and the highest ($\mathrm{Re}_\tau=400$) Reynolds number.
In both cases, the flow fields are filled with tube-like vortices.
Quasi-streamwise vortices are more prevalent in the vortex field, but hairpin vortices can still be observed. Examples of these hairpins are shown in the enlarged views.
Vortices at the higher $\mathrm{Re}$ \RevisedText{displays a high extent of lifting up} and many instances of detached vortices are observed. (A vortex becomes detached when its upstream legs leave the wall and becomes shielded from wall interaction~\citep{perry1995wall,marusic2010wall}).
Meanwhile, most of the vortices at the lower $\mathrm{Re}$ remain attached to the wall with a comparatively weaker \RevisedText{extent} of lifting.

\subsection{\RevisedText{VATIP application in DNS: vortex classification and conformation}}%
\label{Sec:res:DNSVATIP}
Visualization based on the $Q$ field can only provide a cursory glance of the instantaneous vortex fields and lacks both quantitative precision and statistical certainty.
The new VATIP algorithm automatically detects vortices with a variety of shapes without subjective bias.
It thus offers a feasible pathway to the statistical analysis of the population and configurations of vortex structures, facilitating the understanding of their roles in turbulent dynamics.
In this section, VATIP is applied to the DNS results of statistical turbulence. Another algorithm is also proposed to classify vortices according to the topology of the vortex axis-lines identified thereby.
Note that a lower $Q$ threshold of $0.4Q_\text{rms}$ is used for VATIP as discussed previously in \cref{sec:VATIP}.

\begin{figure}
	\centering
	\includegraphics[width=.5\linewidth, trim=0mm 0mm 0mm 0mm, clip]{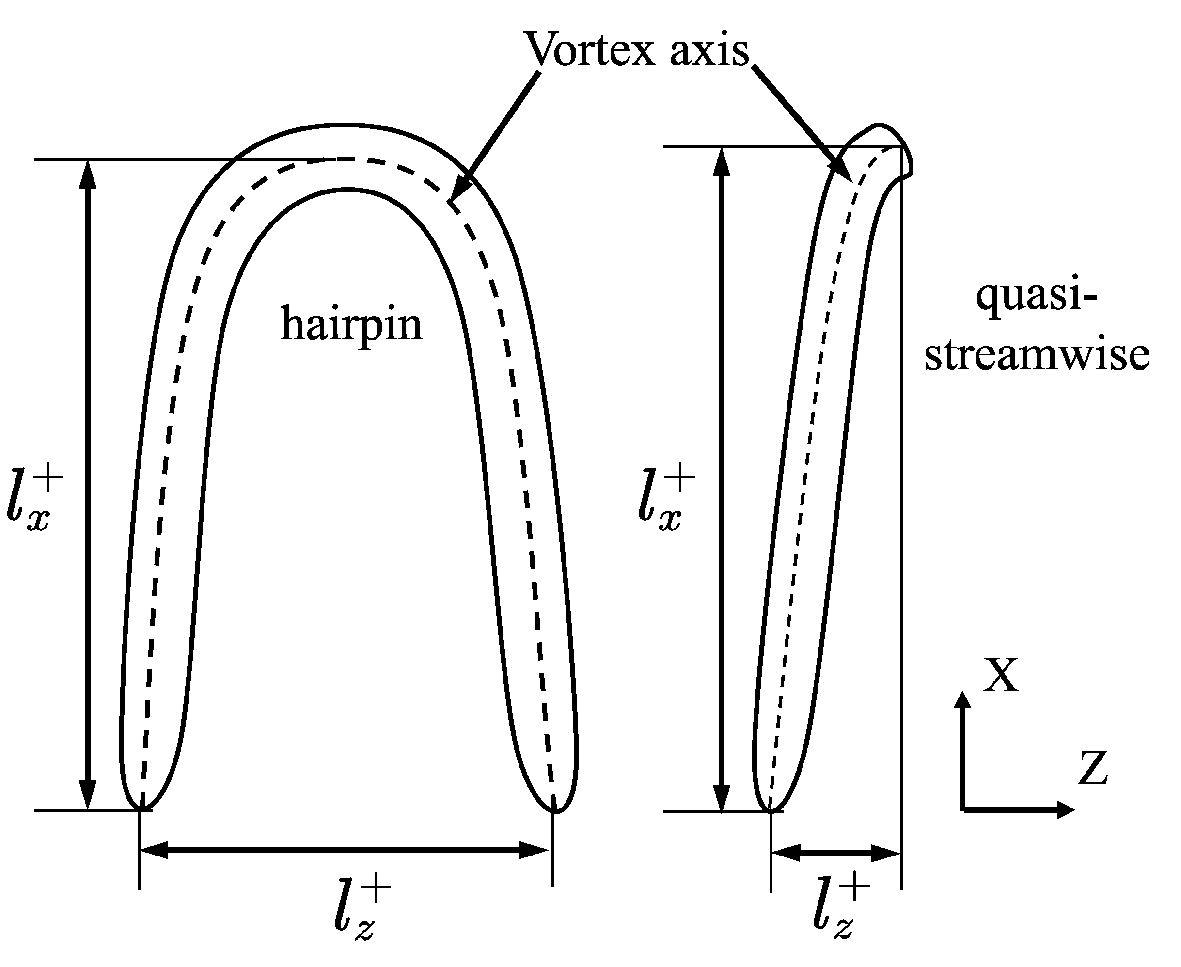}
	\caption{\label{fig:schem_vor_len} Measurements of vortices in $x$ and $z$ dimensions.}
\end{figure}

\begin{figure}
	\centering
	\includegraphics[width=.95\linewidth, trim=0mm 0mm 35mm 8mm, clip]{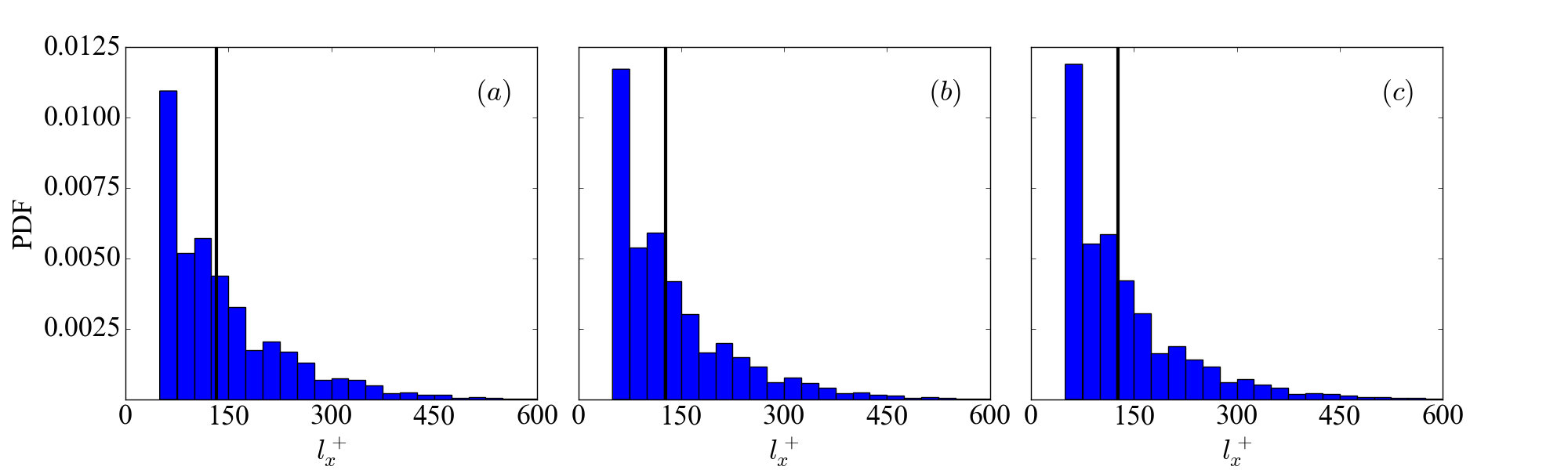}
	\caption{\label{fig:x_length} Probability density function of the streamwise measurement of vortices: (\textit{a}) $\mathrm{Re}_\tau=84.85$, (\textit{b}) $\mathrm{Re}_\tau=169.71$ and (\textit{c}) $\mathrm{Re}_\tau=400$. The vertical line marks the average value.}
\end{figure}

\begin{figure}
	\centering
	\includegraphics[width=.95\linewidth, trim=12.5mm 0mm 35mm 10mm, clip]{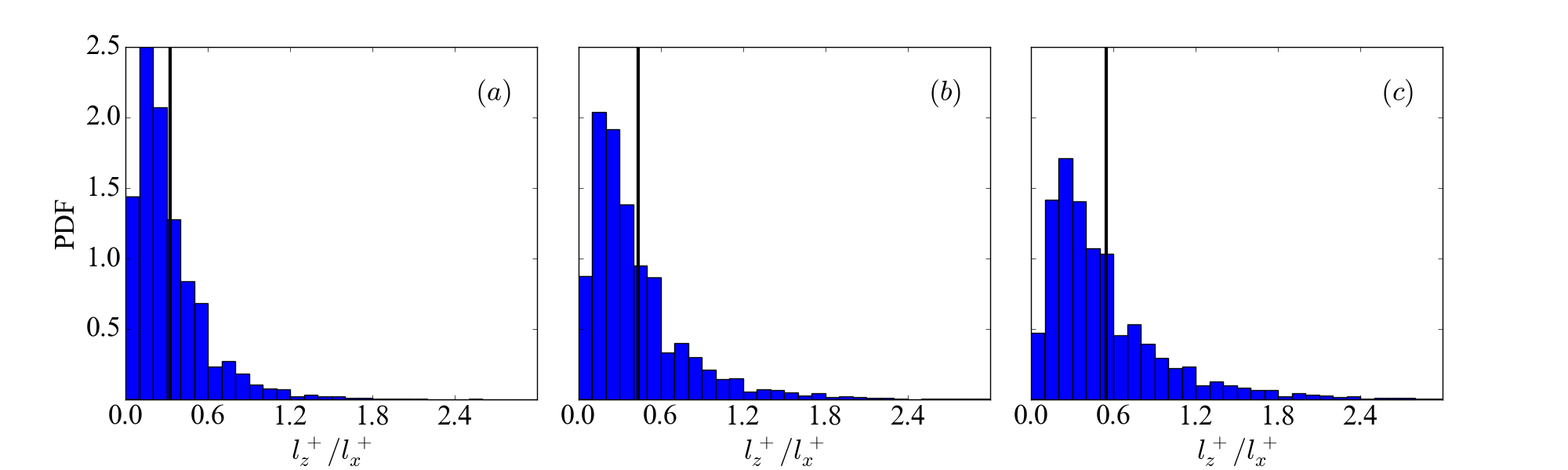}
	\caption{\label{fig:x_z_ratio} Probability density function of the spanwise-streamwise aspect ratio of vortices: (\textit{a}) $\mathrm{Re}_\tau=84.85$, (\textit{b}) $\mathrm{Re}_\tau=169.71$ and (\textit{c}) $\mathrm{Re}_\tau=400$. The vertical line marks the average value.}
\end{figure}

To begin with, vortex size is measured according to \cref{fig:schem_vor_len}: the streamwise and spanwise measurements ($l_x^+$ and $l_z^+$) are defined as the maximal separation between axis-points in these two dimensions, respectively.
The statistical distributions of these measurements are presented in \cref{fig:x_length,fig:x_z_ratio}.
(As discussed below, vortices with $l_x^+<50$ are considered fragments and not included in the statistics.)
At all $\mathrm{Re}$, the \RevisedText{probability density function (PDF)} of $l_x^+$ monotonically decreases with increasing $l^+_x$.
The average $l_x^+$ is about $120$ and is nearly independent of $\mathrm{Re}$. This value is comparable with \citet{jeong1997coherent}'s $200$ (using their streamwise tracking algorithm) and \citet{panton2001overview}'s $100$ (from empirical observation). 
Dependence of this measurement with varying \RevisedText{$Q_\text{threshold}\equiv HQ_\text{rms}$} is rather small: e.g., for $\mathrm{Re}_\tau=84.85$, increasing $H$ from $0.4$ to $1.6$ (well beyond the percolation level), the average $l_x^+$ only decreases from $126$ to $104$. This is well consistent with the earlier (\cref{sec:VATIP}) statement that vortex axis topology is insensitive to the changing $H$ value.

By contrast, $\mathrm{Re}$ has a much stronger effect on the spanwise vortex measurement. The distribution of the vortex aspect ratio $l^+_z/l^+_x$ (\cref{fig:x_z_ratio}) becomes broader and high $l^+_z/l^+_x$ values are more frequently sampled with increasing $\mathrm{Re}$.
The average aspect ratio also increases with $\mathrm{Re}$. 
Because $l_x^+$ is nearly the same, higher $l_z^+/l_x^+$ is solely due to the increasing spanwise measurement of the vortices.
There are two major possible contributions to this increase: (1) streamwise vortices becoming increasingly bent and tilted towards the spanwise direction (see the $t=20$ panel of \cref{fig:tran_grow} for an illustration) and (2) the increasing occurrence of curved and three dimensional vortices such as hairpins.
Quantitative assessment of these changes requires the statistics of vortices of different topologies.

\begin{figure}
	\centering
	\includegraphics[width=.6\linewidth, trim=0mm 0mm 0mm 0mm, clip]{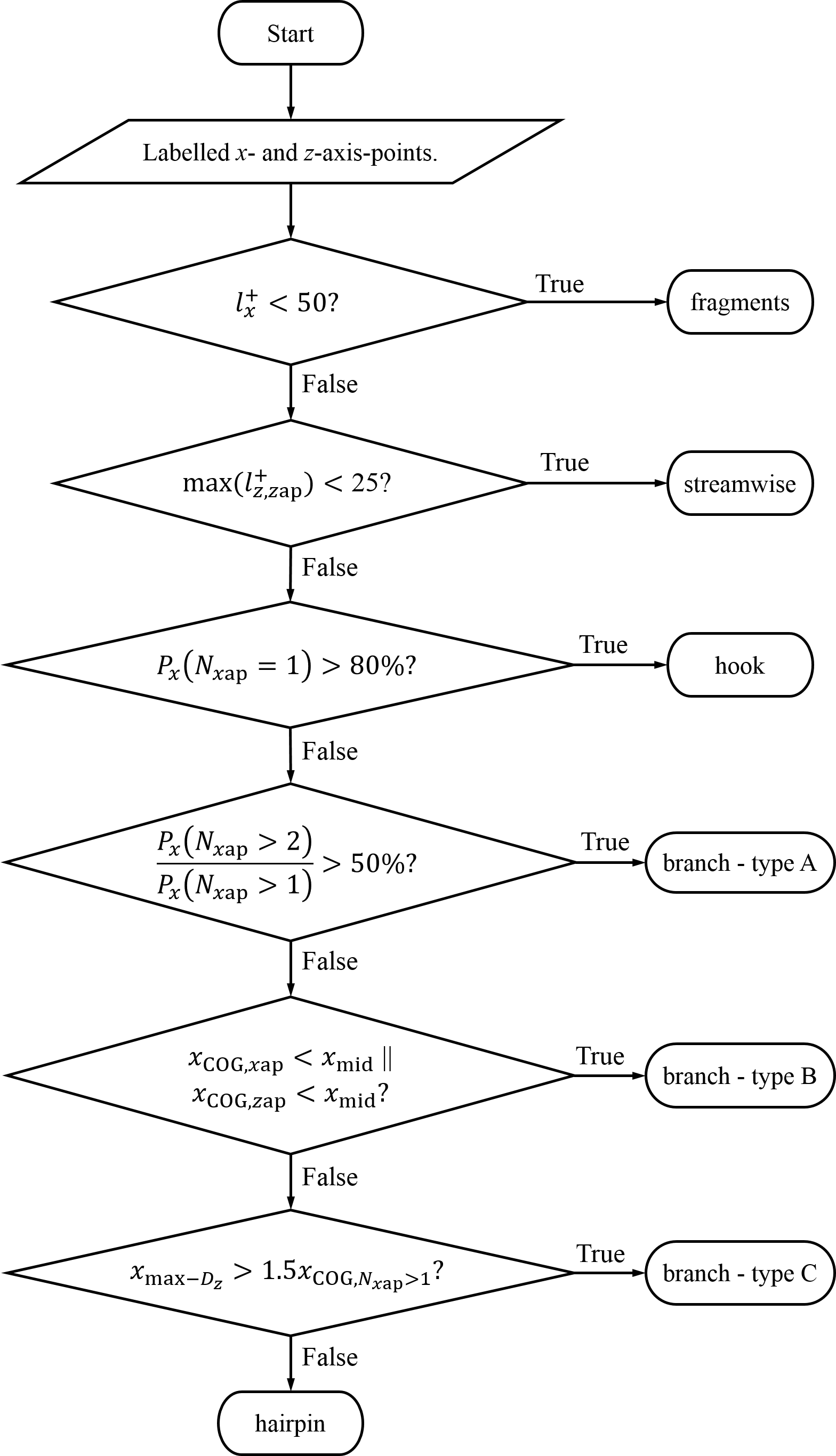}
	\caption{Flow chart of the vortex classification procedure}
	\label{fig:class_flow} 
\end{figure}

\begin{figure}
	\centering
	\includegraphics[width=0.5\linewidth, trim=0mm 0mm 0mm 0mm, clip]{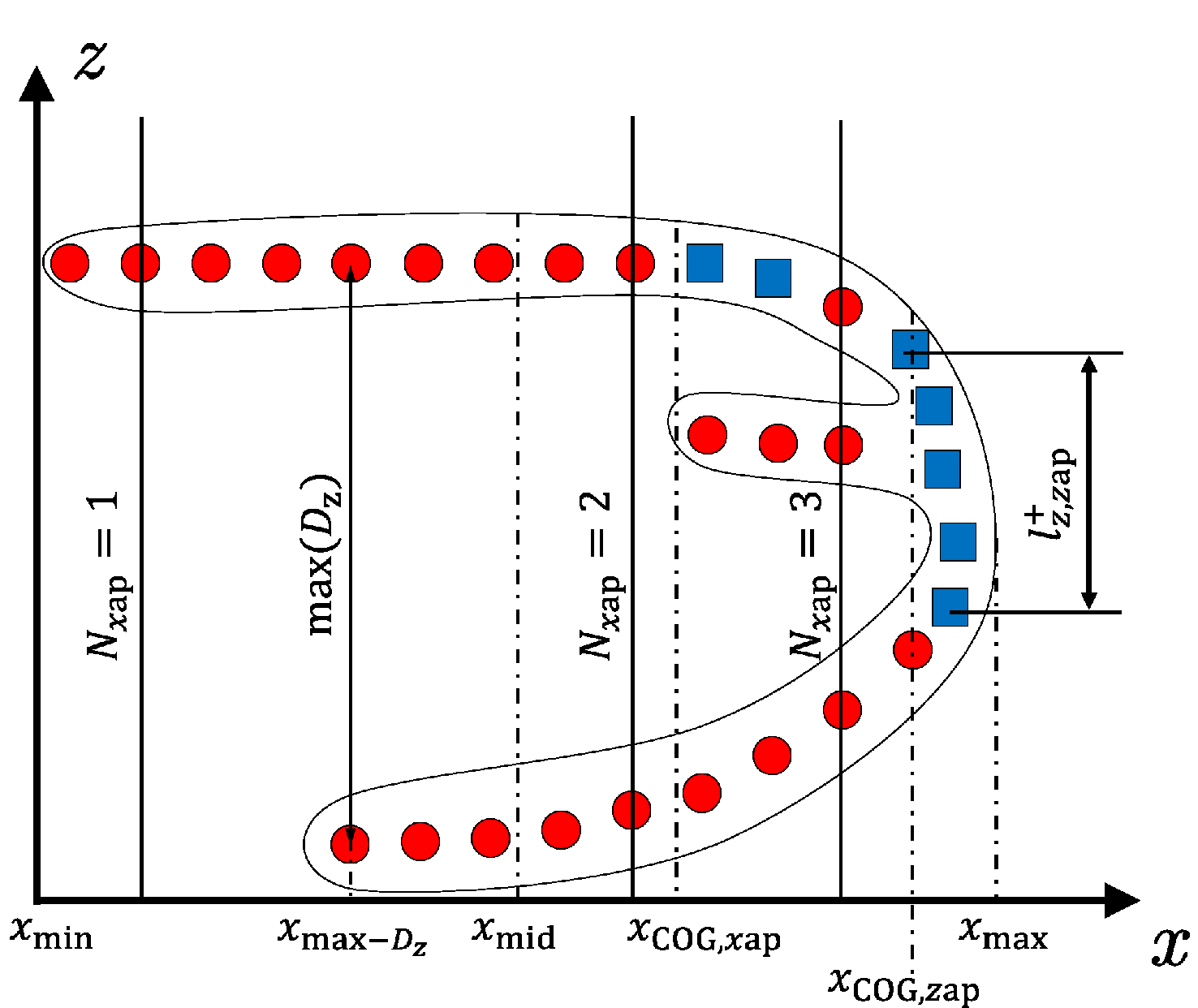}
	\caption{\label{fig:class_para} Definitions of the geometric quantities used in the vortex classification procedure of \cref{fig:class_flow}. Circles and squares represent $x$- and $z$-axis-points, respectively.}
\end{figure}

\begin{figure}
	\centering
	\includegraphics[width=\linewidth, trim=0mm 0mm 0mm 0mm, clip]{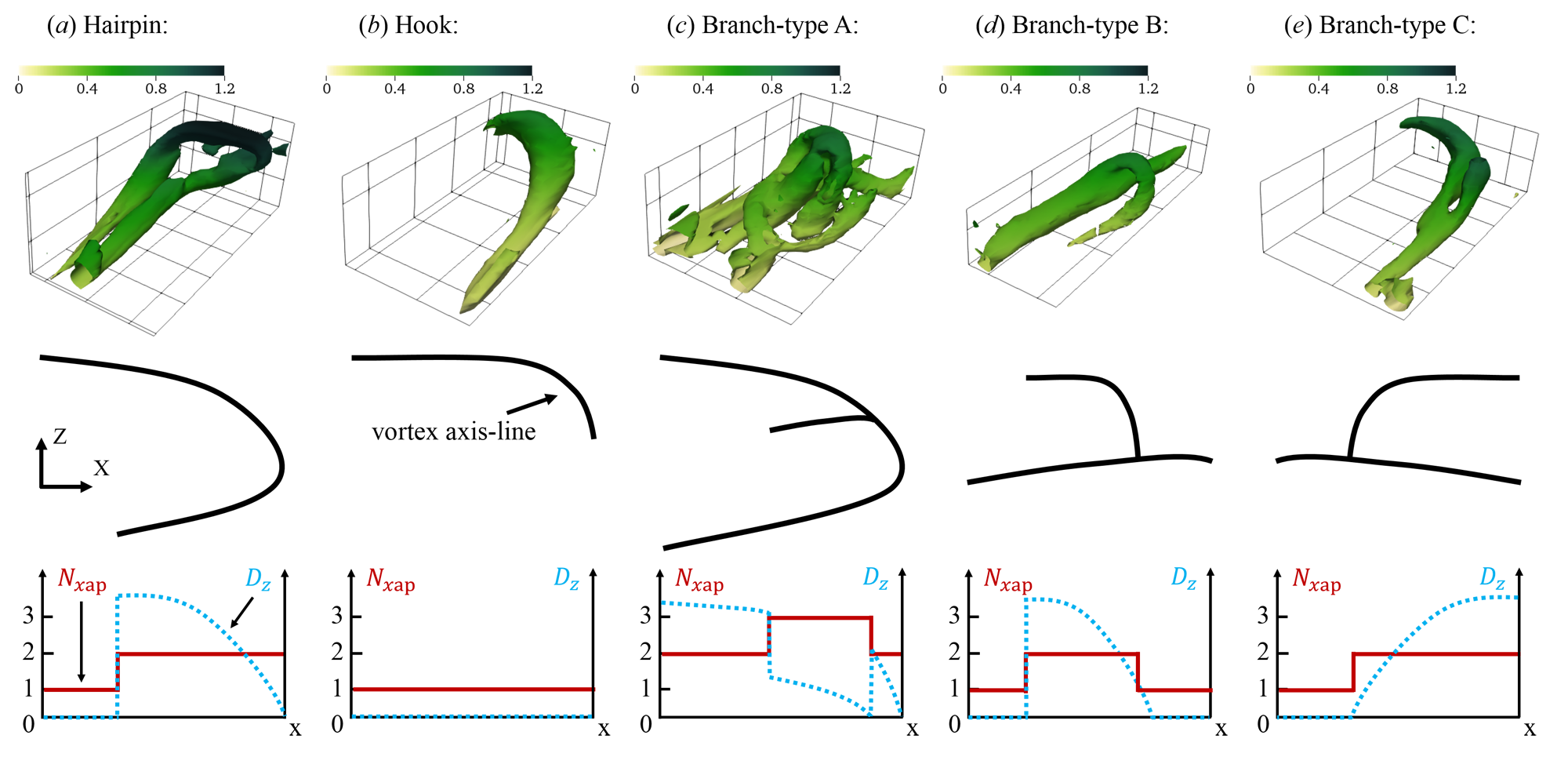}
	\caption{\label{fig:class_Nx}
	Classification of three-dimensional vortices: (a) hairpins, (b) hooks, and (c-e) different types of branches. Top row -- representative examples from DNS; middle row -- schematics of the vortex axis-line; bottom row -- streamwise profiles of the number of $x$-axis-points $N_{x\text{ap}}$ and spanwise span $D_z$.}
\end{figure}

A new procedure is thus proposed to automatically classify the individual vortex axis-lines, obtained from VATIP, according to their dimensions, geometry and topology.
A flow chart of the procedure is provided in \cref{fig:class_flow}; the geometric quantities used in the procedure is shown in \cref{fig:class_para} and typical examples of different types in \cref{fig:class_Nx}.
The procedure consists of a series of binary decisions. First, all axis-lines identified by VATIP are divided into two groups based on the streamwise measurement: those with $l_x^+\geq50$ are considered as clear-cut vortices and smaller pieces are identified as fragments.
This cut-off is smaller than the $150$ wall units used in \citet{jeong1997coherent} because VATIP considers vortices with three-dimensional curvatures and the streamwise dimension does not necessarily account for the full vortex axis length.
Those identified as vortices are further divided into streamwise versus three-dimensional types based on whether a significant spanwise segment can be found in the axis-line. Note that any axis-line identified by VATIP is formed by connecting axis-points in any of the three dimensions (\cref{fig:class_para}). The spanwise spans of all segments consisting of spanwise axis-points only $l^+_{z,z\text{ap}}$ are measured and if the maximum span $\max(l^+_{z,z\text{ap}})\geq 25$, it is determined that the vortex can no longer treated as a streamwise one.

Non-streamwise (or three-dimensional) vortices are further classified into several types based on the axis topology and geometry.
A canonical hairpin is described as a vortex with two largely symmetric streamwise legs conjoining at its downstream head into a spanwise arc (\cref{fig:class_Nx}(a)). Many three-dimensional vortices bear some of the key features of a hairpin but significantly depart from its norm in other aspects. 
The classification procedure relies on two major geometric metrics of the identified vortex axis-line (\cref{fig:class_para}) to differentiate these different types: (1) the number of $x$-axis-points at a given $x$ position $N_{x\text{ap}}$ and (2) the spanwise separation between legs (again) at a given $x$ position $D_z$.
(For irregular vortices with more than two legs, e.g., column (c) of \cref{fig:class_Nx}, $D_z$ is the spanwise separation between the two closest legs.)
Variation of these two metrics with different $x$ positions is sketched for different vortex types in the bottom panels of \cref{fig:class_Nx}.
For a canonical hairpin (column (a)), $N_{x\text{ap}}$ is 2 for the majority of the $x$ range although it may reduce to 1 at the beginning as the legs do not exactly match in length. Its $D_z$ starts high near the leg tips and gradually reduces to 0 as the legs fuse.

For any vortices deemed three-dimensional (versus streamwise) from the previous step, the procedure first checks the percentage of $x$ positions with only one $x$-axis-points $P_x(N_{x\text{ap}}=1)$ ($P_x(C)$ is the percentage of $x$ positions where a specific condition $C$ is satisfied) -- if this quantity is $>80\%$, i.e., for over 80\% of the vortex length it only has one leg, the vortex is a highly asymmetric variant of a hairpin where one of the legs is not clearly developed. This type is termed ``hooks'' in our taxonomy.
The rest vortices have at least two legs, but there are various other branching configurations than the canonical hairpin.
For example, in vortex packets where vortices are highly entangled and dynamically coalescing with one another, multi-legged -- pitchfork-like -- vortices are often observed (\cref{fig:class_Nx}(c)). If a vortex has more regions with three or more legs than those with two, i.e., $P_x(N_{x\text{ap}}>2)/P_x(N_{x\text{ap}}>1)>50\%$, it is identified as a branch type-A.
Even vortices that only branch into two legs may appear significantly different from a canonical hairpin.
For instance, the branch type-B (\cref{fig:class_Nx}(d)) looks more like a fusion between a quasi-streamwise vortex with a partial hairpin (or hook).
This type of vortices was also reported in \citet{robinson1991coherent} and \citet{brooke1993origin} and was believed to result from the spanwise shear dragging a side branch of a quasi-streamwise vortex to form an ``arch'' on its side~\citep{robinson1991coherent}.
The profile of $N_{x\text{ap}}$ for this type shares some similarity with the hairpins, as both start with two legs which gradually merge.
The main difference is that a hairpin ends with the arch where most axis-points are counted, whereas in branch type-B the arch is followed by an extended streamwise segment downstream.
Here, the $x$ projection of the center-of-gravity (COG) of all $x$-axis-points $x_{\text{COG},x\text{ap}}$ and that of all $z$-axis-points  $x_{\text{COG},z\text{ap}}$ are calculated.
A canonical hairpin would be much ``heavier'' at the downstream end, so if both COG's are at the upstream end, i.e., $x_{\text{COG},x\text{ap}}<x_\text{mid}$ and $x_{\text{COG},z\text{ap}}<x_\text{mid}$ ($x_\text{mid}\equiv(x_\text{max}-x_\text{min})/2$ being the $x$ coordinate of the middle point of the vortex axis-line -- see \cref{fig:class_para}), the vortex is classified into branch type-B.
In a similar scenario, when a side branch from a quasi-streamwise vortex protrudes towards the channel center, because of the weaker transverse flows and higher mean velocity, the branch extends substantially downstream before any arch is formed.
This is labeled as branch type-C in this study (\cref{fig:class_Nx}(e)) and identified by the criterion that $x_{\text{max-}D_z}>1.5x_{\text{COG},N_{x\text{ap}}>1}$, where $x_{\text{max-}D_z}$ is the $x$ coordinate of the maximal branch separation $D_z$ and $x_{\text{COG},N_{x\text{ap}}>1}$ is the $x$ coordinate of the COG of of the branched part of the vortex axis-line (where $N_{x\text{-cp}}>1$).
Finally, the remaining vortices -- i.e., those predominated by two legs and with no substantial quasi-streamwise downstream segments -- are classified as hairpins.

The criteria used in this classification procedure are mostly empirical.
For starters, there is no physical ansatz supporting the classification of three-dimensional vortices into these five particular types listed in \cref{fig:class_Nx} -- they are chosen solely based on their empirical observations in our and previous studies.
Likewise, the dividing criteria and cutoff magnitudes used in the procedure (\cref{fig:class_flow}) are all chosen based on a combination of physical intuition and practical experience.
For example, there is no physical basis as to how long a third leg needs to reach for a vortex to be considered a branch type-A (multi-legged) rather than a slightly modified hairpin.
Indeed, the question of whether there is any fundamental difference between various types of branches and the canonical hairpins itself cannot be answered.
The lack of objective vortex classification criteria is an inevitable consequence of the current limited knowledge of the complex vortex dynamics in wall turbulence.
It is for this reason that an algorithm like VATIP is much needed. Future application of VATIP to a wider range of flow systems is anticipated to bring forth better experience and understanding of the characteristics of turbulent vortices, which will lead to a more standardized approach of vortex classification.
Finally, we note that any vagueness in the current classification criteria does not affect the validity of any of the following discussion: e.g., the changes of all three-dimensional vortices show similar $\mathrm{Re}$ dependence (\cref{fig:vor_num}) regardless of the further differentiation between hairpins and different branch types.
In addition, from our test, changing the cut-off magnitudes by up to 50\% does not affect the comparison of vortex statistics between different $\mathrm{Re}$.

\begin{figure}
	\centering
	\includegraphics[width=.8\linewidth, trim=0mm 0mm 0mm 2mm, clip]{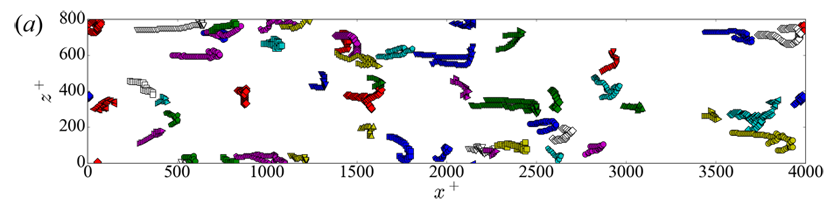}
	\includegraphics[width=.8\linewidth, trim=0mm 0mm 0mm 2mm, clip]{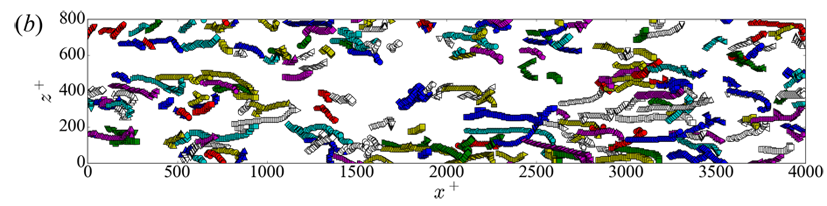}
	\includegraphics[width=.8\linewidth, trim=0mm 0mm 0mm 2mm, clip]{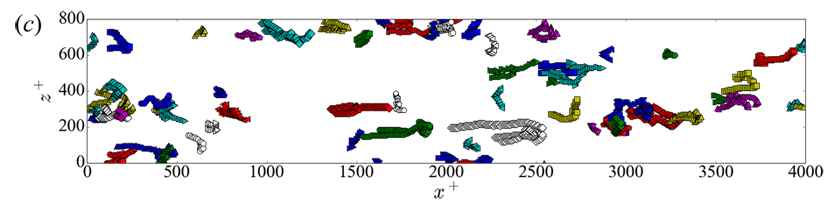}
	\includegraphics[width=.8\linewidth, trim=0mm 0mm 0mm 2mm, clip]{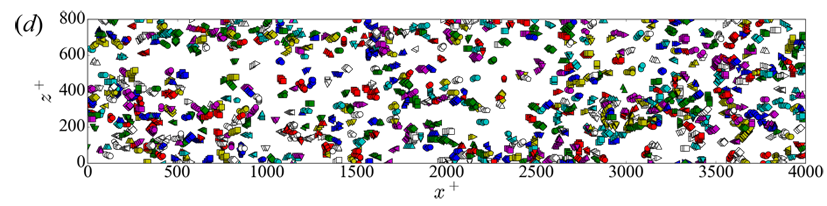}
	\includegraphics[width=.8\linewidth, trim=0mm 0mm 0mm 2mm, clip]{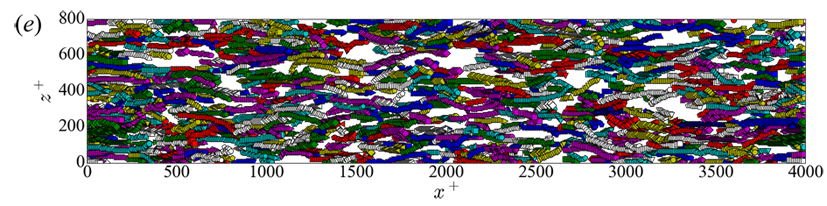}
	\caption{Distribution of vortex axis-lines of different classes in a typical snapshot at $\mathrm{Re}_\tau=84.85$: (\textit{a}) hairpins, (\textit{b}) hooks, (\textit{c}) branches (all types), (\textit{d})fragments, and (\textit{e}) streamwise vortices. Each marker represents one axis point. Individual vortices are differentiated by colors and marker types.}
	\label{fig:vor_class_visu1}
\end{figure}

\begin{figure}
	\centering
	\includegraphics[width=.8\linewidth, trim=0mm 0mm 0mm 2mm, clip]{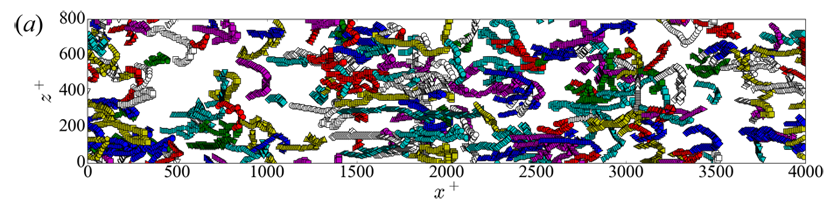}
	\includegraphics[width=.8\linewidth, trim=0mm 0mm 0mm 2mm, clip]{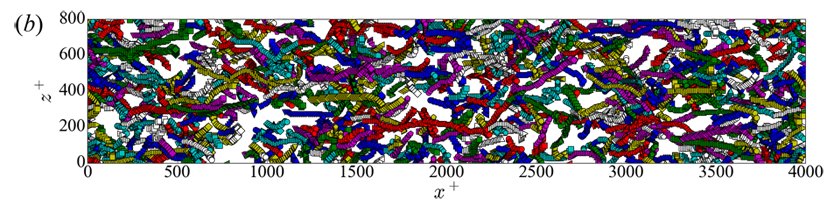}
	\includegraphics[width=.8\linewidth, trim=0mm 0mm 0mm 2mm, clip]{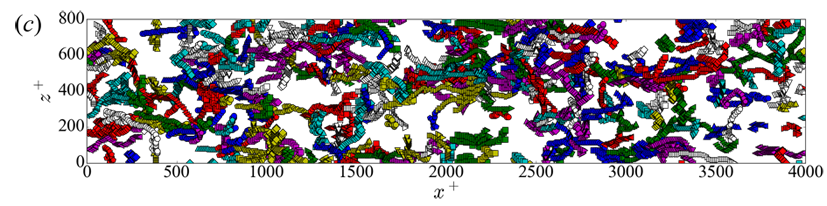}
	\includegraphics[width=.8\linewidth, trim=0mm 0mm 0mm 2mm, clip]{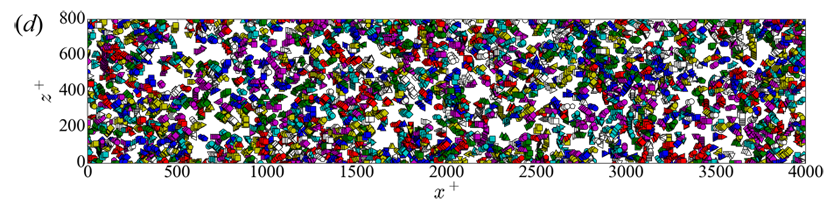}
	\includegraphics[width=.8\linewidth, trim=0mm 0mm 0mm 2mm, clip]{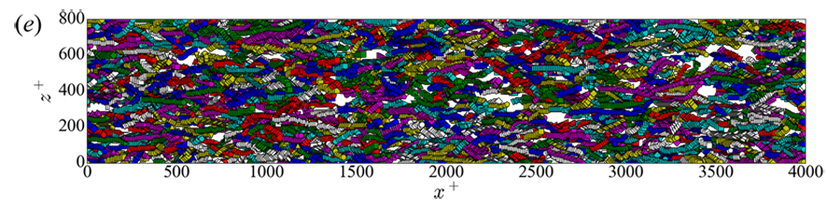}
	\caption{Distribution of vortex axis-lines of different classes in a typical snapshot at $\mathrm{Re}_\tau=400$: (\textit{a}) hairpins, (\textit{b}) hooks, (\textit{c}) branches (all types), (\textit{d})fragments, and (\textit{e}) streamwise vortices. Each marker represents one axis point. Individual vortices are differentiated by colors and marker types.}
	\label{fig:vor_class_visu2} 
\end{figure}

\Cref{fig:vor_class_visu1,fig:vor_class_visu2} show the distributions of vortex axis-lines, as identified by VATIP, of different classes for one typical snapshot of the lowest ($\mathrm{Re}_\tau=84.85$) and the highest $\mathrm{Re}$ ($\mathrm{Re}_\tau=400$), respectively.
Direct visual inspection of these images indicates that the VATIP algorithm together with the vortex classification procedure in \cref{fig:class_flow} has successfully identified and extracted all types of vortices and sorted them properly according to their axis topology.
This resonates with the earlier tests by STG in \cref{fig:ins_vortex}.
Comparing different classes of vortices, streamwise ones still dominate at both $\mathrm{Re}$, but the method has no difficulty in finding all types of three-dimensional vortices.
Unlike the case of boundary layer flow where the so-called ``forest'' of well-organized hairpins were observed~\citep{wu2009direct}, in our DNS results clear-cut hairpins are the minority compared with other three-dimensional configurations. In particular, the asymmetric hook type significantly outnumbers all other three-dimensional vortex types, which validates the earlier empirical notion in the literature about the prevalence of incomplete or one-legged hairpins in plane Poiseuille flow~\citep{robinson1989areview,robinson1991coherent}.
On the other hand, the frequent appearance of various irregular branch types demonstrates the importance of iterative propagation in all three dimensions -- a central element of VATIP.

\begin{figure}
	\centering
	\includegraphics[width=.7\linewidth, trim=0mm 0mm 0mm 0mm, clip]{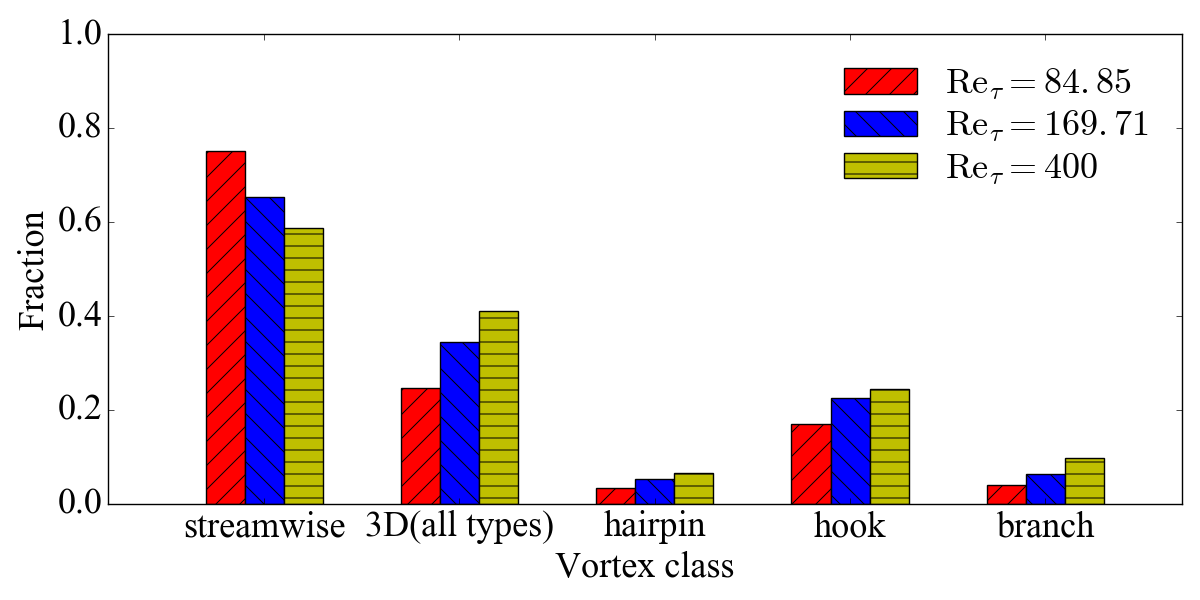}
	\caption{\RevisedText{Fraction} of vortices of different types by vortex numbers. Only vortices with streamwise length $l_x^+>50$ are included.}
	\label{fig:vor_num} 
\end{figure}

Comparing between the two $\mathrm{Re}$, three-dimensional vortices (hairpins, hooks, and branches) grow larger in size at higher $\mathrm{Re}$. This can be attributed to the increasing thickness of the wall layer (more wall units in the wall-normal direction) which allows these vortices to further lift up and develop to a higher altitude. 
They also become more populous at higher $\mathrm{Re}$.
Indeed, even after factoring in the across-the-board increase of all vortices, the percentage share taken by three-dimensional vortices still steadily climbs.
As shown in \cref{fig:vor_num}, with increasing $\mathrm{Re}$, quasi-streamwise vortices take up a lower percentage (despite a net increase in their number) and their share is replaced by all types of three dimension vortices.
From $\mathrm{Re}_\tau=84.85$ to $400$, the share of hooks increases by about 50\% and those of hairpins and branches more than double.
Recall that hooks are often considered as asymmetric or incomplete hairpins, their slower growth (compared with symmetric hairpins and branches) suggests that they are likely the outcome of the insufficient development of hairpins and may become less important at higher $\mathrm{Re}$.
\RevisedText{Finally, in all $\mathrm{Re}$ cases, complete hairpins are significantly outnumbered by its mutants -- hooks and branches.}

\begin{figure}
	\centering
	\includegraphics[width=.8\linewidth, trim=0mm 0mm 0mm 0mm, clip]{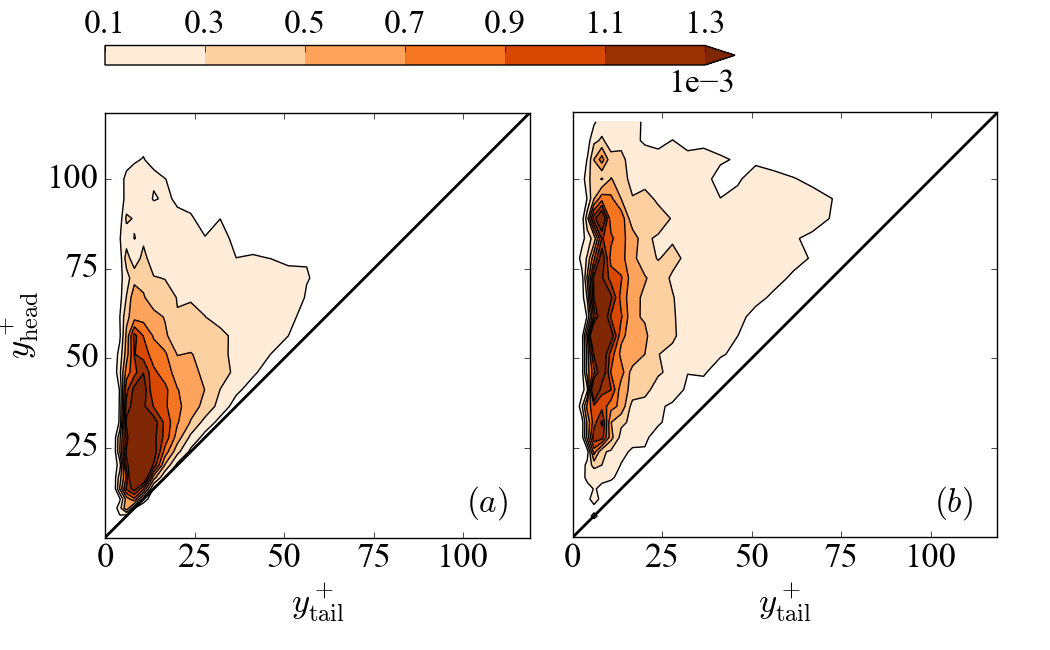}
	\caption{Joint probability density function (PDF) between the $y^+$ positions of the tail and the head of vortices at $\mathrm{Re}_\tau=84.85$: (\textit{a}) quasi-streamwise and (\textit{b}) three-dimensional vortices.}
	\label{fig:vor_height1}
\end{figure}

\begin{figure}
	\centering
	\includegraphics[width=.8\linewidth, trim=0mm 0mm 0mm 0mm, clip]{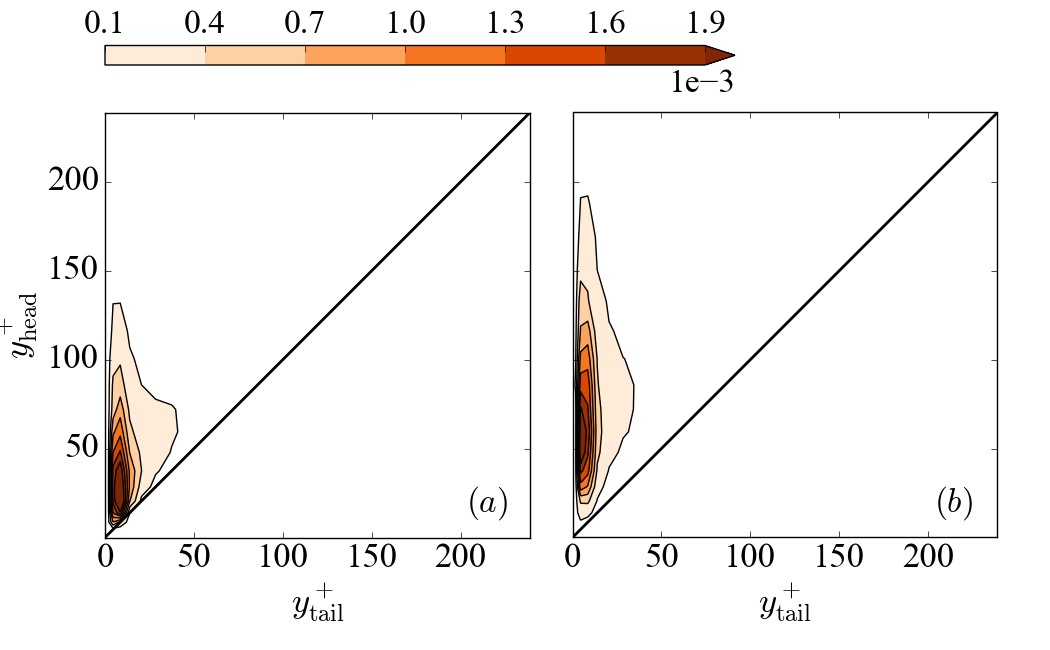}
	\caption{Joint probability density function (PDF) between the $y^+$ positions of the tail and the head of vortices at $\mathrm{Re}_\tau=400$: (\textit{a}) quasi-streamwise and (\textit{b}) three-dimensional vortices.}
	\label{fig:vor_height2}
\end{figure}

Near-wall vortex growth is often characterized as a lift-up process: the downstream end of the vortex becomes detached from the wall and rises towards the outer layer, where it can further burst and generate new disturbances~\citep{hinze1975turbulence,zhou1999mechanisms}.
Lift-up \RevisedText{extent} of vortices at different wall layers can now be statistically analyzed with the axis-lines extracted by VATIP.
\Cref{fig:vor_height1,fig:vor_height2} show the joint PDF between the wall-normal positions of the heads and tails of all quasi-streamwise and three-dimensional vortices at different $\mathrm{Re}$.
The head position $y^+_\text{head}$ is measured as the highest wall-normal position of all axis-points, which is normally found at the downstream end; likewise, the tail position $y^+_\text{tail}$ is the lowest position normally found at the upstream end. 
Obviously, the distribution can only sample the upper-left triangle of the domain.
Vortices \RevisedText{that have not lifted up} are represented by the diagonal where the head position is leveled with the tail and regions closer to the ordinate, i.e., $y^+_\text{head}\gg y^+_\text{tail}$, correspond to highly lift-up vortices.

For the lower $\mathrm{Re}_\tau=84.85$ case (\cref{fig:vor_height1}), both the head and tail positions of quasi-streamwise vortices (panel (a)) concentrate at $10\lesssim y^+\lesssim 50$: i.e., within or near the buffer layer.
Three-dimensional vortices (hairpins, hooks, and branches) have a higher altitude and their distribution peaks at $(15,80)$: i.e., the tail (legs) stretches deep into the buffer layer while the head (arc in the case of hairpins) rises up into the log-law layer.
In terms of distribution, the tails are concentrated at $y^+<25$ whereas the heads are found in a much broader range extending from $y^+=25$ to $y^+>80$.
These observations can all carry over to the higher $\mathrm{Re}_\tau=400$ (\cref{fig:vor_height2}) where, in addition, the larger number of wall units in the $y$ direction allows more room for vortex growth and their lift-up \RevisedText{extent} is easier to observe.
Most quasi-streamwise vortices (\cref{fig:vor_height2}(a)) are lying flat in the buffer layer (concentration peak in the lower-left corner) but two more concentration bands can be spotted: one lies along the ordinate up to $y^+=100$, indicating that a small fraction of streamwise vortices can lift up to the log law layer; the other lies along the diagonal to even higher $y^+$, indicating the existence of flat-lying vortices at higher altitudes.
Both bands are also clearly visible in the three-dimensional case (\cref{fig:vor_height2}(b)) but the vertical one is stronger over a broader range of $y^+$, meaning that these vortices \RevisedText{are more likely to lift up} and their heads can reach various altitudes.
Vortex activities at $y^+>250$ are much weaker and thus not included in \cref{fig:vor_height2}. (Alternatively, following the example of \citet{lozano2012three,lozano2014time}, one may apply non-uniform $Q$ threshold values -- with lower thresholds for the bulk -- for a complete picture.)
Observations from this analysis largely confirm the earlier empirical depiction by \citet{robinson1991coherent} that quasi-streamwise vortices dominate the buffer layer and hairpin-like vortices are more likely to be found in the log-law layer and beyond.
\Citet{robinson1991coherent} conceived the log-law layer to be comprised of a mix of streamwise and hairpin vortices, whereas we are able to more clearly show that streamwise vortices are only concentrated in the lower log-law layer ($y^+<50$) and three-dimensional vortices can rise up to a variety of altitudes.


\RevisedText{%
\subsection{Vortex organization through clustering analysis}
Previous observations of LSMs and VLSMs ignited the immense interest among researchers in understanding the organization patterns of coherent structures~\citep{kim1999very,jimenez1998largest,lee2014spatial}.
Given the specific information, available from VATIP, about the location and conformation of axis-lines representing individual vortices, we adapt the DBSCAN (density-based spatial clustering of applications with noise) algorithm~\citep{ester1996density} -- a widely used clustering analysis method in data mining and machine learning -- to VATIP results for understanding the clustering patterns of vortices.
(Structures classified as fragments according to \cref{fig:class_flow} are not considered in this analysis.)

The standard DBSCAN algorithm groups scattered points in space into clusters based on their spatial proximity and mutual relationship.
Two points that are close to each other (within a cutoff distance $\varepsilon$) are considered as neighbors. Points inside a cluster are known to have many neighbors. Points with at least $N_\text{c,min}$ neighbors are thus labeled as ``core points'' and all interconnected (in the sense of mutually neighboring) core points are grouped into one cluster. (Both $\varepsilon$ and $N_\text{c,min}$ are user-specified parameters.)
If a point does not qualify as a core point by itself but neighbors one or more core points, it is labeled as a ``border point'' which resides on the surface of a cluster. Border points are grouped to the same cluster as their nearest neighboring core point.
Points that do not neighbor any core points and are not core points themselves are isolated outlier not belonging to any cluster.

Since the VATIP output contains not simple size-less points but complex axis-lines representing vortex geometry and topology, the simple distance criterion used for neighbor identification needs to be adapted. We consider two vortices to be neighbors if the minimum distance between any two axis-points -- one on each axis-line -- does not exceed $\varepsilon=4r_\text{v}$ which is only slightly larger than the detection cone diameter used in VATIP tracking ($2\times 1.5r_\text{v}=3r_\text{v}$). We have tested a wide range of $\varepsilon$ and found that for $\varepsilon$ as low as $3.5r_\text{v}$ nearly all vortices in the domain, from both sides of the channel, are interconnected into the same neighbor network: i.e., for $N_\text{c,min}=1$ and any $\varepsilon\geq3.5r_\text{v}$, the DBSCAN algorithm will identify one supersized cluster that includes nearly all vortices. 
The fact that a cutoff distance at the same order of the vortex diameter would connect all vortices is not surprising, considering the level of crowdedness found in their distribution (see \cref{fig:vor_class_visu1,fig:vor_class_visu2}).
Ideally, we would also need to test the $\varepsilon$-dependence at other $N_\text{c,min}$ levels. However, our priority is to understand the importance of multi-vortex cooperation (instead of inter-vortex distance, which we knew would be close). Therefore, given the limited scope of this investigation, we will focus here on the $N_\text{c,min}$-dependence of the clustering results at a constant $\varepsilon=4r_\text{v}$.
}%

\begin{figure}
	\centering
	\includegraphics[width=.5\linewidth, trim=0mm 0mm 0mm 0mm, clip]{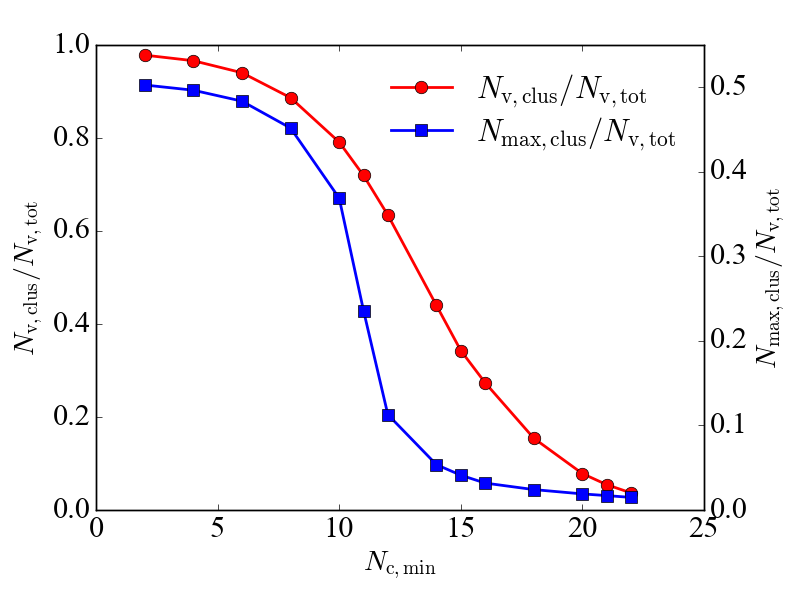}
	\caption{\RevisedText{Dependence of DBSCAN clustering analysis on $N_\text{c,min}$ ($\mathrm{Re}_\tau=169.71$): left/red/circle -- number fraction of all vortices grouped into clusters; right/blue/square -- number fraction of vortices in the largest identified cluster.}}
	\label{fig:MaxCluster_vs_Tot}
\end{figure}

\begin{figure}
	\centering
	\includegraphics[width=.5\linewidth, trim=0mm 0mm 0mm 0mm, clip]{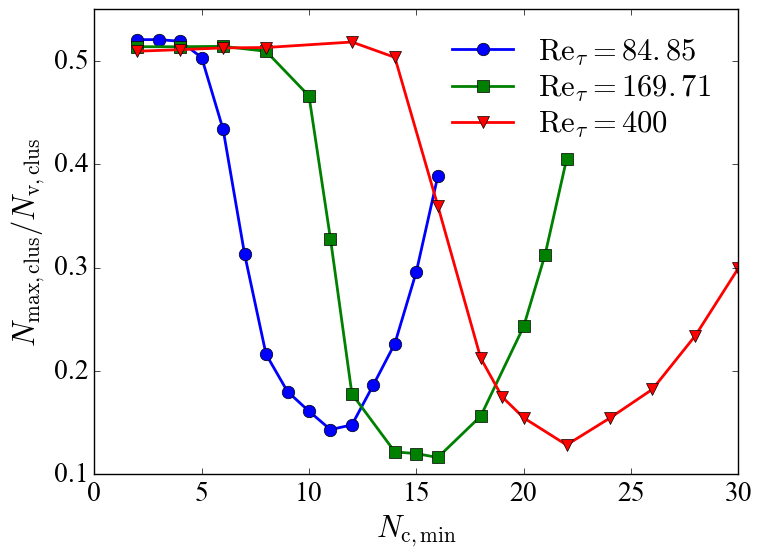}
	\caption{\RevisedText{Number fraction of vortices included in the largest cluster among vortices in all clusters as a function of $N_\text{c,min}$.}}
	\label{fig:MaxClus_vs_AllClus}
\end{figure}

\begin{figure}
	\centering
	\includegraphics[width=.98\linewidth, trim=0mm 0mm 0mm 0mm, clip]{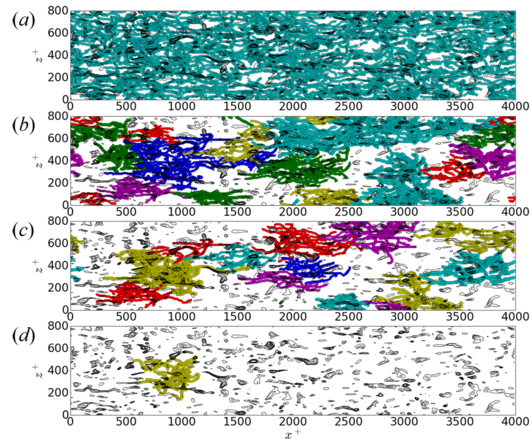}
	\caption{\RevisedText{Distribution of vortex clusters (at one side of the channel and from a typical instantaneous flow image) at $\mathrm{Re}_\tau=169.71$ identified by DBSCAN with (\textit{a}) $N_\text{c,min}=2$, (\textit{b}) $N_\text{c,min}=12$, (\textit{c}) $N_\text{c,min}=15$, and (\textit{d}) $N_\text{c,min}=21$. Individual clusters are differentiated by color.
	Black lines show the contours of $y$-average Reynolds shear stress $\bar\tau_{xy}$ (\cref{eq:aveRSS}) at 11 equispaced levels from \numrange{0.5e-5}{1.25e-4}; higher contour line density corresponds to higher magnitudes.}}
	\label{fig:Clus_ins}
\end{figure}

\RevisedText{%
With increasing $N_\text{c,min}$, less vortices are qualified as core vortices (counterpart to core points in the standard DBSCAN) and more become isolated outliers.
This first leads to the shrinkage of all clusters: for the same total number of vortices $N_\text{v,tot}$, the number of vortices assigned to clusters $N_\text{v,clus}$ decreases (\cref{fig:MaxCluster_vs_Tot}).
At $N_\text{c,min}=2$ (lowest level shown in \cref{fig:MaxCluster_vs_Tot}), $N_\text{v,clus}/N_\text{v,tot}$ starts at close to $1$ (nearly all vortices are grouped into clusters) and steadily drops afterwards towards $0$.
The number of vortices contained in the largest cluster $N_\text{max,clus}$ is also calculated. In \cref{fig:MaxCluster_vs_Tot}, $N_\text{max,clus}/N_\text{v,tot}$ starts at $\approx0.5$, because the channel flow geometry has two boundary layers (near each wall) and at the lowest $N_\text{c,min}$ vortices near each wall are nearly all grouped into one super-cluster.
The decline pattern of this profile is very different from that of $N_\text{v,clus}/N_\text{v,tot}$ -- it drops sharply in a small window of $N_\text{c,min}=8\sim14$ with the steepest slope found between $N_\text{c,min}=10$ and $12$.
This faster decline cannot be solely accounted for by the overall reduction of qualifying core vortices (otherwise $N_\text{max,clus}/N_\text{v,tot}$ would have the same slope as $N_\text{v,clus}/N_\text{v,tot}$).
Indeed, the steeper descent indicates a sudden disintegration of the dominant clusters into smaller pieces. As $N_\text{c,min}$ increases beyond $\approx 8$, some vortices in the structure, which are not as highly intertwined as most others in the vortex cluster network, are disqualified as core vortices. Removing those ``bridge'' vortices dismantles the cluster network into several well-defined and strongly-coupled constituting clusters that are much smaller in size. 
This effect is most clearly seen from the ratio between these two profiles, plotted in \cref{fig:MaxClus_vs_AllClus}. For all three $\mathrm{Re}_\tau$ tested, $N_\text{max,clus}/N_\text{v,clus}$ is initially flat at low $N_\text{c,min}$, indicating that within this regime drops in both profiles in \cref{fig:MaxCluster_vs_Tot} are attributed to the overall reduction of clustered vortices.
Disintegration of the dominant clusters starts when the $N_\text{max,clus}/N_\text{v,clus}$ profile turns downwards, which for $\mathrm{Re}_\tau=169.71$ occurs at $N_\text{c,min}\approx 8$. The process finishes as the curve reaches its minimum at $N_\text{max,clus}/N_\text{v,clus}=0.1\sim0.15$: the domain is now populated by $O(10)$ well-defined clusters with comparable size (see \cref{fig:Clus_ins}(c) for roughly half of the clusters at one side of channel).
After the minimum the profile rises again as a result of smaller clusters gradually being eliminated by the increasingly stringent $N_\text{c,min}$ cutoff.
}

\begin{figure}
	\centering
	\includegraphics[width=.92\linewidth, trim=0mm 0mm 0mm 0mm, clip]{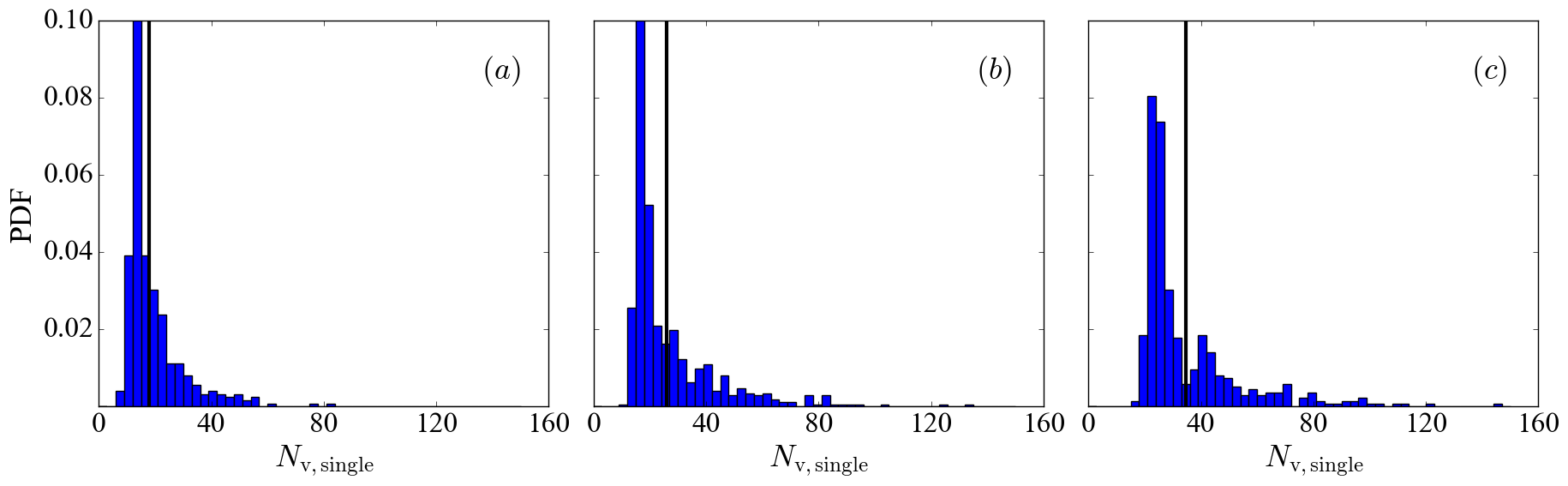}
	\caption{\RevisedText{Probability density function of the number of vortices in a single cluster at (\textit{a}) $\mathrm{Re}_\tau=84.85$, (\textit{b})  $\mathrm{Re}_\tau=169.71$, and (\textit{c})  $\mathrm{Re}_\tau=400$.}}
	\label{fig:Nvclus_dist} 
\end{figure}

\RevisedText{%
The vortex cluster configuration during this disintegration process with increasing $N_\text{c,min}$ is shown in \cref{fig:Clus_ins} for $\mathrm{Re}_\tau=169.71$.
Consistent with our earlier analysis, at $N_\text{c,min}=2$ all vortices at one side of the channel are interconnected (by their neighboring network) into a super-cluster. Disintegration of the network is observed at $N_\text{c,min}=12$ ($>8$ where it starts). At $N_\text{c,min}=15$ (\cref{fig:Clus_ins} (c)), $N_\text{max,clus}/N_\text{v,clus}$ reaches its minimum (\cref{fig:MaxClus_vs_AllClus}) and the disintegration process has completed with a number of clear vortex clusters remaining unbroken. Much space can be found between the clusters where the turbulent flow field is occupied by unclustered vortices.
The Reynolds shear stress, averaged over the wall-normal ($y$) direction,
\begin{gather}
	\bar\tau_{xy}=-\int_0^1 v_x'(x,y,z,t)v_y'(x,y,z,t)dy
	\label{eq:aveRSS}
\end{gather}
is shown with contour lines in the images. Spots with strong $\bar\tau_{xy}$ (dense contour lines) are found within or immediately around these vortex clusters, indicating their strong contribution to the Reynold stress generation.
Interestingly, the characteristic length (in the $x$ direction) of these clusters seems to be between $500\sim1500$ wall units which is at the same level as the typical streamwise length scales of LSMs reported in the literature~\citep{adrian2007hairpin,lee2014spatial}.
The existence of such clusters consisting of a large number ($O(10)$ or higher; see \cref{fig:Nvclus_dist}) of vortices strongly intertwined through multi-body interactions ($>N_\text{c,min}=15$ neighbors, in the case of $\mathrm{Re}_\tau=169.71$, with close contacts between their axis-lines for the core vortices) is consistent with the hypothesis that LSMs are results of the cooperative dynamics involving many vortices organized as ``packets''~\citep{kim1999very}. However, these ``packets'', as discussed below, are not composed of well-aligned hairpin vortices with their classical shape.
In addition, clear evidence for clustering is found in this study for $\mathrm{Re}_\tau$ all the way down to below $100$, suggesting that cooperative dynamics between vortices is a universal feature for wall turbulence not limited to the high-$\mathrm{Re}$ regime.
Meanwhile, VLSMs are often conjectured to occur at a higher level of organization involving the alignment of multiple LSMs~\citep{kim1999very,lee2014spatial}. This would correspond to the cooperative organization involving multiple vortex clusters in this study. The length scale of VLSMs is comparable to or larger than the current domain size and they were previously studied mostly at much higher $\mathrm{Re}_\tau$ ($~O(10^3)$). For these reasons, they are not discussed here.
As $N_\text{c,min}$ further increases to 21, all clusters are now eliminated except the strongest one, which has shrunken in size but still clearly marks the location of strong Reynolds stress activities.
}

\RevisedText{%
Note that the term ``cluster'' has a different meaning here than that in earlier studies of three-dimensional vortex analysis, such as \citet{del2006self} where clusters referred to the interconnected structures with overlapping vortex volumes identified by the scalar identifier ($\Delta$ in that study and $Q$ here), regardless of the individual identities of vortices or their conformation and topology.
In our analysis, a cluster is defined as individual vortices grouped together based on the existence of a mutually interacting (neighboring) network between multiple vortex objects rather than a pure spatial-proximity criterion.
There are likely close connections between these two interpretations, but at this point, a direct comparison is not possible, because, as further discussed in \cref{Sec:limitation}, the current VATIP algorithm can only capture a subset of structures analyzed in \citet{del2006self} that are directly generated from the lift-up-from-wall process.
The strength of the current approach is its access to the information of individual constituting vortex, which we discuss below.
}%

\begin{figure}
	\centering
	\includegraphics[width=.92\linewidth, trim=0mm 0mm 0mm 0mm, clip]{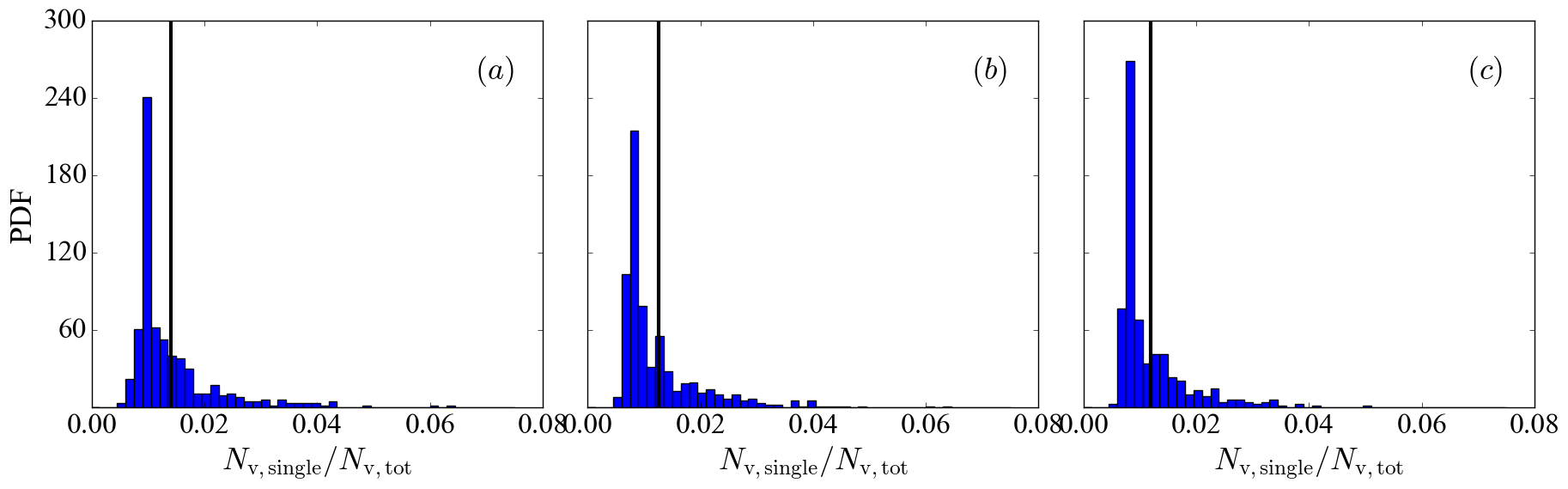}
	\caption{\RevisedText{Probability density function of the number of vortices in a single cluster normalized by the total number of vortices in the domain at (\textit{a}) $\mathrm{Re}_\tau=84.85$, (\textit{b})  $\mathrm{Re}_\tau=169.71$, and (\textit{c})  $\mathrm{Re}_\tau=400$.}}
	\label{fig:Nvclus_Ntot_dist} 
\end{figure}

\RevisedText{%
Unless otherwise noted, we pick the $N_\text{c,min}$ value at the minimum in each $N_\text{max,clus}/N_\text{v,clus}$ curve (\cref{fig:MaxClus_vs_AllClus}) -- i.e., $N_\text{c,min}=11,15$ and $22$ for $\mathrm{Re}_\tau=84.85, 169.71$ and $400$, respectively -- for the DBSCAN analysis, which is the point where the percolating super-cluster has been fully disintegrated into unbreakable clusters while most individual clusters are not yet eliminated. (This choice is in the same spirit as the percolation analysis of \citet{lozano2012three} explained in \cref{Sec:ParamAnalys}).
The PDF of the number of vortices constituting a single cluster $N_\text{v,single}$, shown in \cref{fig:Nvclus_dist}, is clearly skewed to the right with the most probable value at $O(10)$ but some extreme cases with $O(100)$ vortices in each cluster. 
The average $N_\text{v,single}$ increases with $\mathrm{Re}$ and is $\approx 18, 26$ and $35$, respectively, from the lowest to the highest $\mathrm{Re}_\tau$ tested.
Note, however, that the total number of vortices in the domain $N_\text{v,tot}$ also increases with $\mathrm{Re}$.
Indeed, when $N_\text{v,single}$ is normalized by $N_\text{v,tot}$ (\cref{fig:Nvclus_Ntot_dist}), the distribution profile becomes nearly the same between different $\mathrm{Re}$ (mean value at $0.014, 0.013$ and $0.012$ for $\mathrm{Re}_\tau=84.85, 169.71$ and $400$, respectively). 
Since the domain size of different $\mathrm{Re}$ is kept the same in inner units (\cref{tab:numerical}), this observation, that the average cluster size in terms of the number fraction of vortices in each cluster (out of all vortices filling the domain), remains roughly constant, suggests that the cluster size is more or less the same in inner units (i.e, streamwise length within $500\sim1500$; see \cref{fig:Clus_ins}) within the $\mathrm{Re}$ range tested (further demonstrated in \cref{fig:clus_example}).
}%

\begin{figure}
	\centering
	\includegraphics[width=.5\linewidth, trim=0mm 0mm 0mm 0mm, clip]{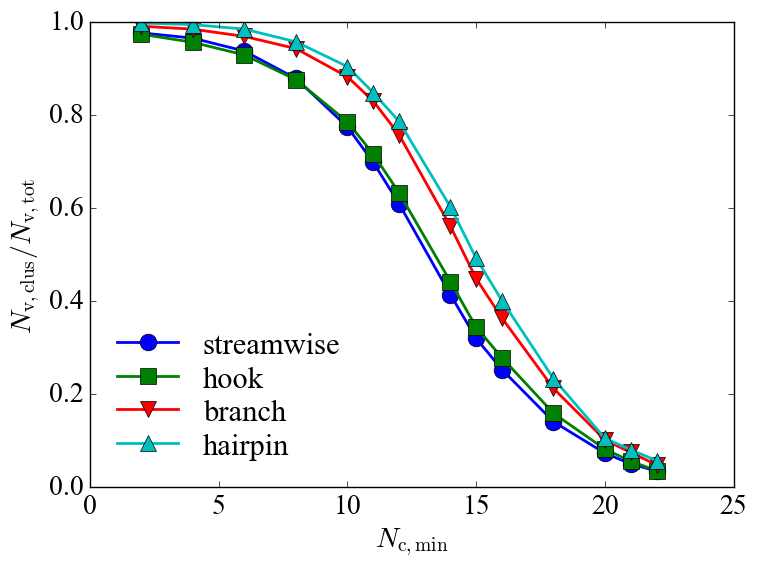}
	\caption{\RevisedText{Number fraction of vortices of different types grouped into clusters out of the total number of all vortices of the same type in the domain ($\mathrm{Re}_\tau=169.71$).}}
	\label{fig:Nvc_o_Nvt} 
\end{figure}

\begin{figure}
	\centering
	\includegraphics[width=.92\linewidth, trim=0mm 0mm 0mm 0mm, clip]{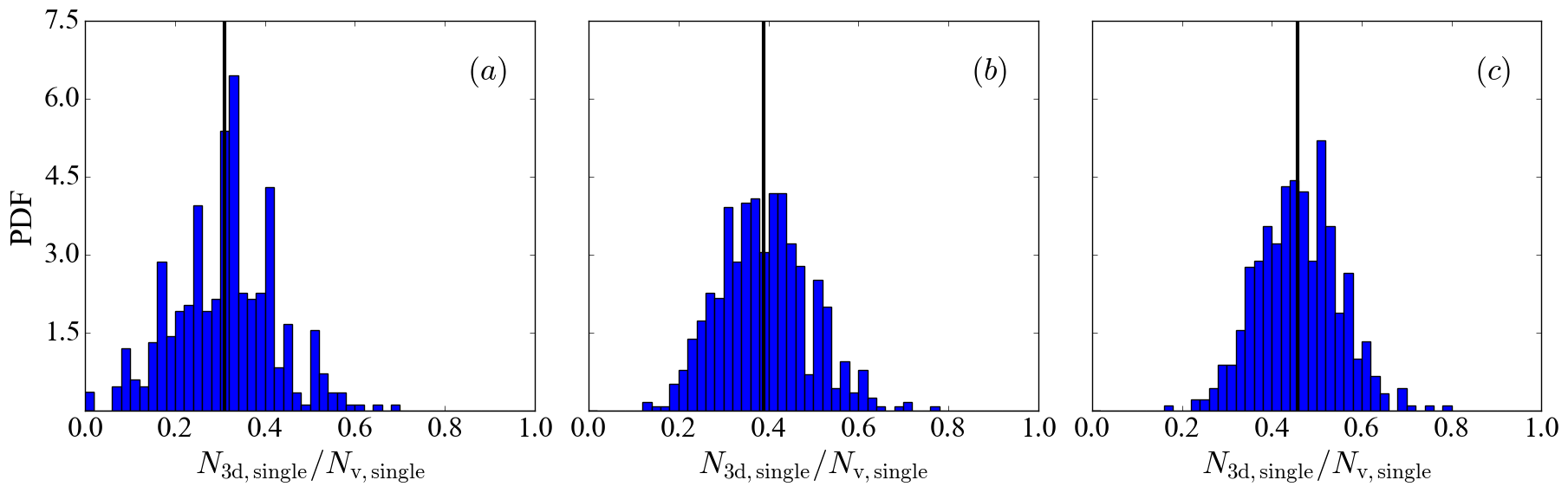}
	\caption{\RevisedText{Probability density function of the number fraction of 3D vortices in a single cluster at (\textit{a}) $\mathrm{Re}_\tau=84.85$, (\textit{b})  $\mathrm{Re}_\tau=169.71$, and (\textit{c})  $\mathrm{Re}_\tau=400$.}}
	\label{fig:N3d_o_Nvt} 
\end{figure}

\RevisedText{%
Comparing vortices of different types (\cref{fig:Nvc_o_Nvt}), hairpins and branches are more likely to be included in a cluster than both streamwise and hook vortices.
Since hooks are essentially incomplete or asymmetric hairpins that are also highly lifted-up, this indicates that the large dimension of hairpins and branches are likely the key factor determining their higher clustering tendency. In particular, their wide span in the $z$ direction exposes them to vortices from a wider flow region, which enables them to play a central role in stitching more vortices into a cluster.
As seen in \cref{fig:N3d_o_Nvt}, within a single cluster, a significant fraction of the vortices belong to the three-dimensional classes (hairpins, branches, or hooks). This fraction increases with $\mathrm{Re}$ and at $\mathrm{Re}_\tau=400$, on average nearly half of the vortices in each cluster are three-dimensional ones.
However, note that hooks and branches significantly outnumber canonical hairpins (\cref{fig:vor_num}), the hypothesized picture of packets of clean-cut hairpins~\citep{adrian2007hairpin} forming the LSMs is not seen at least at the current $\mathrm{Re}$ range.
}

\begin{figure}
	\centering
	\includegraphics[width=.98\linewidth, trim=0mm 0mm 0mm 0mm, clip]{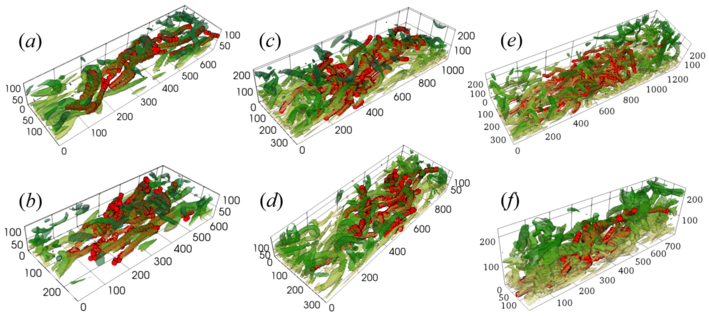}
	\caption{\RevisedText{Representative images of vortex clusters identified by DBSCAN  at (\textit{a},\textit{b})  $\mathrm{Re}_\tau=84.85$, (\textit{c},\textit{d})  $\mathrm{Re}_\tau=169.71$, and (\textit{e},\textit{f})  $\mathrm{Re}_\tau=400$. Isosurfaces show all vortices in the viewable region (color varies from light to dark with $y^+$); red dots show the axis-lines of vortices in the identified cluster.}}
	\label{fig:clus_example}
\end{figure}

\RevisedText{%
Direct images of representative vortex clusters are shown in \cref{fig:clus_example} where vortices forming the particular cluster are highlighted by explicitly showing their axis-lines.
Consistent with our earlier observations, these clusters (at different $\mathrm{Re}$) all have a streamwise length in the range of $500\sim 1500$ wall units.
Two typical organization configurations are observed. In the first (panels (a) and (f)), different vortices forming the cluster have their axis-lines braided together along the streamwise direction. These clusters have a shape of twisted doughnuts and they remain slender (narrow in the $z$ direction) while extending downstream for $O(1000)$ wall units.
For the second type, which is more frequently observed, other than the downstream twisting, the clusters also expand in the spanwise direction by connecting more vortices through the wider vortex types (hairpins and branches).

Finally, we note that the analysis of vortex clustering and organization in this section is still preliminary and limited in scope. It is intended to provide some first insight into how the axis-line information extracted by VATIP can be used to address some of the most important outstanding questions in turbulent dynamics~\citep{Jimenez_JFM2018}. Further research is needed to better connect these observations with the existing conceptual models and results from other structure analysis techniques.
}

\RevisedText{
\subsection{Determination of parameters and settings in VATIP}\label{Sec:ParamAnalys}
After presenting the main results, we are now ready to assess the robustness of VATIP tracking outcomes and discuss the procedure for choosing its parameters and settings.
There are two major adjustable parameters in the method: (1) the threshold magnitude $Q_\text{threshold}$ for vortex identification with the $Q$-criterion and (2) the cutoff cone radius $r_\text{cone}$ used in the axis-line propagation search (\cref{fig:track_method}).
In addition, sensitivity to grid sizes and the selection of the search starting plane will also be examined.

\begin{figure}
	\centering
	\includegraphics[width=0.7\linewidth, trim=0 0 0 0, clip]{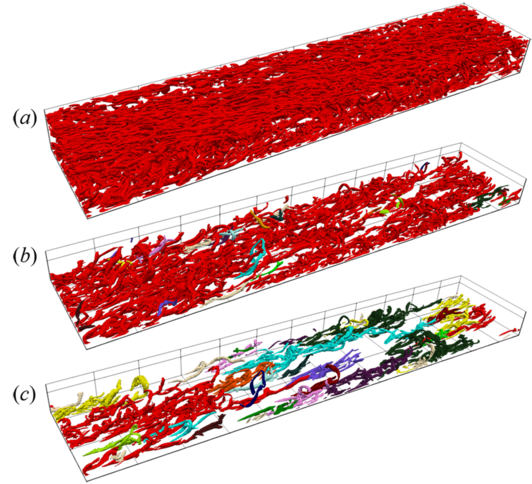}
	\caption{\RevisedText{Vortex disintegration with increasing $Q_\text{threshold}=HQ_\text{rms}$ at $\mathrm{Re}_\tau=169.71$: (a) $H=0.2$, (b) $H=0.4$, and (c) $H=0.7$. 
	Interconnected vortex tube structures are coded with the same color. For clarity, only the largest vortices that cumulatively account for 80\% (for (a) and (b)) or 60\% (for (c)) of the total vortex volume are shown. For (b) and (c), only vortices from the bottom half of the channel are shown.}}
	\label{fig:vortexdisint}
\end{figure}

\begin{figure}
	\centering
	\includegraphics[width=.5\linewidth, trim=0mm 0mm 0mm 0mm, clip]{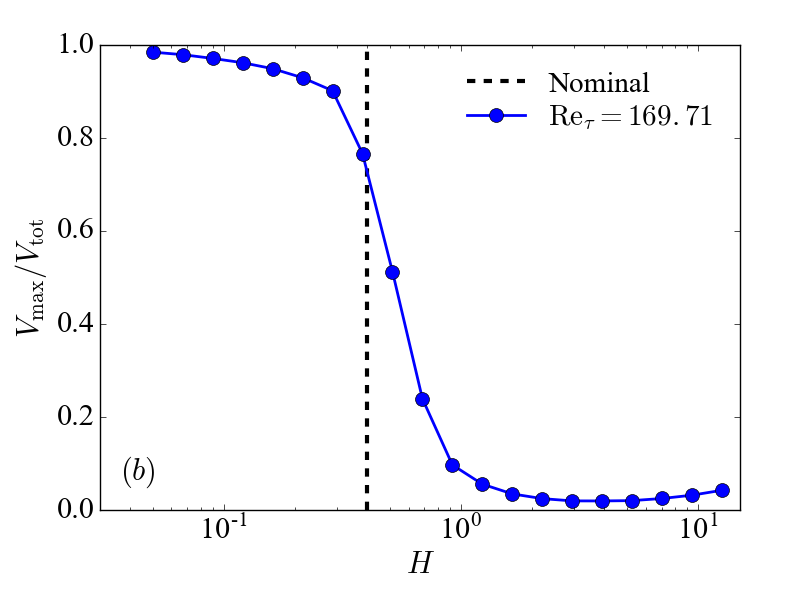}
	\caption{\RevisedText{Percolation diagram for $\mathrm{Re}_\tau=169.71$. The vertical dashed line marks the $HQ = 0.4Q_\text{rms}$ used in this study.}}
	\label{fig:perco_analysis}
\end{figure}

The choice of the threshold for $Q$ (or any other vortex identifier) has been widely discussed in the literature for the purpose of vortex visualization. It is a common practice to choose a threshold in proportion to its RMS value in the flow field
\begin{gather}
	Q_\text{threshold}\equiv HQ_\text{rms}
\end{gather}
where $H$ in this study is chosen based on the percolation analysis, proposed in \citet{lozano2012three} (from which we also borrowed the notation $H$).
When $H$ is low, the identified vortex regions interconnect with one another and form a percolating network across the domain (\cref{fig:vortexdisint}(a)). With increasing $H$, the ``necks'' bridging stronger vortex cores gradually break to reveal individual groups of vortices (\cref{fig:vortexdisint}(b) \& (c)); meanwhile, a higher threshold also erases many weaker vortices from the view.
A percolation diagram (\cref{fig:perco_analysis}) plots the ratio of the volume occupied by the largest interconnected structure ($V_\text{max}$) to that of all vortex regions ($V_\text{tot}$) as a function of $H$.
This value starts at $1$ at the low $H$ end where all structures are interconnected into a complete percolating network.
Increasing $H$ reduces both $V_\text{max}$ and $V_\text{tot}$, whereas the decrease of their ratio $V_\text{max}/V_\text{tot}$ reflects the disintegration of larger interconnected structures into smaller separate pieces.
The latter clearly dominates the window of $0.3\lesssim H\lesssim 0.7$ where the sudden disintegration of the largest structure into multiple objects is reflected in a steep descent in $V_\text{max}/V_\text{tot}$. This window provides the best choices for $H$, in which individual objects are separate and identifiable whereas the most important vortex structures are not yet erased (as what will happen at higher $H$)~\citep{lozano2012three}.

\begin{figure}
	\centering
	\includegraphics[width=.5\linewidth, trim=0mm 0mm 0mm 0mm, clip]{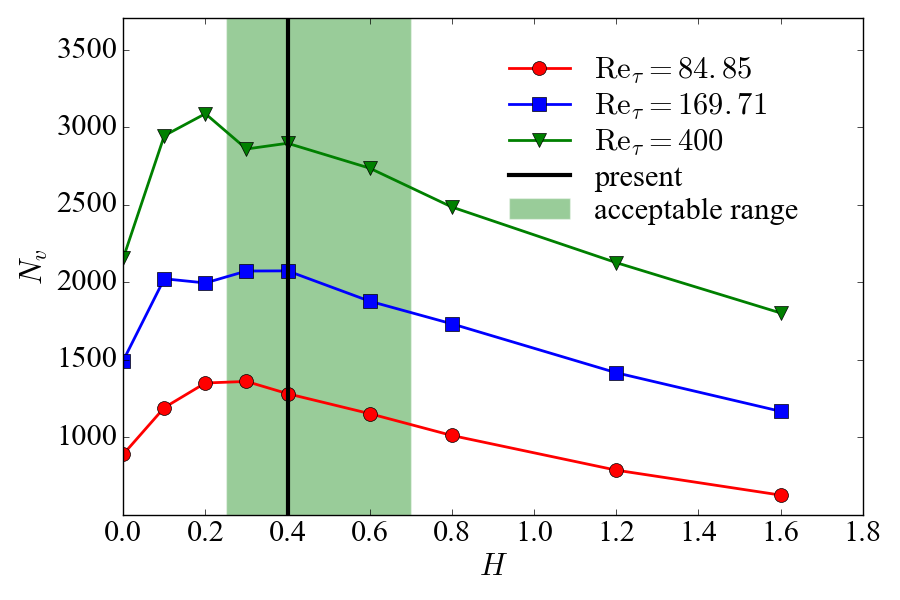}
	\caption{\RevisedText{Average number of vortices (excluding fragments) detected by VATIP in each instantaneous flow field as a function of $H$ (with $r_\text{cone}=1.5r_\text{v}$).}}
	\label{fig:H_Nv}
\end{figure}

Also as a result of the competition between vortex shrinkage and disintegration, the average number of vortices identified by VATIP in each flow domain $N_\text{v}$ displays a non-monotonic dependence on $H$ (\cref{fig:H_Nv}).
$N_\text{v}$ initially increases with $H$, reflecting the splitting of vortex objects. After reaching a maximum at $H\approx 0.2$, $N_\text{v}$ starts to decline because the identified vortices shrink in size with increasing $H$ and are increasingly categorized as fragments. This effect becomes more dominant at higher $H$ after the disintegration of the percolating network.
Within the acceptable range of $H=0.3\sim 0.7$ -- as identified above by the percolation analysis -- the drop of $N_\text{v}$ is relatively mild ($\lesssim 10\%$ for the two higher $\mathrm{Re}_\tau$).
More importantly, as shown later, all major conclusions from the study remain intact within this range of $H$.
It is worth noting that the peak of $N_\text{v}$ is found out of this range at a slightly lower $H$: i.e., the main vortex disintegration events are detected at a slightly lower $H$ using VATIP than the percolation analysis.
This is because VATIP is more sensitive to the breakage between vortex structures: neighboring vortices could well overlap in their shells and be grouped into the same interconnected structure in \cref{fig:vortexdisint} while their axis-lines do not have topological connection. This also explains, in part, why $N_\text{v}$ does not start from $1$ at $H=0$ in \cref{fig:H_Nv}.
(Another reason -- further discussed in \cref{Sec:limitation} -- is that VATIP is designed with wall-generated vortices in mind and may not fully capture the connections between weaker and more isotropic vortex structures in the bulk region, which are only unveiled at very low $H$.)
Taking this into account, the optimal $H$ pick should be slightly lower than that for the steepest descent in \cref{fig:perco_analysis}. Therefore, $H=0.4$ is chosen in this study for VATIP tracking (in comparison with $H=0.7$ used in \citet{Zhu_Xi_JNNFM2018} for vortex visualization).
Note that in \cref{fig:H_Nv}, $N_\text{v}$ is nearly constant in the range of $H=0.3\sim 0.4$.

\begin{figure}
	\centering
	\includegraphics[width=.5\linewidth, trim=0mm 0mm 0mm 0mm, clip]{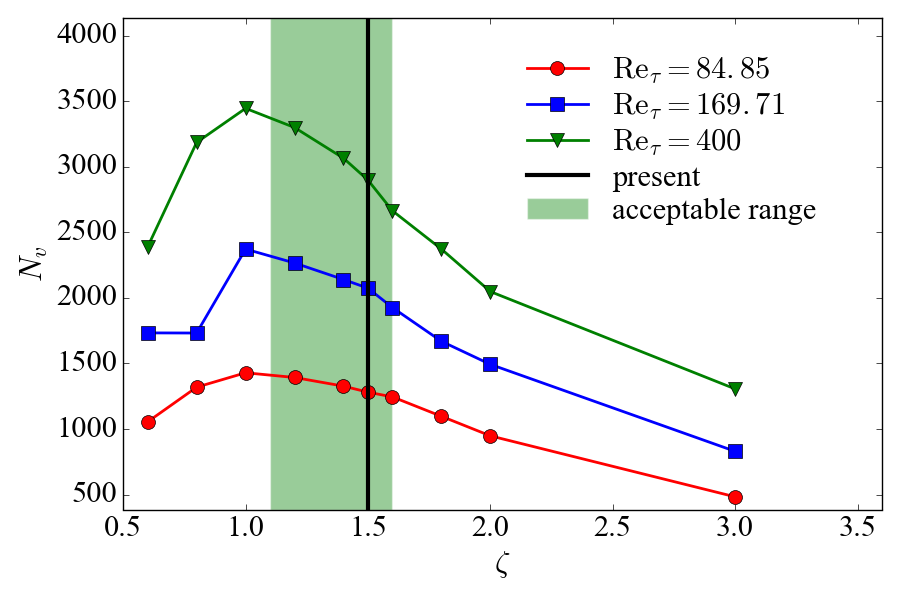}
	\caption{\RevisedText{Average number of vortices (excluding fragments) detected by VATIP in each instantaneous flow field as a function of $\zeta$ (with $Q_\text{threshold}=0.4Q_\text{rms}$).}}
	\label{fig:zeta_Nv}
\end{figure}

The cone size is chosen based on the average radius of the vortex tubes (\cref{Eq:estiRadus})
\begin{gather}
	r_\text{cone}\equiv \zeta r_\text{v}
\end{gather}
where $\zeta$ is expected to be larger than (but in the same order of magnitude of) $1$ to account for vortex size variations.
The average number of vortices identified by VATIP in each flow domain $N_\text{v}$ is also non-monotonic with increasing $\zeta$ (\cref{fig:zeta_Nv}). For $\zeta<1$, many well-defined vortices are broken into pieces and excluded as fragments. Meanwhile, at $\zeta\gg1$, false connection between separate vortices becomes more common and $N_\text{v}$ decreases with $\zeta$.
Interestingly, for all $\mathrm{Re}_\tau$ tested, $N_\text{v}$ reaches maximum at exactly $\zeta=1$, indicating that $r_\text{v}$ calculated by \cref{Eq:estiRadus} does provide an accurate measurement of the vortex radius.
We recommend the range of $\zeta=1.2\sim1.6$ for VATIP where the decline of $N_\text{v}$ is modest (compared with higher $\zeta$) and, more importantly, all major physical observations are consistent with changing $\zeta$ (shown below).
For $\zeta=1.5$ used in this study, the resulting $r_\text{cone}^+$ is about $15$, $16$, and $18$ wall units for $Re_\tau = 84.85$, $169.71$, and $400$, respectively. 
For comparison, \citet{jeong1997coherent} used $r^+_\text{cone}\approx 10$ for their streamwise-only search at $\mathrm{Re_\tau}\approx180$,  which is equivalent to $\zeta\approx1$.
The larger $\zeta$ used in VATIP is necessitated by the expansion of search to all three spatial dimensions.
First, dislocation between successive axis-points is typically larger around the bends or turns of the axis-line, which does not occur in a unidirectional search along nearly straight lines.
Second, inclusion of highly lift-up hairpin-like vortices extends the search deep into the log-law layer, where the vortex diameters are often larger compared with the streamwise vortices in the buffer layer.
Lastly, a streamwise search only looks for new axis-points in the $yz$-plane where the numerical grids are typically more refined (than the $x$ direction) in DNS. Searches in other directions need to accommodate axis-point dislocation in the $x$ direction: with the coarser mesh of \citet{jeong1997coherent}, $r_\text{cone}^+=10$ covers less than one $x$-grid spacing -- $\delta_x^+=17.7$: i.e., no dislocation in $x$ would be allowed.
Our experience also shows that $\zeta=1$ would break well-defined hairpin vortices (such as in STG) into pieces.

\begin{figure}
	\centering
	\includegraphics[width=.92\linewidth, trim=0mm 0mm 0mm 0mm, clip]{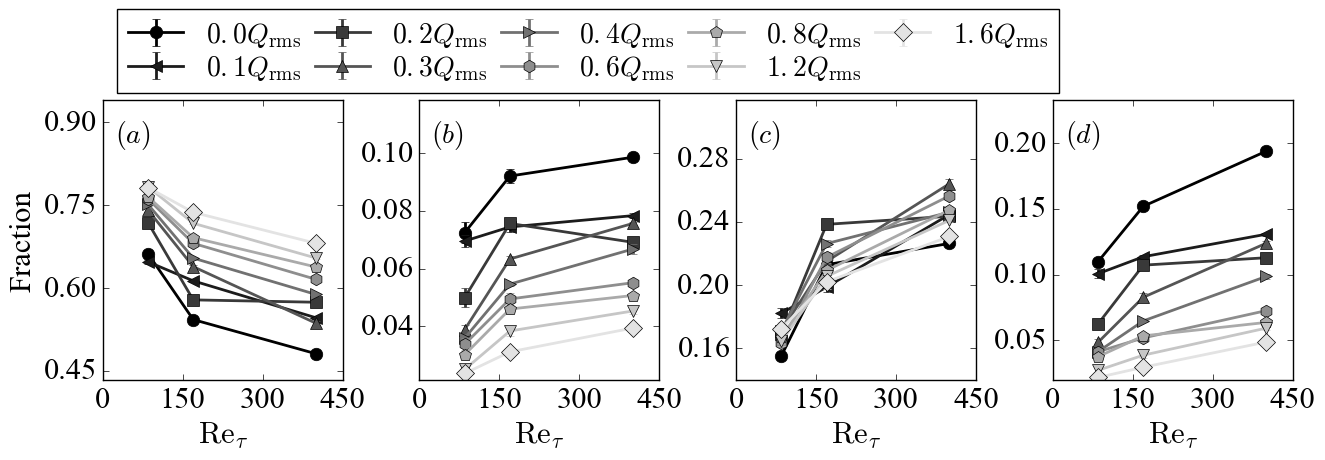}
	\caption{\RevisedText{Number fraction of different vortex types (excluding fragments) as a function $\mathrm{Re}$ identified by VATIP with different $H$ (and a constant $\zeta=1.5$): (a) streamwise, (b) hairpin, (c) hook, and (d) branch.}}
	\label{fig:H_vor_pec}
\end{figure}

\begin{figure}
	\centering
	\includegraphics[width=.92\linewidth, trim=0mm 0mm 0mm 0mm, clip]{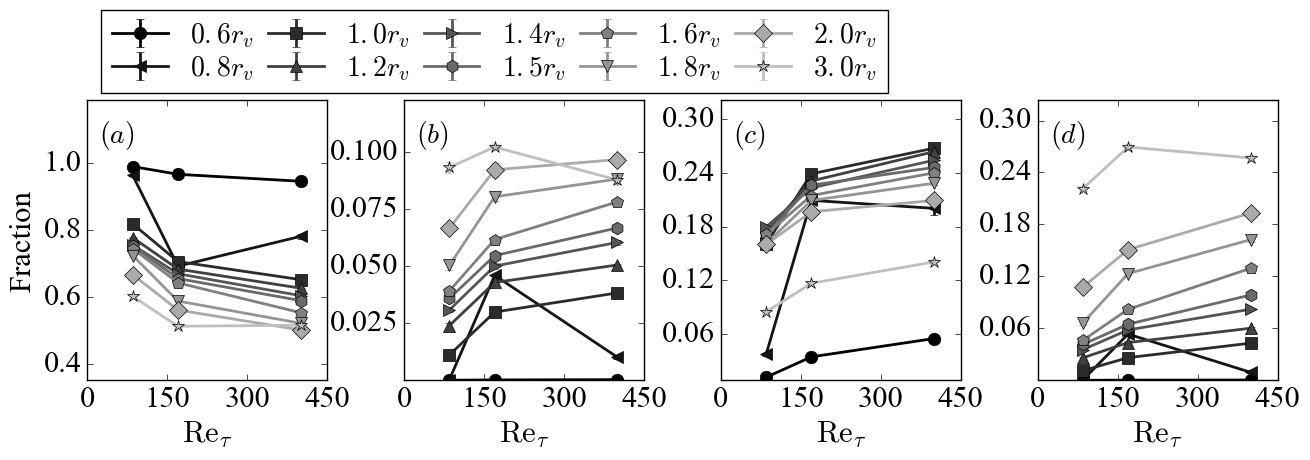}
	\caption{\RevisedText{Number fraction of different vortex types (excluding fragments) as a function $\mathrm{Re}$ identified by VATIP with different $\zeta$ (and a constant $H=0.4$): (a) streamwise, (b) hairpin, (c) hook, and (d) branch.}}
	\label{fig:zeta_vor_pec}
\end{figure}

The similarity between $N_\text{v}$ profiles of different $\mathrm{Re}_\tau$ in both \cref{fig:H_Nv,fig:zeta_Nv} suggests the robustness of VATIP at least within the $\mathrm{Re}$ range tested.
In \cref{fig:H_vor_pec,fig:zeta_vor_pec}, it is clear that within the recommended ranges of $H=0.3\sim 0.7$ and $\zeta = 1.2\sim1.6$, changes in vortices of different types with increasing $\mathrm{Re}$ follow the same consistent trend with different $H$ and $\zeta$.
Clear disruption to the trend is only observed in cases well out of these ranges: most notably $H=0.2$ in \cref{fig:H_Nv} and $\zeta=0.8$ and $3.0$ in \cref{fig:zeta_Nv}.
Quantitative magnitudes of the profiles do depend on $H$ and $\zeta$, which is very much expected. As illustrated in \cref{fig:DNS_vor_det}, adjusting these parameters inevitably changes the lengths of vortex branches and legs, to which the classification scheme of \cref{fig:class_flow} is very sensitive: missing one axis-point at the branch end could result in a vortex being classified as a hook rather than hairpin, or even a fragment rather than a vortex.
Nevertheless, quantitative differences between curves are significantly smaller (mostly contained within a few percentage points) in the ranges of $H=0.3\sim 0.7$ and $\zeta = 1.2\sim1.6$, compared with those out of these ranges.
The physical observation made in \cref{fig:vor_num} are completely robust when acceptable $H$ and $\zeta$ are used.

\begin{figure}
	\centering
	\includegraphics[width=.95\linewidth, trim=0mm 0mm 0mm 0mm, clip]{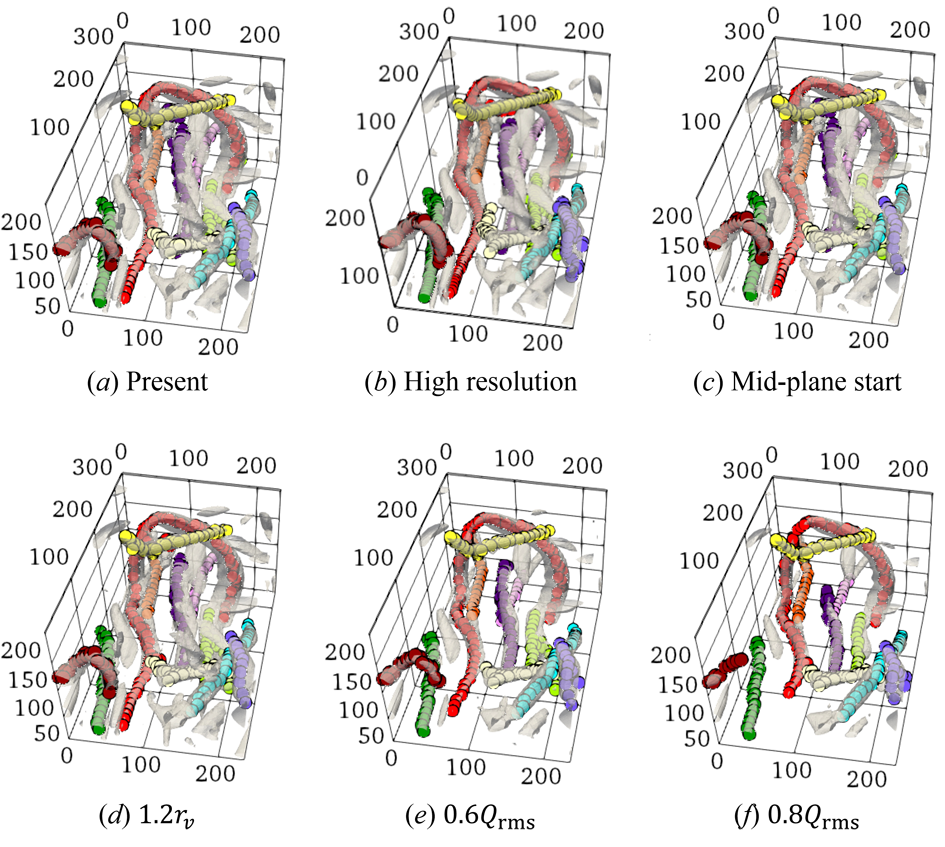}
	\caption{\RevisedText{Effects of parameters and settings on VATIP tracking results in a representative flow region at $\mathrm{Re}_\tau=169.71$ (note: panel (f) is out of the recommended range of $H$). Dots represent axis-points identified by VATIP (different colors for different vortices) and gray tubes are the isosurfaces of $Q=HQ_\text{rms}$ (same $H$ used in VATIP). Partial vortices (with parts extending out of the view box) are not included in VATIP tracking.}}
	\label{fig:DNS_vor_det}
\end{figure}

The robustness of VATIP is most clearly demonstrated in \cref{fig:DNS_vor_det} where different parameters and settings are tested and compared for a same flow region with various vortex configurations. Compared with the standard case (panel (a)) with $H=0.4$ and $\zeta=1.5$, changing $\zeta$ to $1.2$ (panel (d)) or changing $H$ to $0.6$ (panel (e)) brings little noticeable difference.
In panel (f), $H$ is further increased to $0.8$ (beyond the recommended range), which only causes the identified vortex tubes to shrink in size, and VATIP still faithfully captures their axis-lines.
Note that the brown vortex at the lower-left corner has changed from a curved shape to a linear shape at $H=0.8$ owing to the erosion of one of its legs, which well illustrates how changing parameters affect the number of vortices classified into each category (including fragments). These changes, however, do not reflect the reliability of VATIP itself.
We have also doubled the resolutions in all dimensions (panel (b); the flow field is interpolated to the finer grid before VATIP analysis) and changed the starting search plane in $x$ and $z$ directions (first steps in subroutines 1 and 2 of \cref{fig:track_flowchart}; with no translational symmetry, $y$-direction searches always start from the walls -- also see discussion in \cref{Sec:limitation}) from the first planes ($x=0$ or $z=0$) to the middle planes ($x=L_x/2$ or $z=L_z/2$). Both do not lead to any discernible difference in the tracking outcome.
This observation is general: at $\mathrm{Re}_\tau=169.71$, the average number of vortices in each flow domain (excluding fragments) identified by VATIP $N_\text{v}=2178$, whereas the high resolution case has $N_\text{v}=2193$ and the mid-plane start case has $N_\text{v}=2182$. In both cases, the difference is way less than 1\%.

\begin{figure}
	\centering
	\includegraphics[width=.95\linewidth, trim=0mm 0mm 0mm 0mm, clip]{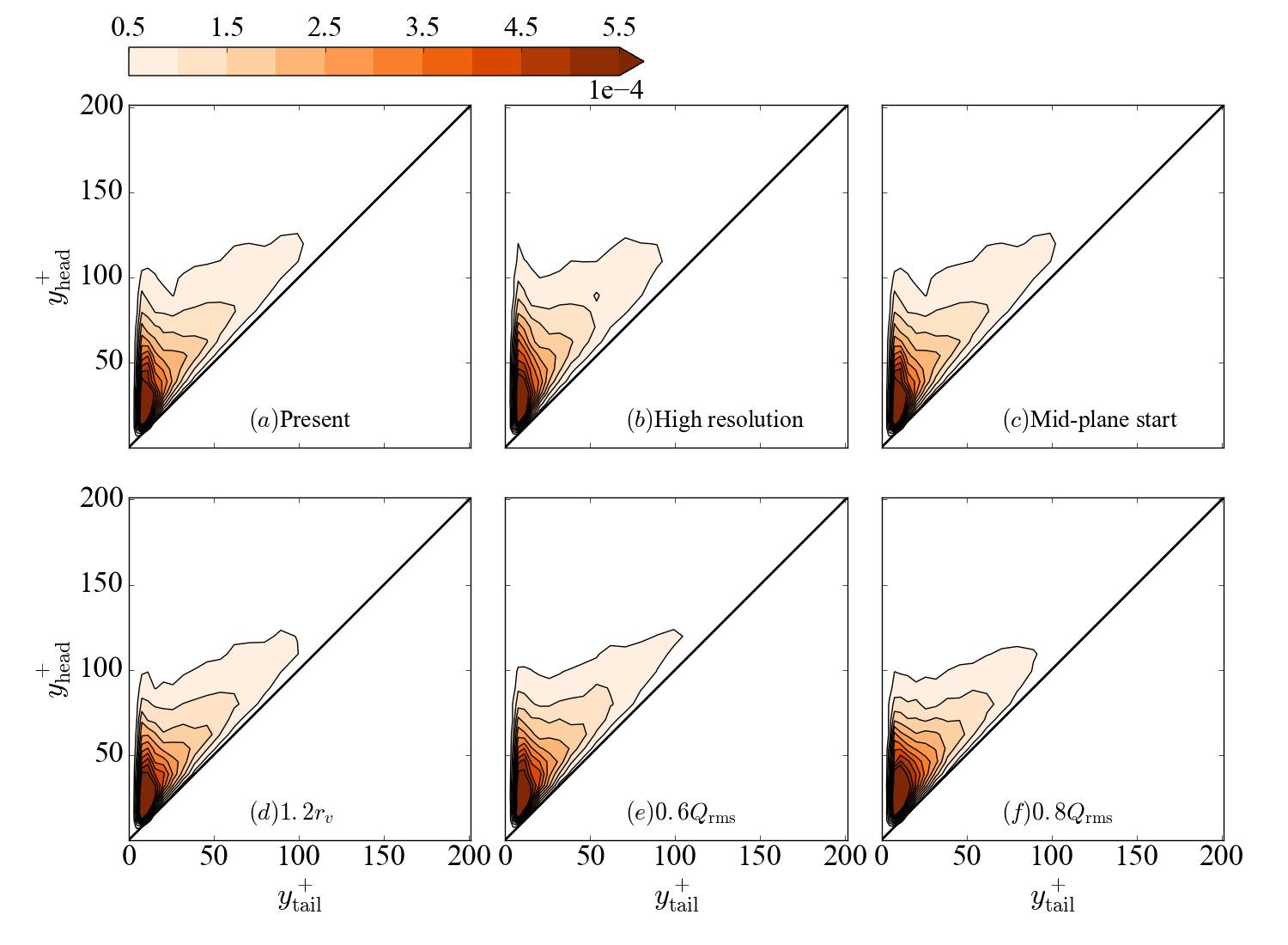}
	\caption{\RevisedText{Joint PDF between the $y^+$ positions of the tail and the head of streamwise vortices at $\mathrm{Re}_\tau=169.71$ under different VATIP parameters and settings (note: panel (f) is out of the recommended range of $H$).}}
	\label{fig:JPDF_Sens_stream}
\end{figure}

\begin{figure}
	\centering
	\includegraphics[width=.95\linewidth, trim=0mm 0mm 0mm 0mm, clip]{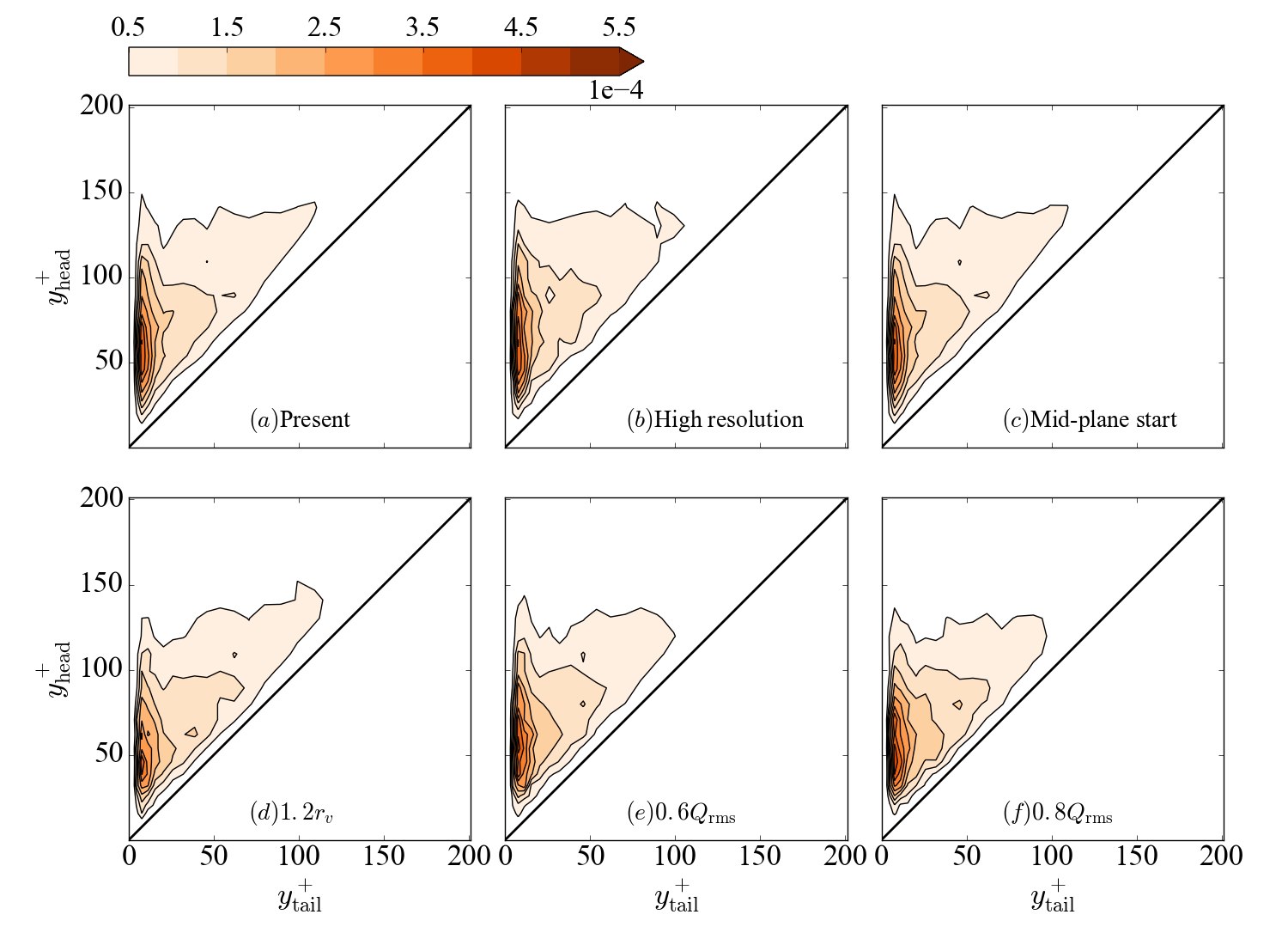}
	\caption{\RevisedText{Joint PDF between the $y^+$ positions of the tail and the head of three-dimensional vortices at $\mathrm{Re}_\tau=169.71$ under different VATIP parameters and settings (note: panel (f) is out of the recommended range of $H$).}}
	\label{fig:JPDF_Sens_curved}
\end{figure}

Finally, echoing the observations in \cref{fig:vor_height1,fig:vor_height2}, we examine the effects of VATIP parameters and settings on vortex conformation statistics -- this time at $\mathrm{Re}_\tau=169.71$.
Despite the small quantitative differences -- which, as discussed above, are inevitable as vortices are classified based on quantitative metrics, the qualitative picture is well preserved for all cases shown, including the $H=0.8$ case which is out of the recommended range.
Similar as the earlier observations at other $\mathrm{Re}_\tau$, the distribution of streamwise vortices is highly concentrated in the lower-left corner corresponding to the buffer layer.
Weaker, but noticeable, concentration bands extend along both the diagonal and the ordinate, reflecting the flat-lying and lifted-up streamwise vortices, respectively.
By contrast, three-dimensional vortices are predominantly lifted-up with their concentration peak found well in the log-law layer.
Changing resolution or the starting plane shows little effect on these distributions, whereas adjusting $H$ or $\zeta$ more directly affects vortex classification and thus causes some subtle changes in the contour shapes, especially at low density levels.


\subsection{Discussion: limitations and future development}\label{Sec:limitation}
Recall from \cref{sec:VATIP}: the algorithm of VATIP is built on the premise that vortices are wall-generated, which start with segments or ``legs'' that align along the $x$ direction (most often in the buffer layer but the algorithm does not impose this restriction) and can lift up to higher-$y^+$ layers to bend, curve, or branch.
VATIP always initiates the propagation points in the $x$-lying legs and later allows them to move away from the walls (in the $y$-direction search) and swing sideways (in the $z$-direction search). For canonical hairpins, the axis-lines initiate from both streamwise legs which rise at the downstream end and merge in the middle along the $z$-direction. Branched vortices are found in a similar manner with one propagation point planted in each streamwise leg and the growing legs (or more appropriately for the branch type -- arms) will eventually merge after a limited number of iterations.
As demonstrated above (comparing the $Q$-isosurfaces and VATIP-identified axis-lines), this algorithm faithfully captures nearly all vortices identified by the $Q$-criterion in wall turbulence for $\mathrm{Re}_\tau\leq400$ tested in this study.

Recent evidences have indicated that at high $\mathrm{Re}$ and large $y^+$, vortices can be generated in the absence of wall interaction~\citep{del2006self,jimenez2013near}. These vortices can deviate significantly from this premise: they are nearly isotropic (segments are equally likely to align with any direction) and often highly branched (multiple arms with complex connection topology).
The current algorithm would not perform as well on those structures. First, the requirement on initialization in the $x$-direction search only will undoubtedly bias the resulting axis-line to have better sampling of the $x$-lying segments. For instance, in a strictly $y$-aligned segment with no connection to any $x$-segment at the bottom, the current algorithm would still capture a point in the middle being an $yz$-planar maximum. It would skip the $x$-direction propagation (for the lack of other $x$-axis-points) and the next step of $y$-search would propagate away from the wall only. The end result is missing one part of the segment below the initial point.
Second, with only one propagation point in each initial $x$-segment, the algorithm would struggle with highly branched configurations because the propagation will only pick one of the directions to proceed after each junction.
This is not much a problem for wall-generated branched vortices at moderate $\mathrm{Re}$ (the focus of this study), because these structures mostly consist of a few conjoining arms, each of which can be traced back to a streamwise leg (i.e., starfish- or octopus-shaped). The current algorithm has been shown to capture these structures well.
However, for detached structures at higher $\mathrm{Re}_\tau$ and higher $y^+$ (see, e.g., fig.~6 of \citet{del2006self}) with more complex configurations, it will likely miss some of the bridges connecting different arms, if they do not happen to align in the $x$ direction, and falsely break them into pieces.

There seems to be a easy remedy in sight, that as long as we relax these constraints to allow a truly multidirectional tracking -- i.e., axis-lines are initiated in all three dimensions and propagation is allowed to follow all branches after each junction, it should be able to capture these isotropic and highly-branched structures.
The problem, however, is its insurmountable side effect: relaxation of the current constraints will inevitably lead to massive false identification and false connection, making the tracking result next to meaningless.
There are two major sources for this problem. First, not all planar $Q$ maxima belong to a vortex axis. Consider a simple linear streamwise vortex as an elongated ellipsoid, the true axis-line aligns with its major axis and consist of $yz$-planar $Q$ maxima. However, the minor axes also contain $Q$ maximum points (in $xy$- and $xz$-planes), which do not belong to any vortex axis-line. This is further compounded by fluctuations in the $Q$ fields, which may create $Q$ maxima unrelated with any actual vortex.
The total number of planar maximum points identified in an $L_x^+\times L_z^+=4000\times 800$ flow domain ranges from $\sim 80,000$ to $\sim 350,000$ (for the lowest and highest $\mathrm{Re}_\tau$ tested, respectively). Only $30\%\sim 40\%$ of them are included in the final axis-lines (counting both vortices and fragments).
Second, in flow fields densely populated by vortices, close encounters between axis-lines of separate vortices are common: spatial proximity does not necessarily indicate connectivity.
The current algorithm takes advantage of the fact that, despite the overall complexity of vortex configuration and distribution, wall-generated vortices can be traced back to the near-wall region where their legs are regularly aligned (largely in parallel) in the streamwise direction. Regularity in their distribution pattern makes tracking easier and connection is usually unambiguous. This is the rationale behind the choice of axis-line initiation in the $x$-search.
Continued propagation in other ($y$ and $z$) directions minimizes false inclusion of points and false connection with other vortices by requiring the new segments to be natural extensions from the growing axis-line. 
By contrast, a general multi-initiation and multi-directional algorithm would indiscriminately connect any points in the vicinity of a growing end.
Extension of the VATIP algorithm to detached vortex structures with no preference to the $x$ direction and more complex branch configurations calls for new physics-based constraints to be incorporated, which is a focus of our future research.

Another potential challenge of extending the method to higher $\mathrm{Re}$ is the determination of parameters. \Cref{Sec:ParamAnalys} thoroughly examined the effects of the adjustable parameters $H$ and $\zeta$ and proposed their recommended ranges of use based on the balance between minimizing false connection and minimizing false disintegration or truncation of vortex axis-lines.
For higher $\mathrm{Re}$, structures in the bulk regions (higher $y^+$) become non-trivial. \citet{del2006self} showed that a lower $H$ is required to capture these detached structures because of the overall lower turbulent intensity in those regions. A $y^+$-dependent $H$ was thus proposed for vortex identification in that study.
It is likely that for VATIP, a similar approach needs to be taken for both $H$ and $\zeta$. Determining the dependence of these parameters on $y^+$ will require trial and error. Moreover, whether a ``sweet-spot'' range still exists for these parameters at the high-$\mathrm{Re}$ and high-$y^+$ regime remains to be seen.
Finally, we note that as a brand new method, its future application and testing in broader parameter regimes and systems will be essential for its continued improvement and generalization. In this sense, the development of VATIP itself is an ``iterative'' process that requires the experience and feedback from its application.  
}

\section{Conclusions}\label{Sec_conclude}
In this study, a new method has been proposed for the identification and extraction of three-dimensional complex vortices from turbulent flow fields.
This method, named VATIP, connects points of vortex axis-lines using the \RevisedText{cone-detective} criterion of \citet{jeong1997coherent} and propagates the growing axis over all three spatial dimensions iteratively in order to accommodate various types of vortex topologies.
Transient simulation based on streak instability (STG) is performed to generate flow fields featuring streamwise, titled/curved, and hairpin vortices and the method is shown to successfully capture all these types.
In addition, a new procedure is proposed to classify the axis-lines obtained by VATIP into different topological types commonly observed in wall turbulence, including quasi-linear vortices, hairpins, hooks (asymmetric/incomplete hairpins), and various branched types.
\RevisedText{%
Tracking outcome from VATIP is shown to be robust with changing parameters and settings. For both adjustable parameters ($H$ and $\zeta$ for the vortex scalar identifier and search cone size, respectively), suitable parameter ranges are identified.
The method is the first that directly extracts the individual axis-lines of typical three-dimensional vortices found in turbulent near-wall layers. Future work will focus on extending this method for complex isotropic vortex configurations at higher $\mathrm{Re}$ and in outer layers, to which the current method is not applicable.
}%

VATIP is applied to analyze the vortex configurations and statistics in statistical turbulence (from DNS) at three different $\mathrm{Re}$, where vortices of all types are successfully identified.
The results show that the streamwise vortex length $l^+_x$ is insensitive to $\mathrm{Re}$ with the distribution nearly identical between all three $\mathrm{Re}$ tested.
The spanwise width $l^+_z$, however, has higher average values at higher $\mathrm{Re}$ as a result of the higher fraction of wide vortices.
The number of vortices increases with $\mathrm{Re}$ (for the same domain size in inner units). Quasi-streamwise vortices are dominant in the low-to-moderate range of $\mathrm{Re}$ ($\mathrm{Re}_\tau$ from $84.85$ to $400$) tested, but their number fraction decreases with $\mathrm{Re}$.
Complex three dimensional vortices of all shapes (hairpins, hooks, and branches) become more prevalent at higher $\mathrm{Re}$, which accounts for the increasing frequency of large $l^+_z$ values.
The number of symmetric hairpins and branched vortices grow faster than asymmetric vortices (hooks), suggesting that the latter is likely an incomplete version of full hairpins occurring more often at lower $\mathrm{Re}$.
Quasi-streamwise vortices populate the buffer layer and the lower log-law layer whereas hairpins and other three-dimensional vortices dominate higher layers (although the legs of these vortices still stretch down to the buffer layer).
The latter is also more likely to be found in a lifted-up state and the head of those vortices can rise to a broad range of distances from the wall.

\RevisedText{%
Clustering analysis is applied to VATIP results for understanding vortex organization patterns. Well-defined vortex clusters consisting of $O(10)\sim O(100)$ individual vortices are consistently identified. These clusters appear at regions with high Reynold shear stress and are reminiscent of the large-scale motions previously observed in the literature. They have a streamwise length scale of $500\sim1500$ wall units, which stays roughly constant (in inner units) for the $\mathrm{Re}$ range tested.

The current study focused on the static analysis of vortex conformation and distribution. On the subject of hairpin vortices, which is heatedly debated in the literature, it reveals the definitive evidence for the existence of such structures in the statistically-steady turbulence of channel flow.
However, it is also shown that canonical hairpin vortices with highly symmetric legs (as reported in the transient boundary layer flow by \citet{wu2009direct}) remain rarities at least within the range of $\mathrm{Re}_\tau\leq400$ tested. They are greatly outnumbered by their asymmetric (hooks) and highly branched mutants.
These latter types seem to have the same level of lift-up and may be formed by the incomplete development of hairpins (for hooks) or their coalescence with other structures (for branches).
Compared with canonical hairpins, branched vortices seem to be equivalently effective at binding vortices into clusters owing to their similar spanwise dimensions, whereas hooks are more similar to streamwise vortices in this aspect.
Important questions on the role of these general hairpin-like vortices on turbulence dynamics, especially, whether they are the cause for or consequence of turbulence generation, cannot be answered until a dynamical tracking approach (such as that of \citet{lozano2014time}) is integrated with VATIP.
}%

The success of VATIP provides access to the detailed statistics on the configuration, topology, and distribution of vortices in \RevisedText{near-wall turbulence}. It thus offers a powerful tool for the study of vortex dynamics and auto-regeneration mechanism of turbulence, as well as other areas such as the vortex development during the bypass transition and the changing vortex dynamics in turbulent drag reduction.

\begin{acknowledgments}
The authors acknowledge the financial support from the Natural Sciences and Engineering Research Council of Canada (NSERC; No.~RGPIN-2014-04903) and the allocation of computing resources awarded by Compute/Calcul Canada.
The computation is made possible by the facilities of the Shared Hierarchical Academic Research Computing Network (SHARCNET: \texttt{www.sharcnet.ca}).
\RevisedText{%
We also express our gratitude towards John~F.~Gibson, Hecke~Schrobsdorff, Tobias~Kreilos, and  Tobias~M.~Schneider for the DNS code.
}%
LX acknowledges the National Science Foundation Grant No.~NSF~PHY11-25915, which partially supported his stay at the Kavli Institute for Theoretical Physics (KITP) at UC Santa Barbara. 
LZ acknowledges the European Research Council H2020 program (ERC-2014-ADG ``COTURB'') which supported his participation in the Third Madrid Summer School on Turbulence.
\end{acknowledgments}

\bibliographystyle{jfm}
\bibliography{Zhu_bibtex,FluidDyn,Polymer}

\end{document}